\pdfoutput=1
\documentclass[11pt, a4paper]{article}

\usepackage{jheppub}
\usepackage{tikz}
\usepackage{relsize}
\usepackage{subcaption}
\usepackage{amsfonts}
\usepackage{mleftright}
\mleftright

\newcommand{\Sf}{\mathfrak{S}}
\newcommand{\Dop}[2]{\Delta_{#1}^{(#2)}}
\newcommand{\Dopj}[3]{\Delta_{#2;\,#1}^{(#3)}}
\newcommand{\modr}[1]{\left[#1\right]}

\newcommand{\diag}{\mathop{\mathrm{diag}}\nolimits}
\newcommand{\Tr}{\mathop{\mathrm{Tr}}\nolimits}
\newcommand{\Res}{\mathop{\mathrm{Res}}\nolimits}
\newcommand{\SU}{\mathrm{SU}}
\newcommand{\SO}{\mathrm{SO}}
\newcommand{\U}{\mathrm{U}}
\newcommand{\Z}{\mathbb{Z}}
\newcommand{\C}{\mathbb{C}}
\newcommand{\T}{\mathbb{T}}
\newcommand{\nn}{\nonumber}

\newcommand{\CI}{{\cal I}}
\newcommand{\CK}{{\cal K}}
\newcommand{\CN}{{\cal N}}
\newcommand{\CS}{{\cal S}}
\newcommand{\CT}{{\cal T}}
\newcommand{\CW}{{\cal W}}

\newcommand{\pc}{\chi}
\newcommand{\po}{\mathrm{o}}

\usetikzlibrary{calc, positioning, decorations, decorations.markings, arrows}
\tikzset{
  baseline={([yshift=-0.75ex]current bounding box.center)},
  node distance=8pt,
}
\newlength{\qsep}
\tikzset{
  x=\qsep, y=\qsep,  font=\smaller,
  >=latex,
  node/.style={draw, fill=white, shape=circle, minimum size=12pt, inner sep=0pt},
  gnode/.style={node},
  fnode/.style={node, shape=rectangle},
  q-/.style={-},
  q->/.style={->, shorten >=1pt, font=\smaller[2]},
  eq-/.style={double, double distance=2pt},
  strike out/.style={
    postaction=decorate,
    decoration={
      markings,
      mark=at position 0.5 with {
        \draw[-] (-2pt, -3pt) -- (2pt, 3pt);
      }
    }
  },
  vev/.style={strike out},
  minp/.style={draw, shape=circle, minimum size=3pt, inner sep=0pt, font=\tiny},
  maxp/.style={minp, double, double distance=1pt, fill=white, minimum size=5pt},
}

\usepackage{etoolbox}
\makeatletter
\patchcmd{\maketitle}{\itshape\small}{\itshape\footnotesize}{}{\PackageWarning{draft}{affiliation font patch failed}}
\makeatother

\title{Surface defects on lens space and transfer matrices}

\author[a,b]{Kazunobu Maruyoshi,}
\author[c,d]{Toshihiro Ota,}
\author[d,e,f,g,h]{and Masahito Yamazaki}

\affiliation[a]{Faculty of Science and Technology, Seikei University\\
  3--3--1 Kichijoji-Kitamachi, Musashino-shi, Tokyo 180--8633, Japan}
\affiliation[b]{Nambu Yoichiro Institute of Theoretical and Experimental Physics, Osaka Metropolitan University\\
  3--3--138, Sugimoto, Sumiyoshi-ku, Osaka 558-8585, Japan}
\affiliation[c]{CyberAgent AI Lab, Shibuya, Tokyo 150--0002, Japan}
\affiliation[d]{Center for Interdisciplinary Theoretical and Mathematical Sciences, RIKEN, Saitama
351-0198, Japan}
\affiliation[e]{Department of Physics, Graduate School of Science, University of Tokyo, Hongo 7-3-1,
Bunkyo-ku, Tokyo 113-0033, Japan}
\affiliation[f]{Kavli Institute for the Physics and Mathematics of the Universe, UTIAS, University
of Tokyo, Kashiwa, Chiba 277-8583, Japan}
\affiliation[g]{Trans-Scale Quantum Science Institute, University of Tokyo, Hongo 7-3-1, Bunkyo-ku,
Tokyo 113-0033, Japan}
\affiliation[h]{Center for Data-Driven Discovery (CD3), Kavli IPMU, University of Tokyo, Chiba
277-8583, Japan}

\emailAdd{maruyoshi@st.seikei.ac.jp}
\emailAdd{ota\_toshihiro@cyberagent.co.jp}
\emailAdd{masahito.yamazaki@ipmu.jp}

\abstract{
We study surface defects in four-dimensional $\CN=2$ theories of class $\CS$ and $\CN=1$ theories of class $\CS_k$ placed on $L(r,1)\times S^1$.
The lens space index depends on the fugacities $p$ and $q$, and the effect of the defects on it is captured by difference operators acting on the fugacities and holonomies of flavor symmetries.
For theories of type $A_1$ we determine in closed form the difference operators for the basic defects, labeled by $p^r$, $q^r$ and $pq$,
and show that for odd $r$ there exist additional elementary defects, whose operators shift the holonomy by one half modulo $r$.
All the operators commute, and the elementary operators generate the others.
For even $r$ the same holds once the holonomy is extended to a twisted sector.
We show that the elementary operators are traces of $L$-operators whose entries generate a Sklyanin algebra, with nome $q^r$ and shift $pq$.
The lens space enters only through the parameters of the algebra and through its representation, which is realized on a doubled holonomy label.
The $p$- and $q$-type algebras form a lens space version of the elliptic modular double.
For class $\CS_k$ theories we propose, and test in several ways, that the defect operators are transfer matrices of inhomogeneous spin chains built from these $L$-operators.
Throughout, we use the TQFT structure of the lens space index, which follows from S-duality, both to fix and to check the operators.
}
\preprint{RIKEN-iTHEMS-Report-26}

\begin{document}
\maketitle
\flushbottom

\section{Introduction}
  The superconformal index \cite{Romelsberger:2005eg,Kinney:2005ej,Dolan:2008qi,Spiridonov:2009za}
  is a supersymmetric partition function on $S^3\times S^1$ and
  has been a powerful tool to study the
  strongly-coupled dynamics of four-dimensional $\CN\ge 1$ theories and dualities among them.
  For four-dimensional $\CN=2$ theories of class $\CS$ \cite{Gaiotto:2009we,Gaiotto:2009hg}, the index can be identified with the partition function of a two-dimensional topological quantum field theory (TQFT) on the Riemann surface defining the theory \cite{Gadde:2009kb,Gadde:2011ik} (see \cite{Rastelli:2014jja} for a review).

  We can enrich this story by incorporating surface defects.
  As shown in \cite{Gaiotto:2012xa}, a surface defect can be engineered by coupling the theory to a larger one and giving a position-dependent vacuum expectation value (vev) to a baryonic operator. This amounts to taking a residue of the index with respect to a flavor fugacity,
  and the net effect of the defect is described by a difference operator acting on the index (see also \cite{Gadde:2013dda,Alday:2013kda,Bullimore:2014nla}).
  These difference operators are elliptic generalizations of the Macdonald operators \cite{Gadde:2011uv} (the elliptic Ruijsenaars--Schneider operators \cite{Ruijsenaars:1986pp}).
  They commute with each other and are simultaneously diagonalized by the wave functions of the TQFT \cite{Razamat:2013qfa,Nazzal:2023wtw}.
  Surface defects of this type have also been studied in the $\CN=1$ theories of class $\CS_k$ \cite{Gaiotto:2015usa,Ito:2016fpl,Arutyunov:2026yjy} (see also \cite{Chen:2020jla} for the corresponding elliptic quantum curves) and in other $\CN=1$ theories obtained from six dimensions \cite{Nazzal:2018brc,Razamat:2018zel,Nazzal:2021tiu,Nazzal:2023bzu}.
  In \cite{Maruyoshi:2016caf} it was found, for brane tiling models including the theories of class $\CS$ and $\CS_k$, that the difference operators are transfer matrices constructed from $L$-operators,
  which for $\SU(2)$ flavor groups are identified with Sklyanin's $L$-operator of the eight-vertex model \cite{Baxter:1972hz,Baxter:1982zz}.
  See also \cite{Yagi:2015lha} for the underlying correspondence between brane tilings and integrable lattice models, \cite{Yagi:2017hmj} for the relation to elliptic quantum groups \cite{Felder:1994be}, and \cite{Costello:2018txb,Maruyoshi:2020cwy} for the analogous description of line defects.

  A natural generalization of the superconformal index is the index on $L(r,1)\times S^1$, the lens space index \cite{Benini:2011nc,Yamazaki:2013nra,Nishioka:2014zpa,Kels:2017toi,Fluder:2017oxm,Jejjala:2022lrm,Pittelli:2024ugf,Imamura:2017wdh}.
  The lens space index is sensitive to the global structure of the gauge group \cite{Aharony:2013hda} and contains information on the holonomies around the torsion cycle of the lens space \cite{Razamat:2013opa,Amariti:2019but,Amariti:2022cxh}.
  It also has a TQFT structure when applied to class $\CS$ theories \cite{Benini:2011nc,Razamat:2013jxa,Alday:2013rs} (see \cite{Rastelli:2014jja} for a review),
  and its building block, the lens elliptic gamma function, is a basic ingredient of the ``rarefied'' elliptic hypergeometric functions \cite{Spiridonov:2016uae,Kels:2017toi,Spiridonov:2019uuw,Sarkissian:2018ppc}, of the lens generalization of the elliptic discrete Painlev\'e equation \cite{Kels:2018xnf}, and of solutions to the Yang--Baxter equation \cite{Kels:2015bda,Gahramanov:2016ilb,Kels:2017toi,Kels:2017vbc}, as studied in the context of the Gauge/YBE correspondence \cite{Yamazaki:2012cp,Terashima:2012cx,Yamazaki:2013nra} (see also \cite{Bazhanov:2010kz,Spiridonov:2010em,Bazhanov:2011mz} and, for a review, \cite{Yamazaki:2018xbx}).

  Surface defects in the lens space index of class $\CS$ theories were studied in \cite{Razamat:2013jxa}.
  There it was shown, using S-duality, that the index is acted upon by commuting difference operators,
  which act on the lattice of holonomies and were identified as matrix-valued generalizations of the elliptic Ruijsenaars--Schneider operators.
  In \cite{Razamat:2013jxa} the holonomy of the $\U(1)$ symmetry whose fugacity is tuned to the pole was set to zero.
  The operator for the pole at $\alpha^2=tpq$ was given explicitly for type $A_1$,
  and the operators for the poles at $\alpha^2=tq^{rn}$ were studied in the limit $p\to0$, where they are related to the Cherednik operators of the double affine Hecke algebra \cite{Cherednik:1995DAHA}.

  In this paper we study surface defects on the lens space, also turning on the holonomy of this $\U(1)$ symmetry.
  We follow the construction of \cite{Gaiotto:2012xa} and compute the residues of the lens space index.
  One of the main technical difficulties is that the holonomies enter the lens elliptic gamma function through their residues modulo $r$,
  so that the computation involves many different cases.
  We organize the computation by writing the coefficients of the difference operators in terms of lens elliptic gamma functions, and reduce them systematically to theta functions.
  We check our results against the TQFT structure of the index:
  the difference operator acting on one puncture of the free trinion must agree with the one acting on another puncture.

  Our main results are as follows.
  First, we find that on the lens space the surface defects carry, in addition to the labels specifying the pole in the fugacity $\alpha$,
  the holonomy $\mu\in\Z_r$ of the $\U(1)_\alpha$ symmetry (set to zero in \cite{Razamat:2013jxa}), and that a pole at $\alpha^2 = tp^aq^b$ contributes only if $2\mu\equiv b-a$ modulo $r$.
  Second, we determine the difference operators for the defects $\CS^{(1,0,0)}$, $\CS^{(0,1,0)}$ and $\CS^{(0,0,1)}$ associated with the poles $\alpha^2=tp^r$, $tq^r$ and $tpq$ for theories of type $A_1$
  (see \eqref{eq:s100}, \eqref{eq:s010} and \eqref{eq:s001}).
  Third, for odd $r$ we find elementary defects $X_p$ and $X_q$ with $\mu\neq 0$, given in \eqref{eq:Xp} and \eqref{eq:Xq},
  whose operators shift the holonomy by ``one half'', namely by $(r\pm1)/2$.
  Fourth, we find that all the operators commute, and that they are generated by the elementary operators:
  $\CS^{(0,0,1)}=(pq/t)^{1/r}X_qX_p$, while $\CS^{(1,0,0)}$ is a polynomial of degree $r$ in $X_p$ determined by a deformed $\SU(2)$ fusion rule \eqref{eq:fusion}.
  As a consequence, the naive generalization $(\CS^{(0,0,1)})^r\propto\CS^{(0,1,0)}\CS^{(1,0,0)}$ of the factorization property known for $r=1$ does not hold for $r>1$,
  except for the extremal terms of the operators.
  For even $r$ the same statements hold with the elementary operators acting on an extended space with a twisted holonomy sector,
  up to the shift $U$ of the $\SU(2)$ holonomy by the center, which enters the fusion formula.
  Fifth, we construct a Sklyanin algebra \cite{Sklyanin:1982tf} on the lens space \eqref{eq:lens-Sklyanin}, whose $L$-operator reproduces the elementary defects.
  The algebra closes only after the holonomy label is doubled from $\Z_r$ to $\Z_{2r}$, and the $p$- and $q$-type algebras commute, forming a lens space elliptic modular double.
  Finally, for class $\CS_k$ theories we propose that the defect operators are transfer matrices built from $k$ copies of the $k=1$ lens $L$-operators \eqref{eq:Sk-lens}.
  For $r=1$ this follows from the TQFT consistency condition together with the residue computation \eqref{eq:Sk-vertex}.
  On the lens space we test the proposal with the TQFT consistency condition and, for $k=2$, with a numerical evaluation of the residue for $r=1,2,3$,
  which also fixes the normalization \eqref{eq:Sk-norm}.

  The rest of this paper is organized as follows.
  In section~\ref{sec:lens} we review the lens space index and set up our notation.
  In sections~\ref{sec:class-S}--\ref{sec:sklyanin} we study the $\CN=2$ theories of class $\CS$:
  in section~\ref{sec:class-S} we describe the construction of surface defects on the lens space and present the difference operators, in section~\ref{sec:algebra} we study their algebraic structure,
  and in section~\ref{sec:sklyanin} we construct the Sklyanin algebra associated with the lens space.
  In section~\ref{sec:class-Sk} we turn to class $\CS_k$ theories with general $k$.
  We conclude in section~\ref{sec:summary}.
  Appendix~\ref{app:formulas} collects definitions and useful formulas, appendix~\ref{app:numerics} explains our numerical checks,
  appendix~\ref{app:factorization} contains the proof of the factorization \eqref{eq:factor-odd}, appendix~\ref{app:fusion} the proof of the fusion formula \eqref{eq:S100-fusion}, and 
  appendix~\ref{app:Sk} outlines the residue computation for the class $\CS_k$ surface defect.

\section{Lens space index}
\label{sec:lens}
  In this section we review the superconformal index on $L(r,1) \times S^1$, which we refer to as the lens space index.
  The lens space $L(r,1)$ is the orbifold of $S^3$ defined by the identification of its complex coordinates
    \begin{align}
    (z_1, z_2) \sim (e^{\frac{2 \pi i}{r}}z_1, e^{-\frac{2 \pi i}{r}}z_2)\;,
    \end{align}
  where $|z_1|^2 + |z_2|^2=1$. 
  The lens index of a four-dimensional $\CN=1$ superconformal theory is defined as the trace over the Hilbert space on the lens space $L(r,1)$ \cite{Benini:2011nc,Razamat:2013jxa,Razamat:2013opa}
    \begin{align}
    \CI
     =     \Tr (-1)^F p^{j_2 + j_1+\frac{R}{2}} q^{j_2-j_1 +\frac{R}{2}} \prod_{i} a_i^{f_i}\;,
        \label{eq:lensindex}
    \end{align}
  where $j_{1,2}$ are the Cartans of the rotation group $\SU(2)_1 \times \SU(2)_2 \sim \SO(4)$, $a_i$ are the fugacities of the flavor symmetries whose charges are $f_i$, and $R$ is the $\U(1)_R$ charge.

  The lens index \eqref{eq:lensindex} receives contributions only from the states satisfying a BPS condition on their quantum numbers.
  It follows that the index does not depend on the continuous parameters of the theory, in particular on the gauge coupling.
  Thus the index can be calculated from the matter content of the theory whenever it has a weakly coupled Lagrangian description.
  
  In contrast to the superconformal index (namely the $r=1$ case), the lens index also depends on the discrete holonomy $V$ of the gauge/flavor symmetry, since $\pi_1(L(r,1))=\Z_r$.
  For an $\SU(N)$ group,
    \begin{align}
    V 
     =     {\rm diag} (e^{\frac{2 \pi i m_1}{r}}, e^{\frac{2 \pi i m_2}{r}}, \ldots, e^{\frac{2 \pi i m_N}{r}})
    \end{align}
  where $m_i$ are integers with $\sum_{i=1}^N m_i \equiv 0$ mod $r$.
  In the following we refer to $m_i$ as holonomies of the gauge/flavor symmetry.

  Let us consider an $\CN=1$ chiral multiplet transforming in the bi-fundamental $(\bar{\square}, \square)$ representation of the $\SU(N)_x \times \SU(N)_z$ global symmetry.
  We represent the lens index of this bi-fundamental multiplet by the following quiver diagram,
  and it is given by
    \begin{align}
      \begin{tikzpicture}
      \node[fnode] (x) at (0,0) {$\mathbf{x}$};
      \node[fnode] (z) at (1,0) {$\mathbf{z}$};
      \draw[q->] (x) -- node[above] {$a$} (z);
      \end{tikzpicture}
        ~ =     \prod_{i, j} \Gamma (x_i^{-1} z_j a, -m_i + \tilde{m}_j)\;,
    \end{align}
  where we have decorated the quiver diagram by $\mathbf{x}$, $\mathbf{z}$ and $a$,
  which are the fugacities of the $\SU(N)_x$, $\SU(N)_z$ and $\U(1)$ symmetries, respectively.
  Here $m_i$ and $\tilde{m}_j$ are the holonomies for the $\SU(N)_x \times \SU(N)_z$ symmetry.
  As in appendix~\ref{app:formulas}, the lens elliptic gamma function is defined by
    \begin{align}
      \Gamma (z, m)
        :=     \Gamma_0(z,m;p,q) \Gamma^p(z,m)\Gamma^q(z,m)\;,
    \end{align}
  where
    \begin{align}
      \Gamma_0(z, m; p,q)
        =     \left( \frac{pq}{z^2} \right)^{\frac{m(r-m)}{4r}} \gamma_0\;,
    \end{align}
  and $\Gamma^{p,q}(z,m)$ are two elliptic gamma functions \eqref{eq:gammapq}.
  Here $m$ stands for its residue $[m]\in\{0,\dots,r-1\}$, and $\gamma_0$, given in appendix~\ref{app:formulas}, depends only on $p$, $q$ and $m$.
  Although $\gamma_0$ is independent of $z$, it cannot be dropped in the following,
  since the difference operators we consider shift the holonomies.

  The index of an $\SU(N)$ vector multiplet is denoted by the circle
    \begin{align}
      \begin{tikzpicture}
      \node[gnode] (z) at (0,0) {$\mathbf{z}$};
      \end{tikzpicture}
      ~ = \CI_V (z, m_i)
        =     \frac{\Xi^{N-1}}{N!} \frac{1}{\prod_{1 \leq i < j \leq N} \Gamma(z_i z_j^{-1}, m_i - m_j) \Gamma(z_i^{-1} z_j, -m_i + m_j)}\;,
    \end{align}
  where $\Xi$ is defined by
    \begin{align}
      \Xi
        =     (p^r; p^r)_\infty (q^r; q^r)_\infty\;,
    \end{align}
  and $(x; q)_\infty$ is the Pochhammer symbol.
  The gauging of the $\SU(N)$ symmetry of a theory $\CT$ whose index is $\CI_\CT$ with $\SU(N)$ holonomies denoted by $m_i$ is given by
    \begin{align}
      \sum_{m_i} \int_{\T^{N-1}}   \prod_{i=1}^{N-1}  \frac{dz_i}{2 \pi i z_i} \CI_V(z, m_i) \CI_{\CT}(z, m_i)\;,
    \end{align}
  where the integration is performed over the unit circle $\T$ for each fugacity $z_i$.

  There are useful identities for the indices.
  The lens elliptic gamma function satisfies the reflection identity
    \begin{equation}
      \Gamma(z, m) \Gamma \left(\frac{pq}{z},-m \right)
        =     1\;.
        \label{eq:mass}
    \end{equation}
  Pictorially, this identity means that a pair of arrows making a loop is canceled if their R-charges add up to two and their flavor charges and holonomies add up to zero
    \begin{equation}
        \label{eq:massterm}
        \begin{tikzpicture}[baseline=(z.base)]
        \node[fnode] (z) at (0,0) {$z$};
        \node[fnode] (w) at (1,0) {$w$};
        \draw[q->] (z) to[bend left] node[above] {$a$} (w);
        \draw[q->] (w) to[bend left] node[below] {$pq/a$} (z);
      \end{tikzpicture}
      \ = \
      \begin{tikzpicture}[baseline=(z.base)]
        \node[fnode] (z) at (0,0) {$z$};
        \node[fnode] (w) at (1,0) {$w$};
      \end{tikzpicture}
      \;.
    \end{equation}
  Physically, this simply means that two massive bi-fundamentals are integrated out.

  The second identity is associated with the fact that the low-energy dynamics of $\SU(N)$ SQCD with $N_f = N$ is described
  by the deformed moduli space of the baryons and the mesons \cite{Seiberg:1994bz}.
  To see this, let us define
    \begin{equation}
      \begin{tikzpicture}[baseline=(z.base)]
      \node[fnode] (z) at (0,0) {$z$};
      \node[fnode] (w) at (1,0) {$w$};
      \draw[eq-] (z) -- (w);
      \end{tikzpicture}
    \end{equation}
  by the relation
    \begin{equation}
      \label{eq:z=w-def}
      \CI_{\CT}(w, \tilde{m}_i)
        =     \sum_{m_i} \int_{\T^{N-1}} \prod_{i=1}^{N-1} \frac{d z_i}{2\pi i z_i} \CI_{V} (z, m_i)  \CI_{\CT}(z, m_i) \
              \begin{tikzpicture}[baseline=(z.base)]
                \node[fnode] (z) at (0,0) {$z$};
                \node[fnode] (w) at (1,0) {$w$};
                \draw[eq-] (z) -- (w);
                \end{tikzpicture} \;,
    \end{equation}
  where $m_i$ and $\tilde{m}_i$ are the holonomies for $\SU(N)_z$ and $\SU(N)_w$.
  Then we have
    \begin{align}
      \begin{tikzpicture}[baseline=(z.base)]
      \node[fnode] (z) at (0,0) {$z$};
      \node[gnode] (x) at (1,0) {$x$};
      \node[fnode] (w) at (2,0) {$w$};
      \draw[q->] (z) -- node[above] {$a$} (x);
      \draw[q->] (x) -- node[above] {$a^{-1}$} (w);
      \end{tikzpicture}
      \ &=  \sum_{n_i} \int_{\T^{N-1}} \prod_{i=1}^{N-1} \frac{d x_i}{2\pi i x_i} \CI_{V} (x, n_i)  \nn \\
      &  \qquad \times    \prod_{i, j} \Gamma (z_i^{-1} x_j a, -m_i + n_j) \Gamma (x_i^{-1} w_j a^{-1}, -n_i + \tilde{m}_j) \nn  \\
      &=    \Gamma\bigl(a^{\pm N}, 0\bigr)  \
                \begin{tikzpicture}[baseline=(z.base)]
                \node[fnode] (z) at (0,0) {$z$};
                \node[fnode] (w) at (1,0) {$w$};
                \draw[eq-] (z) -- (w);
                \end{tikzpicture}
                \;.
                \label{eq:NfN}
    \end{align}
  $\SU(N)$ SQCD with $N_f = N$ has a vacuum where the mesons take nonzero expectation values
  and the flavor symmetry $\SU(N)_w \times \SU(N)_z$ is broken to the diagonal subgroup.
  In this vacuum the fugacities $w$ and $z$ are identified, so we get the quiver on the last line.
  The factor $\Gamma(a^{\pm N}, 0) = \Gamma(a^N, 0) \Gamma(a^{-N}, 0)$ is the contribution from the baryons.

  Finally, the formula found in \cite{Kels:2017toi} describes $\SU(N)$ SQCD with $N_f = N+1$.
  We write down the formula for the $N=2$ case, where the global symmetry is $\SU(6)$ with fugacities $t_I$ and holonomies $l_I$.
  If these fugacities and holonomies satisfy
    \begin{align}
      \prod_I t_I = pq\;, ~~~
      \sum_I l_I = 0 ~({\rm{mod}}~r)\;,
    \end{align}
  then
    \begin{align}
      &  \sum_{n} \int_{\T} \frac{d x}{2\pi i x} \CI_V(x, n) \prod_{I=1}^{6} \Gamma (t_I x, l_I + n) \Gamma (t_I x^{-1}, l_I -n)
       =     \prod_{1 \leq I < J \leq 6} \Gamma (t_I t_J, l_I + l_J)\;.
               \label{eq:NfN+1}
    \end{align}

\section{\texorpdfstring{Surface defects in class $\CS$ theories on lens space}{Surface defects in class S theories on lens space}}
\label{sec:class-S}
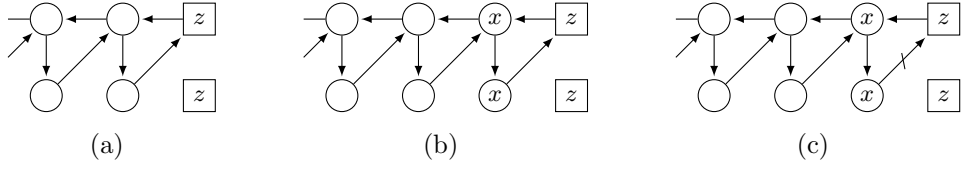
\begin{figure}
  \centering
  \subcaptionbox{\label{fig:TIR}}{
    \begin{tikzpicture}[scale=0.8]
      \node[gnode] (a1) at (1,0) {};
      \node[gnode] (a2) at (1,-1) {};
      \node[gnode] (b1) at (2,0) {};
      \node[gnode] (b2) at (2,-1) {};
      \node[fnode] (c1) at (3,0) {$z$};
      \node[fnode] (c2) at (3,-1) {$z$};

      \draw[q->] (b2) -- (c1);
      \draw[q->] (b1) -- (b2);
      \draw[q->] (c1) -- (b1);

      \draw[q->] (a2) -- (b1);
      \draw[q->] (a1) -- (a2);
      \draw[q->] (b1) -- (a1);

      \draw[q-] (a1) -- ($(a1) + (-0.5,0)$);
      \draw[q->] ($(a1) + (-0.5,-0.5)$) -- (a1);
    \end{tikzpicture}
  } \qquad
  \subcaptionbox{\label{fig:TUV}}{
    \begin{tikzpicture}[scale=0.8]
      \node[gnode] (a1) at (1,0) {};
      \node[gnode] (a2) at (1,-1) {};
      \node[gnode] (b1) at (2,0) {};
      \node[gnode] (b2) at (2,-1) {};
      \node[gnode] (c1) at (3,0) {$x$};
      \node[gnode] (c2) at (3,-1) {$x$};
      \node[fnode] (d1) at (4,0) {$z$};
      \node[fnode] (d2) at (4,-1) {$z$};

      \draw[q->] (c2) -- (d1);
      \draw[q->] (c1) -- (c2);
      \draw[q->] (d1) -- (c1);

      \draw[q->] (b2) -- (c1);
      \draw[q->] (b1) -- (b2);
      \draw[q->] (c1) -- (b1);

      \draw[q->] (a2) -- (b1);
      \draw[q->] (a1) -- (a2);
      \draw[q->] (b1) -- (a1);

      \draw[q-] (a1) -- ($(a1) + (-0.5,0)$);
      \draw[q->] ($(a1) + (-0.5,-0.5)$) -- (a1);
    \end{tikzpicture}
  } \qquad
  \subcaptionbox{\label{fig:TUVvev}}{
    \begin{tikzpicture}[scale=0.8]
      \node[gnode] (a1) at (1,0) {};
      \node[gnode] (a2) at (1,-1) {};
      \node[gnode] (b1) at (2,0) {};
      \node[gnode] (b2) at (2,-1) {};
      \node[gnode] (c1) at (3,0) {$x$};
      \node[gnode] (c2) at (3,-1) {$x$};
      \node[fnode] (d1) at (4,0) {$z$};
      \node[fnode] (d2) at (4,-1) {$z$};

      \draw[q->,vev] (c2) -- (d1);
      \draw[q->] (c1) -- (c2);
      \draw[q->] (d1) -- (c1);

      \draw[q->] (b2) -- (c1);
      \draw[q->] (b1) -- (b2);
      \draw[q->] (c1) -- (b1);

      \draw[q->] (a2) -- (b1);
      \draw[q->] (a1) -- (a2);
      \draw[q->] (b1) -- (a1);

      \draw[q-] (a1) -- ($(a1) + (-0.5,0)$);
      \draw[q->] ($(a1) + (-0.5,-0.5)$) -- (a1);
    \end{tikzpicture}
  }
  \caption{(a) An $\CN=2$ linear quiver theory $\CT_{\rm IR}$ with the flavor node specified by the fugacity $z$. The upper and the lower gauge/flavor groups are identified.
  (b) The theory $\CT_{\rm UV}$, obtained by gauging the $\SU(2)_x$ symmetry of $\CT_{\rm IR}$ and coupling it to a bi-fundamental hypermultiplet of $\SU(2)_x\times\SU(2)_z$.
  (c) The vev is given to the baryon made of the bi-fundamental chiral multiplet marked by the stroke.}
    \label{fig:N=2linear}
\end{figure}

  In this section we consider the lens space index of the $\CN=2$ class $\CS$ theories and the surface defect insertion.
  A typical $\CN=2$ linear quiver gauge theory, which we will refer to as $\CT_{\rm IR}$, is depicted in our notation in figure~\ref{fig:TIR},
  where the upper and the lower nodes are identified.
  For simplicity we focus on the case where all the gauge and flavor groups are $\SU(2)$.
  In the $\CN=1$ language, the hypermultiplet consists of two bi-fundamental chiral multiplets with fugacities $\sqrt{t}/\alpha$ and $\sqrt{t}\,\alpha$,
  where $\alpha$ is the fugacity for the $\U(1)_\alpha$ flavor symmetry acting on the hypermultiplet,
  while the $\CN=2$ vector multiplet of $\SU(2)_x$ contains an adjoint chiral multiplet with fugacity $pq/t$.
  
  Following \cite{Gaiotto:2012xa}, surface defects are engineered by renormalization group (RG) flows.
  To see this, we introduce the theory $\CT_{\rm UV}$ depicted in figure~\ref{fig:TUV},
  which is obtained from $\CT_{\rm IR}$ by gauging an $\SU(2)$ flavor symmetry and coupling it to a new $\SU(2)$ flavor symmetry through a bi-fundamental hypermultiplet.
  We denote the fugacity of the newly gauged group by $x$ and that of the new flavor group by $z$.
  We then give a vev to the baryon built out of the bi-fundamental with fugacity $\sqrt{t}/\alpha$, as indicated in figure~\ref{fig:TUVvev}.
  The RG flow triggered by a constant vev brings us back to $\CT_{\rm IR}$,
  while the RG flow triggered by a position-dependent vev produces $\CT_{\rm IR}$ in the presence of a surface defect.

\subsection{Poles and labels of the defects}
\label{subsec:poles}

  In terms of the index, giving a vev to the baryon with fugacity $t/\alpha^2$ amounts to computing the residue of the index of $\CT_{\rm UV}$ at a pole in $\alpha$ \cite{Gaiotto:2012xa}.
  A new feature on the lens space is that
  one can turn on a holonomy $\mu\in\Z_r$ for the $\U(1)_\alpha$ symmetry along with the fugacity $\alpha$, as well as the $\SU(2)$ holonomies.
  Including $\mu$, the index of the bi-fundamental which acquires the vev reads
    \begin{align}
      \begin{tikzpicture}
      \node[fnode] (x) at (0,0) {$x$};
      \node[fnode] (z) at (1,0) {$z$};
      \draw[q->] (x) -- node[above] {$\frac{\sqrt{t}}{\alpha}$} (z);
      \end{tikzpicture}
        ~ =  \prod_{i,j=1}^2 \Gamma_0 \, \Gamma^p \left( \frac{\sqrt{t}}{\alpha} z_i x_j^{-1}, m_i - n_j -\mu \right) \Gamma^q \left( \frac{\sqrt{t}}{\alpha} z_i x_j^{-1}, m_i - n_j -\mu \right)\;,
        \label{eq:bifund}
    \end{align}
  where $x_1 = x$, $x_2 = x^{-1}$, $z_1=z$, $z_2=z^{-1}$, and $(n_1,n_2)=(n,-n)$, $(m_1,m_2)=(m,-m)$ are the holonomies of $\SU(2)_x$ and $\SU(2)_z$.
  $\Gamma_0$ will be referred to as the prefactor.
  The factor $\Gamma^p$ has poles in the $x$-plane at
    \begin{align}
      x^{(p)}
      =    \frac{\sqrt{t}}{\alpha}\, z_i\, p^{[m_i - n - \mu]} (pq)^{j} (p^r)^{\ell}\;,
      \qquad j,\ell\in\Z_{\geq 0}\;,
    \end{align}
  while $\Gamma^q$ has poles at
    \begin{align}
      x^{(q)}
      =    \frac{\sqrt{t}}{\alpha}\, z_i\, q^{r-[m_i - n - \mu]} (pq)^{j} (q^r)^{\ell}\;,
    \end{align}
  with $[m]\in\{0,1,\dots,r-1\}$ denoting the residue of $m$ modulo $r$.
  These poles lie inside the unit circle for $|\alpha|$ close to $|\sqrt{t}|$,
  and the factors with $x_2=x^{-1}$ provide their images under $x\to x^{-1}$ outside the unit circle.
  When $\alpha$ is tuned such that a pole inside collides with a pole outside, the $x$-contour is pinched and the index of $\CT_{\rm UV}$ develops a pole in $\alpha$.
  There are four types of collisions, according to whether the colliding poles come from $\Gamma^p$ or $\Gamma^q$,
  which we call the $(p,p)$, $(q,q)$, $(p,q)$ and $(q,p)$ sectors.
  For example, in the $(p,p)$ sector the collision takes place at
    \begin{align}
      (p,p): \quad
      \frac{\alpha^2}{t}
        =     p^{[m-n-\mu]+[-m+n-\mu]}\,(pq)^{j_1+j_2}\,(p^r)^{\ell_1+\ell_2}\;,
              \label{eq:pp}
    \end{align}
  and similarly for the other sectors.
  All in all, the poles are located at
    \begin{align}
      \alpha^2
        =     t\, p^{a} q^{b}\;,
        \qquad a,b\in\Z_{\geq 0}\;,
        \label{eq:poles-ab}
    \end{align}
  and the holonomy $\mu$ is constrained as follows.
  The poles of $\Gamma(y,M)$ lie at $y=p^{-a'}q^{-b'}$ with $a',b'\in\Z_{\geq0}$ and $a'-b'\equiv M \pmod r$:
  for $\Gamma^p(y,M)$ one has $a'-b'=[M]+r\ell$, and for $\Gamma^q(y,M)$ one has $a'-b'=[M]-r(\ell+1)$.
  A pinching involves a pole of the factor with $x_j=x$ and holonomy $m_i-n-\mu$, at $x=u\,z_i\,p^{a_1}q^{b_1}$ with $u=\sqrt{t}/\alpha$,
  and a pole of the factor with $x_j=x^{-1}$, $z_{i'}=z_i^{-1}$ and holonomy $-m_i+n-\mu$, at $x=u^{-1}z_i\,p^{-a_2}q^{-b_2}$.
  The two collide when $\alpha^2=tp^aq^b$ with $a=a_1+a_2$ and $b=b_1+b_2$, and then
  $a-b\equiv(m_i-n-\mu)+(-m_i+n-\mu)=-2\mu$.
  Conversely, if this condition holds, the pinching with $a_1=b_1=0$ occurs in the sector $n\equiv m_i-\mu$ of the $\SU(2)_x$ holonomy.
  Therefore a pole at \eqref{eq:poles-ab} in the sector with $\U(1)_\alpha$ holonomy $\mu$ exists if and only if
    \begin{align}
      2\mu
        \equiv  b - a \pmod{r}\;.
        \label{eq:pole-condition}
    \end{align}
  For $\mu=0$ this is the condition $a\equiv b \pmod r$ found in \cite{Razamat:2013jxa}.
  In \cite{Gaiotto:2012xa} ($r=1$) the defects are labeled by the pair $(a,b)$.
  For $r>1$ it is natural to write
    \begin{align}
      \alpha^2
        =     t\, (p^r)^{n_1} (q^r)^{n_2} (pq)^{n_3}\;,
    \end{align}
  and to label the corresponding defects, for which $2\mu\equiv0$, by $(n_1,n_2,n_3)$, with the understanding that $\mu=0$.
  For even $r$ the condition $2\mu\equiv0$ also admits $\mu=r/2$, and the defects of the two sectors differ by the holonomy shift $U$ introduced in section~\ref{subsec:elementary}.
  The condition \eqref{eq:pole-condition} shows, however, that there are more poles, with $2\mu\not\equiv0$, which are not of this form.
  In particular, the pole $\alpha^2 = tp$ exists for odd $r$ with $\mu = (r-1)/2$,
  and it gives rise to the ``elementary'' defect of section~\ref{subsec:elementary}.

\subsection{Difference operators representing surface defects}
\label{sec:operators}
  Let us write the index of $\CT_{\rm IR}$ as $\CI_{\rm IR}(z,m)$, where $(z,m)$ are the fugacity and the holonomy of the $\SU(2)$ flavor symmetry on which the defect is constructed.
  By the Weyl symmetry, $\CI_{\rm IR}(z,m)=\CI_{\rm IR}(z^{-1},-m)$.
  We define the shift operators
    \begin{align}
      \bigl(\Dop{h}{\delta} f\bigr)(z,m)
        :=    f(h z, m+\delta)\;,
    \end{align}
  where $h$ is a monomial in $p^{1/2}$ and $q^{1/2}$ and $\delta\in\Z_r$.
  When we evaluate the residue of the index of $\CT_{\rm UV}$ at a pole \eqref{eq:poles-ab}, the contribution of each pinching is obtained by evaluating the rest of the integrand at the pinching point $x = x_*$, $n=n_*$.
  This gives
    \begin{align}
      2\Xi \Res_{\alpha=\alpha_*} \frac{1}{\alpha}\, \CI_{\rm UV}(z,m;\alpha)
        = \bigl(\Sf\, \CI_{\rm IR}\bigr)(z,m)\;,
      \qquad
      \Sf
        =     \sum_{(h,\delta)} c_{h,\delta}(z,m)\, \Dop{h}{\delta}\;,
    \end{align}
  where the sum runs over the pinchings $x_* = hz$, $n_* = m+\delta$, and the coefficients are given by
    \begin{align}
      c_{h,\delta}(z,m)
        &=   \CN_{h,\delta}(m)\,
              \frac{\Gamma\bigl(\frac{pq}{t}x_*^{\pm2},\pm 2n_*\bigr)}{\Gamma\bigl(x_*^{\pm 2},\pm 2n_*\bigr)}\,
              \Gamma\bigl(u_*\, z x_*, m+n_*-\mu\bigr)\,
              \Gamma\bigl(u_*\, z^{-1}x_*^{-1}, -m-n_*-\mu\bigr) \nn\\
        &    \times \prod_{\epsilon,\epsilon'=\pm1}
              \Gamma\bigl(t u_*^{-1} z^{\epsilon} x_*^{\epsilon'}, \epsilon m + \epsilon' n_* + \mu\bigr)\;,
              \label{eq:gamma-form}
    \end{align}
  with $u_* = \sqrt{t}/\alpha_*$ and $x_* = hz$.
  Here $\Gamma(y^{\pm 2},\pm 2n)$ stands for $\Gamma(y^2,2n)\Gamma(y^{-2},-2n)$.
  The ratio of gamma functions in the first line comes from the vector multiplet of $\SU(2)_x$ (including the adjoint chiral multiplet),
  the second line from the bi-fundamental chiral multiplet with fugacity $\sqrt{t}\alpha$,
  and the remaining factors from the two components of the bi-fundamental with fugacity $\sqrt{t}/\alpha$ which are not involved in the pinching.
  The $z$-independent constants $\CN_{h,\delta}(m)$ collect the residues of the elliptic gamma functions involved in the pinching.

  The coefficients \eqref{eq:gamma-form} can be reduced to ratios of theta functions
  by repeated use of the quasi-periodicity of the elliptic gamma function \eqref{eq:gamma-quasi} and the reflection identity \eqref{eq:mass}.
  This is where the computation becomes cumbersome,
  mainly because the holonomies enter through the residues $[\,\cdot\,]$ modulo $r$ and the prefactors $\Gamma_0$ of the lens elliptic gamma functions.
  We have carried out the reduction systematically, and the results are summarized in the following.

\paragraph{TQFT consistency.}
  Before presenting the results, let us describe the consistency condition that we use to fix and check the operators.
  The index of an $\CN=2$ theory of class $\CS$ of type $A_1$ on $L(r,1)\times S^1$ is computed by a two-dimensional TQFT \cite{Benini:2011nc,Razamat:2013jxa,Alday:2013rs}.
  Its basic building block is the free trinion, a half-hypermultiplet in the tri-fundamental representation of $\SU(2)_a\times\SU(2)_b\times\SU(2)_c$,
  whose index is
    \begin{align}
      \CI_{\rm tri}(a,m_a;b,m_b;c,m_c)
        =     \prod_{\epsilon_a,\epsilon_b,\epsilon_c=\pm1}
              \Gamma\bigl(\sqrt{t}\,a^{\epsilon_a} b^{\epsilon_b} c^{\epsilon_c}, \epsilon_a m_a + \epsilon_b m_b + \epsilon_c m_c\bigr)\;.
              \label{eq:trinion}
    \end{align}
  Since the surface defect can be moved from one puncture to another \cite{Gaiotto:2012xa},
  the difference operator $\Sf$ must satisfy
    \begin{align}
      \Sf_{(a)}\, \CI_{\rm tri}
        =     \Sf_{(b)}\, \CI_{\rm tri}\;,
        \label{eq:TQFT}
    \end{align}
  where $\Sf_{(a)}$ acts on $(a,m_a)$ and $\Sf_{(b)}$ on $(b,m_b)$.
  We have verified numerically that all the operators presented below satisfy \eqref{eq:TQFT} in all the holonomy sectors $(m_a,m_b,m_c)\in\Z_r^3$ (appendix~\ref{app:numerics}).
  Moreover, once the $z$-dependence of the coefficients is known, \eqref{eq:TQFT} determines the constants $\CN_{h,\delta}(m)$ up to an overall normalization,
  which we specify for each operator below.

\paragraph{$(0,0,1)$ defect.}
  Consider the defect of type $(0,0,1)$, namely the pole at $\alpha = \sqrt{tpq}$ with $\mu=0$.
  In the $(p,p)$ sector the pole in the $x$-plane is at $x = \frac{\sqrt{t}pq}{\alpha}z$ (and its Weyl images),
  which collides with its image when $\alpha = \sqrt{t pq}$.
  The resulting shift is $x_*=(pq)^{\pm1/2}z$ with $n_*=m$.
  The $(p,q)$ and $(q,p)$ sectors give rise to the shifts $x_*=(q/p)^{\pm 1/2}z$ with $n_* = m\pm1$.
  After the reduction we obtain
    \begin{multline}
      \CS^{(0,0,1)}
      = \frac{\theta(t^{-1}; p^r) \theta(t; q^r)}{\theta((pq)^{-1}; p^r)\theta(pq; q^r)}
        \sum_{s=\pm1} \Bigg[ \frac{\theta(\frac{pq}{t} p^{\modr{2sm}} z^{2s}; p^r)\theta(\frac{pq}{t} q^{r-\modr{2sm}} z^{2s}; q^r)}{\theta(z^{2s} p^{\modr{2sm}}; p^r)\theta(z^{2s} q^{r-\modr{2sm}}; q^r)}\, \Dop{(pq)^{s/2}}{0}  \\
        + \left( \frac{pq}{t} \right)^{\frac{2 - r + 2\modr{2sm}}{r}} \frac{\theta(\frac{pq}{t} p^{\modr{2sm}} z^{2s}; p^r)\theta(\frac{pq}{t} q^{\modr{2sm}} z^{-2s}; q^r)}{\theta(z^{2s} p^{\modr{2sm}}; p^r)\theta(z^{-2s} q^{\modr{2sm}}; q^r)}\, \Dop{(q/p)^{s/2}}{s} \Bigg]\;.
      \label{eq:s001}
    \end{multline}
  The exponent of $pq/t$ in the second line depends on $m$ only through $\modr{2sm}$,
  so that \eqref{eq:s001} is well defined on $m\in\Z_r$.
  For the $(p,p)$ sector the contributions from the prefactors $\Gamma_0$ cancel out.
  The relative factor $(pq/t)^{(2-r+2[2sm])/r}$ between the $(p,p)$ sector and the $(p,q)$, $(q,p)$ sectors
  has been fixed by the TQFT consistency condition \eqref{eq:TQFT}.
  The operator \eqref{eq:s001} coincides with the one obtained for type $A_1$ in \cite{Razamat:2013jxa} (eqs.~(4.8) and (4.9) there), including the relative factor.
  Its constant prefactor is $\Gamma((pq)^{\pm1};p^r,q^r)/\Gamma(t^{\pm1};p^r,q^r)$, as in \cite{Razamat:2013jxa}, where the remaining constant depends on the normalization of the vector multiplet.
  For $r=1$ the normalization agrees with the operator of \cite{Gaiotto:2012xa} (see \eqref{eq:factorS1S3} below),
  and for any $r$ the factorization \eqref{eq:factor-odd} holds with the normalization \eqref{eq:Xp}, \eqref{eq:Xq} of the elementary operators,
  which agrees with the residue computation of section~\ref{subsec:Sk-Sklyanin}.

\paragraph{$(1,0,0)$ defect.}
  Next consider the defect of type $(1,0,0)$, namely the pole at $\alpha = \sqrt{t p^r}$ with $\mu=0$.
  Only the $(p,p)$ sector contributes.
  From \eqref{eq:pp} the collisions occur either for $[m-n]=[n-m]=0$ with $\ell_1+\ell_2=1$,
  or for $[m-n]=M$, $[n-m]=r-M$ with $1\leq M\leq r-1$ and $\ell_1=\ell_2=0$.
  The pinching points are $x_* = p^{(r-2M)/2}z$ and $n_* = m + M$ with $M=0,1,\dots,r$, together with their Weyl images.
  Reducing \eqref{eq:gamma-form} we obtain
    \begin{align}
      \CS^{(1,0,0)}
        =     \sum_{M=0}^{r} \left(\frac{pq}{t}\right)^{\frac{(M+[2m])(2M-r)}{r}}
              K_M(q^r)\; c_M\bigl(W_m(z)\bigr)\; \Dop{p^{(r-2M)/2}}{M}\;,
              \label{eq:s100}
    \end{align}
  where
    \begin{align}
      W_m(z)
        :=    z^2 q^{-[2m]}\;,
    \end{align}
  the constants $K_M$ are
    \begin{align}
      K_M(q^r)
        =     \prod_{i=0}^{r-M-1}\frac{\theta(t(pq)^i; q^r)}{\theta((pq)^{i+1}; q^r)}
              \prod_{i=0}^{M-1} \frac{\theta(t(pq)^i; q^r)}{\theta((pq)^{i+1}; q^r)}\;,
    \end{align}
  and the $z$-dependent part is given by
    \begin{align}
      c_M(W)
        =     \frac{\theta(W(pq)^{r-2M};q^r)}{\theta(W(pq)^{-M};q^r)}
              \prod_{i=1-M}^{r-2M} \frac{\theta(t^{-1}W(pq)^i;q^r)}{\theta(W(pq)^i;q^r)}
              \prod_{i=r-2M}^{r-M-1} \frac{\theta(t W(pq)^{i};q^r)}{\theta(W(pq)^{i+1};q^r)}\;.
    \end{align}
  The $M=0$ and $M=r$ terms, as well as their normalization $K_0(q^r)$, follow directly from the residue in the case $\ell_1+\ell_2=1$,
  and the relative constants of the other terms have been fixed by the TQFT consistency condition \eqref{eq:TQFT}.

  Since $\CI_{\rm IR}(z,m)=\CI_{\rm IR}(z^{-1},-m)$, the shift of the $(r-M)$-th term acting at $(z,m)$ equals the shift of the $M$-th term acting at $(z^{-1},-m)$.
  Accordingly, the coefficient of the $(r-M)$-th term at $(z,m)$ agrees with that of the $M$-th term at $(z^{-1},-m)$,
  and $\CS^{(1,0,0)}$ commutes with the Weyl reflection $(z,m)\to(z^{-1},-m)$.
  This is not manifest in \eqref{eq:s100}, but follows from \eqref{eq:theta-refl} with nome $q^r$.
  Since $[-2m]=r-[2m]$ for $[2m]\neq0$, we have
    \begin{align}
      W_{-m}(z^{-1})
        =     \begin{cases}
                W_m(z)^{-1}\;, & [2m]=0\;,\\
                q^{-r}\,W_m(z)^{-1}\;, & [2m]\neq0\;.
              \end{cases}
    \end{align}
  For instance, $K_0=K_r$, and \eqref{eq:theta-refl} gives
  $c_r\bigl(W_{-m}(z^{-1})\bigr) = (t/pq)^{r+[-2m]+[2m]}\, c_0\bigl(W_m(z)\bigr)$,
  which is compensated by the ratio $(pq/t)^{r+[-2m]}/(pq/t)^{-[2m]}$ of the prefactors in \eqref{eq:s100}.
  All theta functions in \eqref{eq:s100} have the nome $q^r$, while the shifts are in powers of $p^{1/2}$.
  This is the lens space counterpart of the fact that for $r=1$ the $(1,0)$ defect of \cite{Gaiotto:2012xa} is a $q$-elliptic difference operator acting by $p$-shifts.
  Since the $M$-th term shifts the holonomy by $M$, the sum over $M$ does not factorize into a sum over the shifts of $z$ and a sum over the shifts of $m$.

\paragraph{$(0,1,0)$ defect.}
  The lens elliptic gamma function is symmetric under $p\leftrightarrow q$ together with $m\to -m$, up to phases which cancel in the operators.
  Correspondingly, the $(0,1,0)$ defect is obtained from the $(1,0,0)$ defect by this exchange:
    \begin{align}
        \CS^{(0,1,0)}
          &=     \sum_{M=0}^{r} \left(\frac{pq}{t}\right)^{\frac{(M+[-2m])(2M-r)}{r}}
                                K_M(p^r)\; \tilde{c}_M\bigl(\tilde{W}_m(z)\bigr)\; \Dop{q^{(r-2M)/2}}{-M}\;,
                \label{eq:s010}
                \\
        \tilde{W}_m(z)
                    &:=    z^2 p^{-[-2m]}\;,
                \nonumber
   \end{align}
  where $K_M(p^r)$ and $\tilde{c}_M$ are obtained from $K_M(q^r)$ and $c_M$ by replacing $q^r$ with $p^r$ in the nome of all theta functions.
  In the limit $p\to0$ the operators for the poles at $\alpha^2=tq^{rn}$ were obtained in \cite{Razamat:2013jxa} and related to the Cherednik operators.
  We expect \eqref{eq:s010} to reduce to their operator for $n=1$ in this limit, but we have not checked this in detail.

\subsection{Elementary defects}
\label{subsec:elementary}
  As discussed in section~\ref{subsec:poles}, for odd $r$ there is a pole at $\alpha^2=tp$ in the sector with $\U(1)_\alpha$ holonomy $\mu=(r-1)/2$.
  The pinchings are at $x_*=p^{\pm 1/2}z$ with $n_* = m\mp h_r$, where
    \begin{align}
      h_r
        :=    \frac{r+1}{2}\;,
        \qquad 2h_r\equiv 1 \pmod{r}\;.
    \end{align}
  Thus the holonomy is shifted by ``one half'', which is well defined since $2$ is invertible modulo odd $r$.
  The corresponding difference operator is
    \begin{align}
      X_p
        =     \frac{\theta(t;q^r)}{\theta(pq;q^r)} \sum_{s=\pm1}
              \left(\frac{pq}{t}\right)^{\frac{1-s}{2}-\frac{s[2m]}{r}}
              \frac{\theta\bigl((\frac{pq}{t})^{s}\, W_m(z);q^r\bigr)}{\theta\bigl(W_m(z);q^r\bigr)}\,
              \Dop{p^{s/2}}{-s h_r}\;.
              \label{eq:Xp}
    \end{align}
  Similarly, the pole at $\alpha^2=tq$ with $\mu = (r+1)/2$ gives
    \begin{align}
      X_q
        =     \frac{\theta(t;p^r)}{\theta(pq;p^r)} \sum_{s=\pm1}
              \left(\frac{pq}{t}\right)^{\frac{1-s}{2}-\frac{s[-2m]}{r}}
              \frac{\theta\bigl((\frac{pq}{t})^{s}\, \tilde{W}_m(z);p^r\bigr)}{\theta\bigl(\tilde{W}_m(z);p^r\bigr)}\,
              \Dop{q^{s/2}}{s h_r}\;.
              \label{eq:Xq}
    \end{align}
  Both $X_p$ and $X_q$ satisfy the TQFT consistency condition \eqref{eq:TQFT}.
  For $r=1$ their normalization reduces to that of $\Sf_{(1,0)}$ and $\Sf_{(0,1)}$,
  and for general $r$ it agrees with the $k=1$ case of the normalization obtained from the residue computation in section~\ref{subsec:Sk-Sklyanin}.
  As we will see in section~\ref{sec:algebra}, they generate all the other defects discussed so far.

\paragraph{Extended holonomy label.}
  The coefficients of all our operators depend on the holonomy $m$ only through $[2m]$.
  It is convenient to trade $m$ for the label
    \begin{align}
      \varkappa
        :=    2m \in \Z_{2r}\;,
        \label{eq:kappa-def}
    \end{align}
  which, unlike $[2m]\in\Z_r$, retains the full information on $m$ also for even $r$, where $[2m]$ does not distinguish $m$ from $m+r/2$.
  The operators \eqref{eq:s001}, \eqref{eq:s100} and \eqref{eq:s010} then act on functions of $(z,\varkappa)$.
  The shift $m\to m+\delta$ becomes $\varkappa\to\varkappa+2\delta$, and the coefficients depend on $\varkappa$ only modulo $r$.
  In terms of $\varkappa$ the operator \eqref{eq:Xp} reads
    \begin{align}
      X_p
        =     \frac{\theta(t;q^r)}{\theta(pq;q^r)} \sum_{s=\pm1}
              \left(\frac{pq}{t}\right)^{\frac{1-s}{2}-\frac{s\varkappa}{r}}
              \frac{\theta\bigl((\frac{pq}{t})^{s}\, z^2q^{-\varkappa};q^r\bigr)}{\theta\bigl(z^2q^{-\varkappa};q^r\bigr)}\,
              \Delta_{p^{s/2}}^{(\varkappa\to\varkappa-s)}\;,
              \label{eq:Xp-kappa}
    \end{align}
  where the shift operator acts on functions of $(z,\varkappa)$ by $(z,\varkappa)\to(p^{s/2}z,\varkappa-s)$, and the exponent $-s\varkappa/r$ is evaluated with $\varkappa$ unreduced.
  The coefficients of \eqref{eq:Xp-kappa} are invariant under $\varkappa\to\varkappa+r$:
  the change of the monomial factor is compensated by the quasi-periodicity of the theta functions (see appendix~\ref{app:factorization}).
  Hence \eqref{eq:Xp-kappa} is well defined on $\varkappa\in\Z_{2r}$ for any $r$.
  Similarly, using $[-2m]=r-[2m]$ for $[2m]\neq0$ and \eqref{eq:theta-refl}, the operator \eqref{eq:Xq} becomes
    \begin{align}
      X_q
        =     \frac{\theta(t;p^r)}{\theta(pq;p^r)} \sum_{s=\pm1}
              \left(\frac{pq}{t}\right)^{\frac{1-s}{2}+\frac{s\varkappa}{r}}
              \frac{\theta\bigl((\frac{pq}{t})^{s}\, z^2p^{\varkappa};p^r\bigr)}{\theta\bigl(z^2p^{\varkappa};p^r\bigr)}\,
              \Delta_{q^{s/2}}^{(\varkappa\to\varkappa+s)}\;.
              \label{eq:Xq-kappa}
    \end{align}
  We also define the holonomy shift
    \begin{align}
      (U f)(z,\varkappa)
        :=    f(z,\varkappa+r)\;,
        \qquad U^2 = 1\;,
        \label{eq:U-def}
    \end{align}
  which commutes with all the operators of this section, since their coefficients depend on $\varkappa$ only modulo $r$.
  
  The relation between the label $\varkappa$ and the $\SU(2)$ holonomy depends on the parity of $r$.
  For odd $r$, every $\varkappa\in\Z_{2r}$ is of the form $2m$ or $2m+r$ with a unique $m=[h_r\varkappa]\in\Z_r$,
  and the $\SU(2)$ index is embedded as $f(z,\varkappa)=\CI_{\rm IR}(z,[h_r\varkappa])$, which is $U$-invariant.
  Lifted to $\Z_{2r}$, the shift $m\to m-sh_r$ of \eqref{eq:Xp} becomes $\varkappa\to\varkappa-s(r+1)\equiv\varkappa-s+r$,
  so that \eqref{eq:Xp} is $U$ times \eqref{eq:Xp-kappa}, and the two operators agree on $U$-invariant functions.
  The same holds for \eqref{eq:Xq} and \eqref{eq:Xq-kappa}.
  From now on $X_p$ and $X_q$ denote the operators \eqref{eq:Xp-kappa} and \eqref{eq:Xq-kappa} acting on functions of $(z,\varkappa)$ with $\varkappa\in\Z_{2r}$.

  For even $r$, the $\SU(2)$ holonomies give only even $\varkappa=2m$, and $X_p$ maps them to odd ones.
  Thus for even $r$ the elementary operators are not defects of the $\SU(2)$ theory by themselves,
  in agreement with the fact that \eqref{eq:pole-condition} cannot be satisfied for $a+b$ odd.
  They act on an extended space which includes a ``twisted'' sector with odd $\varkappa$, namely half-integer $m$.
  Since $H^2(L(r,1);\Z_2)=\Z_2$ for even $r$, it is tempting to identify this sector with the $\SO(3)$ bundles with nontrivial $w_2$ \cite{Razamat:2013opa,Tachikawa:2014mna}.
  The twisted sector also appears in the residue computation itself.
  For $r=2$ the pole at $\alpha^2=tq$ requires $\mu=1/2$ by \eqref{eq:pole-condition}.
  In our numerical residue computation for class $\CS_2$ (appendix~\ref{app:numerics}) this sector is realized with half-integer holonomies of the $\SU(2)$ gauge groups adjacent to the defect,
  while all the chiral multiplets carry integer holonomies.
  Such configurations are bundles for the quotient of the gauge group times $\U(1)_\alpha$ by a diagonal $\Z_2$ subgroup of their centers,
  so that for even $r$ the elementary defects involve this global structure.

  The operator $U$, on the other hand, preserves the $\SU(2)$ sector and acts on it as $m\to m+r/2$,
  which multiplies the $\SU(2)$ holonomy $\diag(e^{2\pi i m/r},e^{-2\pi i m/r})$ by the central element $-1$.
  It is not the identity on the $\SU(2)$ index, which depends on $m$ and not only on $[2m]$, as the trinion index \eqref{eq:trinion} shows.
  Correspondingly, for even $r$ the condition \eqref{eq:pole-condition} determines $\mu$ only modulo $r/2$, and the operators of the two sectors at the same pole are related by
    \begin{align}
      \Sf_{\mu+r/2}
        =     U\, \Sf_{\mu}\;,
        \label{eq:U-sector}
    \end{align}
  since under $(n_*,\mu)\to(n_*+r/2,\mu+r/2)$ the pinching conditions and all the holonomy arguments in \eqref{eq:gamma-form} change by multiples of $r$, while $n_*=m+\delta$ shifts by $r/2$.
  The labels $(n_1,n_2,n_3)$ of section~\ref{subsec:poles} refer to $\mu=0$.
  As we will see, even for even $r$ the defects of the $\SU(2)$ theory are generated by $X_p$, $X_q$ and $U$ acting on the extended space.
  
\paragraph{The case $r=1$.}
  For $r=1$ we have $m=0$, $h_1=1$ and $W_0(z)=z^2$, and the operators reduce to those of \cite{Gaiotto:2012xa}.
  In particular $\CS^{(1,0,0)}=X_p$ and $\CS^{(0,1,0)}=X_q$ with
    \begin{align}
      \CS^{(1,0,0)}
        &= \frac{\theta(t; q)}{\theta(pq; q)} \sum_{s=\pm 1} \frac{\theta(\frac{pq}{t} z^{2s}; q)}{\theta(z^{2s}; q)} \Dop{p^{s/2}}{0}
        =  \Sf_{(1,0)}\;,  \\
      \CS^{(0,1,0)}
        &= \frac{\theta(t; p)}{\theta(pq; p)} \sum_{s=\pm 1} \frac{\theta(\frac{pq}{t} z^{2s}; p)}{\theta(z^{2s}; p)} \Dop{q^{s/2}}{0}
        =  \Sf_{(0,1)}\;,
    \end{align}
  where we used $\theta(x;q)=\theta(q/x;q)$ and $\theta(qx;q)=-x^{-1}\theta(x;q)$ to rewrite the $s=-1$ terms.
  Similarly,
    \begin{multline}
      \CS^{(0,0,1)}
        = \frac{\theta(t^{-1}; p) \theta(t; q)}{\theta((pq)^{-1}; p) \theta(pq; q)}
          \sum_{s=\pm 1} \Bigg[ \frac{\theta(\frac{pq}{t} z^{2s}; p) \theta(\frac{pq}{t} z^{2s} q; q)}{\theta(z^{2s}; p) \theta(z^{2s} q; q)} \Dop{(pq)^{s/2}}{0}  \\
            + \frac{pq}{t} \frac{\theta(\frac{pq}{t} z^{2s}; p) \theta(\frac{pq}{t} z^{-2s}; q)}{\theta(z^{2s}; p) \theta(z^{-2s}; q)} \Dop{(q/p)^{s/2}}{0}  \Bigg]
        =  \Sf_{(1,1)}\;,
    \end{multline}
  and one verifies the factorization
    \begin{equation}
      \CS^{(0,0,1)} = \frac{pq}{t}\, \CS^{(0,1,0)} \CS^{(1,0,0)}\;.
      \label{eq:factorS1S3}
    \end{equation}

\section{Algebraic structure}
\label{sec:algebra}
  We now investigate the algebra generated by the difference operators obtained in the previous section.

\paragraph{Commutativity.}
  All the operators obtained above commute with each other:
    \begin{align}
      \bigl[\CS^{(1,0,0)},\CS^{(0,1,0)}\bigr]
        =     \bigl[\CS^{(1,0,0)},\CS^{(0,0,1)}\bigr]
        =     \bigl[\CS^{(0,1,0)},\CS^{(0,0,1)}\bigr]
        =     0\;,
        \qquad
      \bigl[X_p, X_q\bigr]
        =     0\;.
        \label{eq:commute}
    \end{align}
  They also commute with the holonomy shift $U$ of \eqref{eq:U-def}, since their coefficients depend on $\varkappa$ only modulo $r$.
  This is expected physically, since the surface defects can be separated in the direction transverse to them \cite{Gaiotto:2012xa},
  and all the operators are diagonalized by the same eigenfunctions of the TQFT.
  For any $r$, \eqref{eq:commute} follows from the results below:
  $[X_p,X_q]=0$ is proven in appendix~\ref{app:factorization}, $\CS^{(0,0,1)}$ is proportional to $X_qX_p$ by \eqref{eq:factor-odd},
  and $\CS^{(1,0,0)}$ and $\CS^{(0,1,0)}$ are $U$ times polynomials in $X_p$ and $X_q$, respectively, by \eqref{eq:S100-fusion}, which is proven in appendix~\ref{app:fusion}.
  All the relations of this section have also been checked numerically (appendix~\ref{app:numerics}).

\paragraph{Factorization.}
  The $(0,0,1)$ defect factorizes into the elementary ones:
    \begin{align}
      \CS^{(0,0,1)}
        =     \left(\frac{pq}{t}\right)^{\frac{1}{r}} X_q\, X_p
        =     \left(\frac{pq}{t}\right)^{\frac{1}{r}} X_p\, X_q\;.
        \label{eq:factor-odd}
    \end{align}
  Indeed, the shifts match: $X_qX_p$ contains the shift $(z,\varkappa)\to(q^{s/2}p^{s'/2}z,\varkappa+s-s')$,
  which is $m\to m$ for $s=s'$ and $m\to m+s$ for $s=-s'$, reproducing the holonomy shifts $0$ and $s$ in \eqref{eq:s001} for any $r$.
  For $r=1$ this reduces to \eqref{eq:factorS1S3}.
  For even $r$ the intermediate states have odd $\varkappa$, i.e.~they lie in the twisted sector of section~\ref{subsec:elementary}, while $X_qX_p$ maps even $\varkappa$ to even $\varkappa$.
  No factor of $U$ appears here, since two shifts of $\varkappa$ by $\pm1$ return to the $\SU(2)$ sector.
  We prove \eqref{eq:factor-odd} in appendix~\ref{app:factorization}.
  In the language of the ``transfer matrices'' of \cite{Maruyoshi:2016caf},
  $X_p$ and $X_q$ are the lens space analogues of $\Tr L_p$ and $\Tr L_q$,
  and \eqref{eq:factor-odd} states that the $(0,0,1)$ defect is the product of two commuting transfer matrices.

\paragraph{Fusion and the $(1,0,0)$ defect.}
  The $(1,0,0)$ defect is \emph{not} proportional to $(X_p)^r$, even though the two operators contain the same set of shifts.
  Instead, it is obtained by a fusion rule.
  Define the operators $\hat{\CS}_n$ recursively by
    \begin{align}
      \hat{\CS}_0 = 1\;, \qquad
      \hat{\CS}_1 = X_p\;, \qquad
      \hat{\CS}_{n+1}
        =     X_p\, \hat{\CS}_n - \omega_n\, \hat{\CS}_{n-1}\;,
              \label{eq:fusion}
    \end{align}
  with the constants
    \begin{align}
      \omega_n
        :=    \left(\frac{pq}{t}\right)^{1-\frac{1}{r}}
              \frac{\theta(t;q^r)^2}{\theta(pq;q^r)^2}\,
              \frac{\theta((pq)^n;q^r)\,\theta(t^2(pq)^{n-1};q^r)}{\theta(t(pq)^{n-1};q^r)\,\theta(t(pq)^n;q^r)}\;.
              \label{eq:omega}
    \end{align}
  Then $\hat{\CS}_n$ is proportional to the defect at $\alpha^2 = tp^n$ with $2\mu\equiv -n \pmod r$:
  its shifts $(z,\varkappa)\to(p^{(n-2k)/2}z,\varkappa-n+2k)$, $k=0,\dots,n$, are those of the pinchings at this pole.
  For even $r$ the condition $2\mu\equiv-n$ has the two solutions $\mu$ and $\mu+r/2$, and the shifts of $\hat\CS_r$ select the sector $\mu\equiv r/2$,
  whereas $\CS^{(1,0,0)}$ in \eqref{eq:s100} is the defect with $\mu=0$.
  By \eqref{eq:U-sector} the two differ by $U$, and for $n=r$ we find
    \begin{align}
      \CS^{(1,0,0)}
        =     \left(\frac{pq}{t}\right)^{-\frac{r-1}{2}}
              \frac{K_0(q^r)}{\bigl(\theta(t;q^r)/\theta(pq;q^r)\bigr)^{r}}\; U\, \hat{\CS}_r\;.
              \label{eq:S100-fusion}
    \end{align}
  The factor $U$ is also visible in the shifts:
  $\hat\CS_r$, a combination of $(X_p)^r,(X_p)^{r-2},\dots$, shifts $(z,\varkappa)\to(p^{j/2}z,\varkappa-j)$,
  while the $M$-th term of \eqref{eq:s100} shifts $(z,\varkappa)\to(p^{j/2}z,\varkappa+2M)=(p^{j/2}z,\varkappa+r-j)$ with $j=r-2M$.
  For odd $r$ the operator $U$ acts trivially on the $\SU(2)$ index (section~\ref{subsec:elementary}), so that on it \eqref{eq:S100-fusion} holds with $U$ replaced by $1$.
  For even $r$ the factor $U$ is needed, since $U$ acts on the $\SU(2)$ index as $m\to m+r/2$.
  In appendix~\ref{app:fusion} we prove \eqref{eq:S100-fusion} for any $r$ by deriving the closed form \eqref{eq:Shat-closed} of $\hat\CS_n$.
  Thus, up to the holonomy shift $U$, $\CS^{(1,0,0)}$ is a polynomial of degree $r$ in $X_p$ with constant coefficients,
    \begin{align}
      \CS^{(1,0,0)} \propto U \Bigl[ (X_p)^r - \Bigl(\sum_{n=1}^{r-1}\omega_n\Bigr) (X_p)^{r-2} + \cdots \Bigr]\;,
    \end{align}
  and the same holds for $\CS^{(0,1,0)}$ and $X_q$ with $p\leftrightarrow q$.
  
  The recursion \eqref{eq:fusion} is a deformation of the $\SU(2)$ fusion rule $\mathbf{2}\otimes\mathbf{n}=(\mathbf{n+1})\oplus(\mathbf{n-1})$,
  with the $(1,0,0)$ defect playing the role of the spin $r/2$ representation.
  In fact the coefficient functions of $X_p$ coincide, up to monomial factors, with those of the $r=1$ operator $\Sf_{(1,0)}$ of \cite{Gaiotto:2012xa} under the substitution
    \begin{align}
      (p,\,q,\,t)
        \;\to\; (pq,\,q^r,\,t q^r)\;,
        \qquad
        z^2 \to W_m(z)\;,
        \label{eq:map}
    \end{align}
  and correspondingly $\omega_n = (pq/t)^{-(2r+1)/r}\,\omega_n^{r=1}\big|_{(p,q,t)\to(pq,q^r,tq^r)}$,
  where
  \begin{align}
    \omega^{r=1}_n = \frac{\theta(t;q)^2}{\theta(pq;q)^2} \frac{t\,\theta(p^n;q)\theta(t^2p^{n-1};q)}{\theta(tp^{n-1};q)\theta(tp^n;q)}
  \end{align}
  are the fusion constants for $r=1$.

\paragraph{On the factorization of $(\CS^{(0,0,1)})^r$.}
  For $r>1$, one might have expected the generalization of \eqref{eq:factorS1S3} to be $(\CS^{(0,0,1)})^r \propto \CS^{(0,1,0)}\,\CS^{(1,0,0)}$,
  since both sides contain the same $(r+1)^2$ shift operators $\Dop{h}{\delta}$, with $h=p^{(r-2M)/2}q^{(r-2M')/2}$ and $\delta=M-M'$.
  The results above show that this is not the case.
  By \eqref{eq:factor-odd}, $(\CS^{(0,0,1)})^r = \frac{pq}{t} (X_q)^r (X_p)^r$,
  while by \eqref{eq:S100-fusion} $\CS^{(0,1,0)}\CS^{(1,0,0)}$ contains, besides $(X_q)^r(X_p)^r$, the lower terms of the fusion polynomials
  (the two factors of $U$ cancel, since $U^2=1$ and $U$ commutes with $X_p$ and $X_q$).
  Only the four ``extremal'' terms, with shifts $\Dop{p^{\pm r/2}q^{\pm r/2}}{\cdot}$, agree:
    \begin{align}
      \Bigl[(\CS^{(0,0,1)})^r\Bigr]_{\rm extremal}
        &=     \frac{C^r}{K_0(p^r)K_0(q^r)}\,
              \Bigl[\CS^{(0,1,0)}\CS^{(1,0,0)}\Bigr]_{\rm extremal}\;,
              \nn \\
      C
        &=     \frac{\theta(t^{-1}; p^r) \theta(t; q^r)}{\theta((pq)^{-1}; p^r)\theta(pq; q^r)}\;,
              \label{eq:extremal}
    \end{align}
  which reduces to \eqref{eq:factorS1S3} for $r=1$.
  This holds also for even $r$ (we checked $r=2,4$),
  while the non-extremal terms of the two sides differ by $z$-dependent factors.

\paragraph{Interpretation.}
  The structure found above can be interpreted in terms of the $\Z_r$ orbifold of the elliptic modular double \cite{Spiridonov:2016uae}.
  The elementary operators $X_p$ and $X_q$, which shift the holonomy by ``one half'', are the orbifold images of the $(1,0)$ and $(0,1)$ operators of \cite{Gaiotto:2012xa},
  and \eqref{eq:factor-odd} is the orbifold image of \eqref{eq:factorS1S3}.
  For even $r$ the generators are supplemented by the central holonomy shift $U$, which enters the fusion formula \eqref{eq:S100-fusion}.
  On the other hand, the $(1,0,0)$ defect, whose residue is taken at $\alpha^2=tp^r$, is the orbifold image of the $(r,0)$ defect,
  which for $r=1$ is generated by fusion from $\Sf_{(1,0)}$ \cite{Bullimore:2014nla}
  (we have checked that the $r=1$ defects at $\alpha^2 = tp^n$ satisfy the recursion \eqref{eq:fusion} with $\omega_n$ replaced by $\omega^{r=1}_n$).

\section{Sklyanin algebra from the lens space index}
\label{sec:sklyanin}

  For $r=1$ the difference operators of \cite{Gaiotto:2012xa} are traces of $L$-operators whose entries generate the Sklyanin algebra \cite{Maruyoshi:2016caf}.
  In this section we show that the same is true on the lens space, provided the holonomy is extended appropriately.
  We use additive variables, $p=e^{2\pi i\sigma}$, $q=e^{2\pi i\tau}$, $z=e^{2\pi i x}$,
  and the Jacobi theta functions $\theta_a(\zeta|\tau)$ of appendix~\ref{app:formulas}.

\subsection{\texorpdfstring{Sklyanin algebra and $L$-operators}{Sklyanin algebra and L-operators}}

  Let $\eta$ and $\tau$ be complex parameters.
  The Sklyanin algebra is generated by $S^0,\dots,S^3$ subject to the relations which follow from the RLL relation
    \begin{align}
      R_{12}(u-v)\,L_1(u)\,L_2(v)
        =     L_2(v)\,L_1(u)\,R_{12}(u-v)\;,
        \label{eq:RLL}
    \end{align}
  where
    \begin{align}
     & L(u)
        =     \sum_{a=0}^3 \CW_a(u)\, S^a\, \boldsymbol{\sigma}^a\;,
        \qquad
      R(u)
        =     \sum_{a=0}^3 \CW_a(u)\, \boldsymbol{\sigma}^a\otimes\boldsymbol{\sigma}^a\;,
        \\
      &\qquad \CW_a(u)
        =     \displaystyle\frac{\theta_{a+1}(u+\eta|\tau)}{\theta_{a+1}(\eta|\tau)}\;,
    \end{align}
  with $\boldsymbol{\sigma}^0=1$ and $\boldsymbol{\sigma}^{1,2,3}$ the Pauli matrices (in bold face, to distinguish them from $\sigma$).
  The algebra has a representation by difference operators \cite{Sklyanin:1983ig},
    \begin{align}
      S^a
        =     \frac{i^{\delta_{a,2}}\,\theta_{a+1}(\eta|\tau)}{\theta_1(2x|\tau)}
              \Bigl[\theta_{a+1}(2x-c|\tau)\, e^{\eta\partial_x} - \theta_{a+1}(-2x-c|\tau)\, e^{-\eta\partial_x}\Bigr]\;,
              \label{eq:Sklyanin-rep}
    \end{align}
  where $c$ parametrizes the spin.
  For $r=1$, with $\eta=\sigma/2$ and $e^{-2\pi i c}=pq/t$, one finds $\Tr L(u) = 2\CW_0(u)S^0 \propto \Sf_{(1,0)}$.

\subsection{Lens Sklyanin generators}

  Consider functions $f(x,\varkappa)$ where the extended holonomy label $\varkappa$ takes values in $\Z_{2r}$, and set
    \begin{align}
      Y_\varkappa(x)
        :=    2x-\varkappa\tau\;,
        \qquad
      \eta'
        :=    \frac{\sigma+\tau}{2}\;,
        \qquad
      \tau'
        :=    r\tau\;,
        \qquad
      g_\varkappa
        :=    \left(\frac{pq}{t}\right)^{-\varkappa/r}\;,
    \end{align}
  so that $e^{2\pi i Y_\varkappa}=z^2q^{-\varkappa}$ generalizes $W_m(z)$ and $e^{4\pi i\eta'}=pq$.
  We define
    \begin{align}
      \bigl(S^a f\bigr)(x,\varkappa)
        &=   \frac{i^{\delta_{a,2}}\,\theta_{a+1}(\eta'|\tau')}{\theta_1(Y_\varkappa|\tau')}
              \Bigl[\theta_{a+1}(Y_\varkappa-c|\tau')\, g_\varkappa\, f\bigl(x+\tfrac{\sigma}{2},\varkappa-1\bigr) \nn\\
        &    \qquad\qquad\qquad
              - \theta_{a+1}(-Y_\varkappa-c|\tau')\, g_{\varkappa+1}^{-1}\, f\bigl(x-\tfrac{\sigma}{2},\varkappa+1\bigr)\Bigr]\;,
              \label{eq:lens-Sklyanin}
    \end{align}
  with $e^{-2\pi i c}=pq/t$.
  The step $(x,\varkappa)\to(x+\sigma/2,\varkappa-1)$ shifts $Y_\varkappa$ by $2\eta'$, i.e.~$W\to pqW$,
  and the factors $g_\varkappa$ compensate the quasi-periodicity of the coefficients when $\varkappa$ winds around $\Z_{2r}$.
  These operators have the following properties, which we have also checked numerically (appendix~\ref{app:numerics}).

  The operators \eqref{eq:lens-Sklyanin} satisfy the RLL relation \eqref{eq:RLL} with $(\eta,\tau)$ replaced by $(\eta',\tau')$.
  Hence they generate the Sklyanin algebra with parameters $(\eta',\tau')$.
  The RLL relation is an identity among the coefficients of the shifts at a given point $(x,\varkappa)$, which involves only neighboring values of $\varkappa$.
  Locally in $\varkappa$, the factors $g_\varkappa$ are removed by the conjugation with $\psi(\varkappa)=(pq/t)^{\varkappa(\varkappa+1)/(2r)}$.
  The coefficients then become those of \eqref{eq:Sklyanin-rep} as functions of $Y_\varkappa$, with $(\eta,\tau)$ replaced by $(\eta',\tau')$,
  and the RLL relation follows from that of \cite{Sklyanin:1983ig}.
  Globally, the coefficients including $g_\varkappa$ are periodic under $\varkappa\to\varkappa+2r$ (see below), and the relation holds on $\Z_{2r}$.

  If $\varkappa$ is restricted to $\Z_r$ instead of $\Z_{2r}$, the RLL relation fails.
  The reason is that under $\varkappa\to\varkappa+r$ the coefficients of $S^a$ change by the signs $(\epsilon_0,\epsilon_1,\epsilon_2,\epsilon_3)=(+,-,-,+)$, coming from the quasi-periodicity of the Jacobi theta functions under $Y\to Y-\tau'$.
  Consequently, with $U: f(x,\varkappa)\mapsto f(x,\varkappa+r)$,
    \begin{align}
      U S^a U^{-1}
        =     \epsilon_a\, S^a\;.
    \end{align}
  $S^0$ and $S^3$ preserve the subspace of $U$-invariant functions, while $S^1$ and $S^2$ map it to the $U$-anti-invariant subspace.
  For odd $r$ the $U$-invariant subspace contains the index of the $\SU(2)$ theory (section~\ref{subsec:elementary}).
  For even $r$ the $\SU(2)$ index is supported on even $\varkappa$, on which $U$ acts as the central holonomy shift $m\to m+r/2$.
  Every $S^a$ maps it to the twisted sector with odd $\varkappa$, so that only products of an even number of generators act within the $\SU(2)$ theory.

  Finally, $X_p$ in \eqref{eq:Xp-kappa} coincides with $S^0$, up to normalization and a conjugation by an exponential function of $x$:
    \begin{align}
      X_p
        =     \frac{\theta(t;q^r)}{\theta(pq;q^r)}\,
              \frac{(pq/t)^{\frac{r-1}{2r}}}{\theta_1(\eta'|\tau')}\;
              \phi^{-1} S^0 \phi\;,
        \qquad
      \phi(x)
        =     \left(\frac{pq}{t}\right)^{\frac{x}{r\sigma}}\;.
                \label{eq:Xp-S0}
    \end{align}
  Thus the elementary defect is the trace of the lens $L$-operator, $\Tr L(u)=2\CW_0(u)S^0$.

  In this way $\Z_r$ acts on the holonomy label, as anticipated in section~\ref{sec:algebra}, and its extension to $\Z_{2r}$ is needed for the algebra to close.
  The parameters of the lens Sklyanin algebra are related to those for $r=1$ by \eqref{eq:map}: the shift $\eta=\sigma/2$ is replaced by $\eta'$, corresponding to $W\to pqW$, and the nome $\tau$ by $r\tau$.

\paragraph{Remark on the case $r=1$.}
  Even for $r=1$, the construction with $\varkappa\in\Z_2$ realizes $\Sf_{(1,0)}$ as $S^0$ of the Sklyanin algebra with $\eta'=\sigma/2+\tau/2$ rather than $\eta=\sigma/2$,
  since the elementary shift acts on $W=z^2q^{-\varkappa}$ by $pq$ rather than by $p$.
  The two algebras are not isomorphic: a shift of $\eta$ by a half-period is not an automorphism of the pair consisting of the elliptic curve and the point $\eta$,
  and their structure constants are not related by a permutation of the generators.
  It would be interesting to decompose the representation \eqref{eq:lens-Sklyanin} into irreducible components of the Sklyanin algebra.

\subsection{Lens elliptic modular double}

  The $q$-type generators $\tilde{S}^a$ are defined analogously, with $\tau'$ replaced by $\tau''=r\sigma$, $Y_\varkappa$ by $\tilde{Y}_\varkappa = 2x+\varkappa\sigma$,
  the steps $(x,\varkappa)\to(x\pm\tau/2,\varkappa\pm1)$, and $g_\varkappa$ by $\tilde{g}_\varkappa=(pq/t)^{\varkappa/r}$.
  The same $\eta'$ appears in both algebras.
  The $p$-steps leave $\tilde{Y}_\varkappa$ invariant and the $q$-steps leave $Y_\varkappa$ invariant.
  We find
    \begin{align}
      \bigl[S^a, \tilde{S}^b\bigr]
        =     0\;,
        \qquad a,b=0,1,2,3\;,
    \end{align}
  which we have checked numerically.
  The two commuting Sklyanin algebras with parameters $(\eta',r\tau)$ and $(\eta',r\sigma)$ form a lens space version of the elliptic modular double,
  which is closely related to the rarefied elliptic hypergeometric functions of \cite{Spiridonov:2016uae} (see also \cite{Sarkissian:2018ppc,Kels:2018xnf}).
  See \cite{Faddeev:1999fe} for the modular double, \cite{Spiridonov:2008Essays,Derkachov:2013uca,Lodin:2017lrc} for the elliptic modular double and its role in 4d partition functions, and \cite{Bulbul:2025ryz} for a lens version of the hyperbolic modular double.

\section{\texorpdfstring{Surface defects in class $\CS_k$ theories on lens space}{Surface defects in class S\_k theories on lens space}}
\label{sec:class-Sk}
  In this section we consider surface defects in class $\CS_k$ theories, which are $\CN=1$ SCFTs generalizing the class $\CS$ theories.
  We will see that these theories have a structure similar to that of class $\CS$, so that the surface defects can again be expressed in terms of the lens Sklyanin generators, more precisely as transfer matrices built from the $L$-operators.
  We first review the class $\CS_k$ theories in section~\ref{sec:Sk-review}.
  We then compute the lens index of the class $\CS_k$ theory with a surface defect in section~\ref{subsec:surfaceSk},
  and relate the result to transfer matrices in section~\ref{subsec:Sk-Sklyanin}.

\subsection{\texorpdfstring{Class $\CS_k$ theories}{Class S\_k theories}}
\label{sec:Sk-review}
  In this subsection we review the $\CN=1$ theories of class $\CS_k$ \cite{Gaiotto:2015usa,Franco:2015jna,Hanany:2015pfa}.
  The $\CN=2$ theories of class $\CS$ correspond to $k=1$.
  In this review we keep $N$ general, and we will set $N=2$ when we actually carry out the index computation.

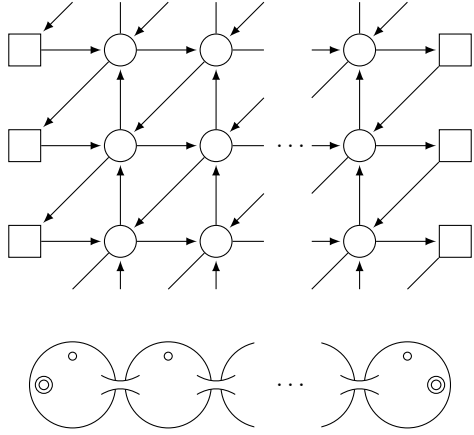
\begin{figure}
  \centering
  \begin{tikzpicture}[baseline=(x)]
    \node[fnode] (a0) at (0,0) {};
    \node[fnode] (b0) at (0,-1) {};
    \node[fnode] (c0) at (0,-2) {};
    \node[gnode] (a1) at (1,0) {};
    \node[gnode] (b1) at (1,-1) {};
    \node[gnode] (c1) at (1,-2) {};
    \node[gnode] (a2) at (2,0) {};
    \node[gnode] (b2) at (2,-1) {};
    \node[gnode] (c2) at (2,-2) {};

    \begin{scope}[shift={(0.5,0)}]
      \node[gnode] (a3) at (3,0) {};
      \node[gnode] (b3) at (3,-1) {};
      \node[gnode] (c3) at (3,-2) {};
      \node[fnode] (a4) at (4,0) {};
      \node[fnode] (b4) at (4,-1) {};
      \node[fnode] (c4) at (4,-2) {};
    \end{scope}

    \draw[q->] (a1) -- (b0);
    \draw[q->] (b1) -- (c0);

    \draw[q->] (a2) -- (b1);
    \draw[q->] (b2) -- (c1);

    \draw[q->] (a4) -- (b3);
    \draw[q->] (b4) -- (c3);

    \draw[q->] (a0) -- (a1);
    \draw[q->] (a1) -- (a2);
    \draw[q->] (a3) -- (a4);

    \draw[q->] (b0) -- (b1);
    \draw[q->] (b1) -- (b2);
    \draw[q->] (b3) -- (b4);

    \draw[q->] (c0) -- (c1);
    \draw[q->] (c1) -- (c2);
    \draw[q->] (c3) -- (c4);

    \draw[q->] (b1) -- (a1);
    \draw[q->] (c1) -- (b1);

    \draw[q->] (b2) -- (a2);
    \draw[q->] (c2) -- (b2);

    \draw[q->] (b3) -- (a3);
    \draw[q->] (c3) -- (b3);

    \draw[q->] ($(a0) + (0.5,0.5)$) -- (a0);
    \draw[q->] ($(a1) + (0.5,0.5)$) -- (a1);
    \draw[q->] ($(a2) + (0.5,0.5)$) -- (a2);
    \draw[q->] ($(a3) + (0.5,0.5)$) -- (a3);
    \draw[q->] ($(b2) + (0.5,0.5)$) -- (b2);
    \draw[q->] ($(c2) + (0.5,0.5)$) -- (c2);

    \draw[q-] (a3) -- ($(a3) + (-0.5,-0.5)$);
    \draw[q-] (b3) -- ($(b3) + (-0.5,-0.5)$);
    \draw[q-] (c1)  -- ($(c1) + (-0.5,-0.5)$);
    \draw[q-] (c2)  -- ($(c2) + (-0.5,-0.5)$);
    \draw[q-] (c3)  -- ($(c3) + (-0.5,-0.5)$);
    \draw[q-] (c4)  -- ($(c4) + (-0.5,-0.5)$);

    \draw[q-] (a1) -- ($(a1) + (0,0.5)$);
    \draw[q-] (a2) -- ($(a2) + (0,0.5)$);
    \draw[q-] (a3) -- ($(a3) + (0,0.5)$);

    \draw[q->] ($(c1) + (0,-0.5)$) -- (c1);
    \draw[q->] ($(c2) + (0,-0.5)$) -- (c2);
    \draw[q->] ($(c3) + (0,-0.5)$) -- (c3);

    \draw[q-] (a2) -- ($(a2) + (0.5,0)$);
    \draw[q-] (b2) -- ($(b2) + (0.5,0)$);
    \draw[q-] (c2) -- ($(c2) + (0.5,0)$);

    \draw[q->] ($(a3) + (-0.5,0)$) -- (a3);
    \draw[q->] ($(b3) + (-0.5,0)$) -- (b3);
    \draw[q->] ($(c3) + (-0.5,0)$) -- (c3);

    \node (x) at (2.8,-1) {$\dots$};

  \begin{scope}[shift={(0,-3.5)}]
    \node[maxp] at (0.2,0) {};
    \node[minp] (min1) at (0.5,0.3) {};
    \node[minp] (min2) at (1.5,0.3) {};
    \node[minp] (min5) at (4,0.3) {};
    \node[maxp] (maxt) at (4.3,0) {};

    \node at (2.8,0) {$\dots$};

    \clip (0,-0.5) rectangle (2.4,0.5) (3.1,-0.5) rectangle (4.5,0.5);
    \draw (0.5,0) circle (0.45);
    \draw (1.5,0) circle (0.45);
    \draw (2.5,0) circle (0.45);
    \draw (3,0) circle (0.45);
    \draw (4,0) circle (0.45);
    \fill[white] (0.8,-0.1) to[bend left] (1.2, -0.1) -- (1.2,0.1)
    to[bend left]  (0.8,0.1) -- cycle;
    \draw (0.8,-0.1) to[bend left] (1.2, -0.1) (1.2,0.1)  to[bend left]  (0.8,0.1);
    \fill[white] (1.8,-0.1) to[bend left] (2.2, -0.1) -- (2.2,0.1)
    to[bend left]  (1.8,0.1) -- cycle;
    \draw (1.8,-0.1) to[bend left] (2.2, -0.1) (2.2,0.1)  to[bend left]  (1.8,0.1);
    \fill[white] (3.3,-0.1) to[bend left] (3.7, -0.1) -- (3.7,0.1)
    to[bend left]  (3.3,0.1) -- cycle;
    \draw (3.3,-0.1) to[bend left] (3.7, -0.1) (3.7,0.1)  to[bend left]  (3.3,0.1);
  \end{scope}
  \end{tikzpicture}
  \caption{A brane tiling model as a class $\CS_k$ theory associated
    with a punctured sphere.}
  \label{fig:class-Sk}
\end{figure}

  A typical class $\CS_k$ theory is represented by the quiver diagram in figure~\ref{fig:class-Sk},
  where the vertical direction is periodic, whereas the horizontal direction is either a finite interval or periodic.
  Each node corresponds to an $\SU(N)$ gauge or flavor group.
  The theory is obtained by the twisted compactification of the six-dimensional $\CN=(1,0)$ theory on the Riemann surface $\Sigma$ (in this case a punctured sphere).
  For the quiver in figure~\ref{fig:class-Sk}, $k$ is the number of nodes in the vertical direction.

  The global symmetry of the six-dimensional theory is $\SU(2)_R \times \U(1)_t \times \SU(k)_\beta \times \SU(k)_\gamma$.
  By performing the topological twist along $\Sigma$ by the Cartan of $\SU(2)_R$, we can introduce Wilson lines for $\SU(k)_\beta \times \SU(k)_\gamma$, preserving their maximal tori.
  Thus the preserved symmetry in $4d$ is $\U(1)_R \times \U(1)_t \times \mathrm{S}[\prod_{i=1}^k \U(1)_{\beta_i}] \times \mathrm{S}[\prod_{i=1}^k \U(1)_{\gamma_i}]$.

  A puncture of $\Sigma$ gives rise to an additional global symmetry in the four-dimensional theory.
  The minimal puncture gives the $\U(1)_\alpha$ symmetry, whose fugacity we denote by $\alpha_i$.
  The maximal puncture gives rise to a set of $k$ $\SU(N)$ flavor symmetries, represented in the quiver in the figure by a column of $k$ flavor nodes
  at each end.
  In addition, a maximal puncture carries labels called ``color'' $\pc \in \Z_k$ and ``orientation'' $\po \in \{\pm 1\}$.

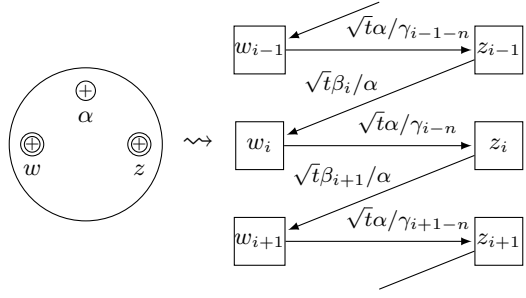
\begin{figure}
  \centering
  \begin{tikzpicture}[scale=0.8, baseline=(x.base)]
    \node (x) at (0,0) {\vphantom{x}};

    \draw (0,0) circle (1);

    \node[maxp, label={below:$w$}] (w) at (-0.7,0) {$+$};
    \node[maxp, label={below:$z$}] (z) at (0.7,0) {$+$};
    \node[minp, label={below:$\alpha$}] (min) at (0,0.7) {$+$};
  \end{tikzpicture}
  \
  {\large $\leadsto$}
  \
  \begin{tikzpicture}[xscale=2.5, baseline=(wi.base)]
    \node[fnode, minimum size=18pt] (wi-1) at (0,1) {$w_{i-1}$};
    \node[fnode, minimum size=18pt] (wi) at (0,0) {$w_i$};
    \node[fnode, minimum size=18pt] (wi+1) at (0,-1) {$w_{i+1}$};
    \node[fnode, minimum size=18pt] (zi-1) at (1,1) {$z_{i-1}$};
    \node[fnode, minimum size=18pt] (zi) at (1,0) {$z_i$};
    \node[fnode, minimum size=18pt] (zi+1) at (1,-1) {$z_{i+1}$};

    \draw[q->] (0.5,1.5) -- (wi-1);
    \draw[q->] (zi-1) -- (wi)
    node[above=3pt, pos=0.7]{$\sqrt{t} \beta_i/\alpha$};
    \draw[q->] (zi) -- (wi+1)
    node[above=4pt, pos=0.7] {$\sqrt{t} \beta_{i+1}/\alpha$};
    \draw[q-] (zi+1) -- (0.5,-1.5);

    \draw[q->] (wi-1) -- (zi-1)
    node[above, pos=0.65] {$\sqrt{t} \alpha/\gamma_{i-1-n}$};
    \draw[q->] (wi) -- (zi)
    node[above, pos=0.65] {$\sqrt{t} \alpha/\gamma_{i-n}$};
    \draw[q->] (wi+1) -- (zi+1)
    node[above, pos=0.65] {$\sqrt{t} \alpha/\gamma_{i+1-n}$};
  \end{tikzpicture}

  \caption{The building block of class $\CS_k$ theories associated with a trinion with two maximal punctures and one minimal puncture. The left maximal puncture has $(\pc, \po) = (n+1, +1)$ and the right one has $(\pc, \po) = (n, -1)$.}
  \label{fig:trinion+++Sk}
\end{figure}

  A building block of the quiver is a strip of bi-fundamental chiral multiplets depicted in figure~\ref{fig:trinion+++Sk}, associated with a trinion with one minimal and two maximal punctures.
  The orientation simply distinguishes the two maximal punctures in the trinion.
  In our pictures, we will always place the maximal puncture with positive orientation on the left and the one with negative orientation on the right.
  The color and the orientation are defined by the relation between fugacities such that arrows with fugacities $\sqrt{t} \beta_i/\alpha$ and $\sqrt{t} \alpha/\gamma_{i-\pc+\po}$ start from
  or end at the same node in the column of nodes corresponding to a maximal puncture labeled $(\pc, \po)$.
  The color of the positively oriented puncture is greater than that of the negatively oriented puncture by one.

  Two maximal punctures can be glued if they have opposite orientations and the same color. 
  This operation corresponds to the gauging of the diagonal combination of the $\SU(N)^k$ flavor symmetries of the maximal punctures, and at the same time to adding bi-fundamental chiral multiplets corresponding to arrows going upward between the gauged nodes.
  The restriction on the color and orientation ensures that the mixed anomalies for $\U(1)_{\beta_i}$ and $\U(1)_{\gamma_i}$ cancel.

  When compactifying the six-dimensional theory, one can turn on flux for the abelian part of its flavor symmetry, which produces a family of theories associated with the same $\Sigma$ but with different flux backgrounds for $\U(1)_t$, $\U(1)_{\beta_i}$ and $\U(1)_{\gamma_i}$ \cite{Razamat:2016dpl,Bah:2017gph}.
  For example, turning on the flux for $\U(1)_{\beta_i}$ proceeds as follows \cite{Gaiotto:2015usa}: let us create flux for $\U(1)_{\beta_*}$ in a class $\CS_k$ theory associated with $\Sigma$, where $* \in \Z_k$ is a fixed index. 
  To do that, we glue a trinion theory to a maximal puncture of $\Sigma$. 
  The new surface thus obtained has one more minimal puncture than $\Sigma$, and hence one more flavor symmetry $\U(1)_\alpha$. 
  Then, we close this puncture by giving a constant vev to the baryon made of the bi-fundamental chiral multiplet with fugacity $\sqrt{t} \beta_*/\alpha$, and flip the other baryons in the same column whose fugacities are $(\sqrt{t} \beta_i/\alpha)^N$ with $i \neq *$.  
  (By flipping a chiral operator $X$, we mean coupling $X$ to a free chiral multiplet $\phi_X$ through a superpotential $\phi_X X$.) 
  After closing the puncture, we obtain a theory associated with $\Sigma$ with the color of the maximal puncture shifted by one.  
  The result is interpreted as a theory with one unit of $\U(1)_{\beta_*}$ flux turned on. 
  Similarly, one can turn on (minus) one unit of $\U(1)_{\gamma_*}$ flux by giving a vev to the baryon with fugacity $(\sqrt{t} \alpha/\gamma_*)^N$.

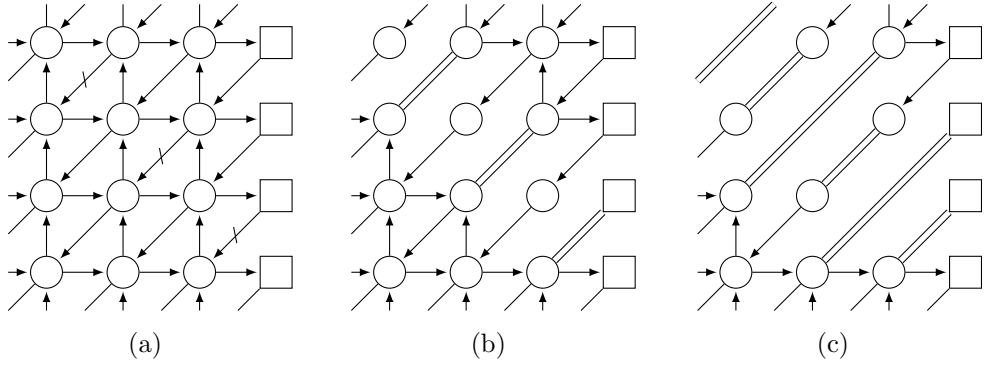
\begin{figure}
  \centering
  \subcaptionbox{\label{fig:closing-all-a}}{
    \begin{tikzpicture}[scale=0.8]
      \node[gnode] (a1) at (1,0) {};
      \node[gnode] (b1) at (1,-1) {};
      \node[gnode] (c1) at (1,-2) {};
      \node[gnode] (d1) at (1,-3) {};
      \node[gnode] (a2) at (2,0) {};
      \node[gnode] (b2) at (2,-1) {};
      \node[gnode] (c2) at (2,-2) {};
      \node[gnode] (d2) at (2,-3) {};
      \node[gnode] (a3) at (3,0) {};
      \node[gnode] (b3) at (3,-1) {};
      \node[gnode] (c3) at (3,-2) {};
      \node[gnode] (d3) at (3,-3) {};
      \node[fnode] (a4) at (4,0) {};
      \node[fnode] (b4) at (4,-1) {};
      \node[fnode] (c4) at (4,-2) {};
      \node[fnode] (d4) at (4,-3) {};

      \draw[q->,vev] (a2) -- (b1);
      \draw[q->] (b2) -- (c1);
      \draw[q->] (c2) -- (d1);

      \draw[q->] (a3) -- (b2);
      \draw[q->,vev] (b3) -- (c2);
      \draw[q->] (c3) -- (d2);

      \draw[q->] (a4) -- (b3);
      \draw[q->] (b4) -- (c3);
      \draw[q->,vev] (c4) -- (d3);

      \draw[q->] (a1) -- (a2);
      \draw[q->] (a2) -- (a3);
      \draw[q->] (a3) -- (a4);

      \draw[q->] (b1) -- (b2);
      \draw[q->] (b2) -- (b3);
      \draw[q->] (b3) -- (b4);

      \draw[q->] (c1) -- (c2);
      \draw[q->] (c2) -- (c3);
      \draw[q->] (c3) -- (c4);

      \draw[q->] (d1) -- (d2);
      \draw[q->] (d2) -- (d3);
      \draw[q->] (d3) -- (d4);

      \draw[q->] (b1) -- (a1);
      \draw[q->] (c1) -- (b1);
      \draw[q->] (d1) -- (c1);

      \draw[q->] (b2) -- (a2);
      \draw[q->] (c2) -- (b2);
      \draw[q->] (d2) -- (c2);

      \draw[q->] (b3) -- (a3);
      \draw[q->] (c3) -- (b3);
      \draw[q->] (d3) -- (c3);

      \draw[q->] ($(a1) + (0.5,0.5)$) -- (a1);
      \draw[q->] ($(a2) + (0.5,0.5)$) -- (a2);
      \draw[q->] ($(a3) + (0.5,0.5)$) -- (a3);

      \draw[q-] (a1) -- ($(a1) + (0,0.5)$);
      \draw[q-] (a2) -- ($(a2) + (0,0.5)$);
      \draw[q-] (a3) -- ($(a3) + (0,0.5)$);

      \draw[q->] ($(d1) + (0,-0.5)$) -- (d1);
      \draw[q->] ($(d2) + (0,-0.5)$) -- (d2);
      \draw[q->] ($(d3) + (0,-0.5)$) -- (d3);

      \draw[q->] ($(a1) + (-0.5,0)$) -- (a1);
      \draw[q->] ($(b1) + (-0.5,0)$) -- (b1);
      \draw[q->] ($(c1) + (-0.5,0)$) -- (c1);
      \draw[q->] ($(d1) + (-0.5,0)$) -- (d1);

      \draw[q-] (a1) -- ($(a1) + (-0.5,-0.5)$);
      \draw[q-] (b1) -- ($(b1) + (-0.5,-0.5)$);
      \draw[q-] (c1) -- ($(c1) + (-0.5,-0.5)$);
      \draw[q-] (d1) -- ($(d1) + (-0.5,-0.5)$);

      \draw[q-] (d2) -- ($(d2) + (-0.5,-0.5)$);
      \draw[q-] (d3) -- ($(d3) + (-0.5,-0.5)$);
      \draw[q-] (d4) -- ($(d4) + (-0.5,-0.5)$);
    \end{tikzpicture}
  }
  \quad
  \subcaptionbox{\label{fig:closing-all-b}}{
    \begin{tikzpicture}[scale=0.8]
      \node[gnode] (a1) at (1,0) {};
      \node[gnode] (b1) at (1,-1) {};
      \node[gnode] (c1) at (1,-2) {};
      \node[gnode] (d1) at (1,-3) {};
      \node[gnode] (a2) at (2,0) {};
      \node[gnode] (b2) at (2,-1) {};
      \node[gnode] (c2) at (2,-2) {};
      \node[gnode] (d2) at (2,-3) {};
      \node[gnode] (a3) at (3,0) {};
      \node[gnode] (b3) at (3,-1) {};
      \node[gnode] (c3) at (3,-2) {};
      \node[gnode] (d3) at (3,-3) {};
      \node[fnode] (a4) at (4,0) {};
      \node[fnode] (b4) at (4,-1) {};
      \node[fnode] (c4) at (4,-2) {};
      \node[fnode] (d4) at (4,-3) {};

      \draw[eq-] (a2) -- (b1);
      \draw[q->] (b2) -- (c1);
      \draw[q->] (c2) -- (d1);

      \draw[q->] (a3) -- (b2);
      \draw[eq-] (b3) -- (c2);
      \draw[q->] (c3) -- (d2);

      \draw[q->] (a4) -- (b3);
      \draw[q->] (b4) -- (c3);
      \draw[eq-] (c4) -- (d3);

      \draw[q->] (a2) -- (a3);
      \draw[q->] (a3) -- (a4);

      \draw[q->] (b3) -- (b4);

      \draw[q->] (c1) -- (c2);

      \draw[q->] (d1) -- (d2);
      \draw[q->] (d2) -- (d3);
      \draw[q->] (d3) -- (d4);

      \draw[q->] (c1) -- (b1);
      \draw[q->] (d1) -- (c1);

      \draw[q->] (d2) -- (c2);

      \draw[q->] (b3) -- (a3);

      \draw[q->] ($(a1) + (0.5,0.5)$) -- (a1);
      \draw[q->] ($(a2) + (0.5,0.5)$) -- (a2);
      \draw[q->] ($(a3) + (0.5,0.5)$) -- (a3);

      \draw[q-] (a2) -- ($(a2) + (0,0.5)$);
      \draw[q-] (a3) -- ($(a3) + (0,0.5)$);

      \draw[q->] ($(d1) + (0,-0.5)$) -- (d1);
      \draw[q->] ($(d2) + (0,-0.5)$) -- (d2);
      \draw[q->] ($(d3) + (0,-0.5)$) -- (d3);

      \draw[q->] ($(b1) + (-0.5,0)$) -- (b1);
      \draw[q->] ($(c1) + (-0.5,0)$) -- (c1);
      \draw[q->] ($(d1) + (-0.5,0)$) -- (d1);

      \draw[q-] (a1) -- ($(a1) + (-0.5,-0.5)$);
      \draw[q-] (b1) -- ($(b1) + (-0.5,-0.5)$);
      \draw[q-] (c1) -- ($(c1) + (-0.5,-0.5)$);
      \draw[q-] (d1) -- ($(d1) + (-0.5,-0.5)$);

      \draw[q-] (d2) -- ($(d2) + (-0.5,-0.5)$);
      \draw[q-] (d3) -- ($(d3) + (-0.5,-0.5)$);
      \draw[q-] (d4) -- ($(d4) + (-0.5,-0.5)$);
    \end{tikzpicture}
  }
  \quad
  \subcaptionbox{\label{fig:closing-all-c}}{
    \begin{tikzpicture}[scale=0.8]
      \node (a1) at (1,0) {};
      \node[gnode] (b1) at (1,-1) {};
      \node[gnode] (c1) at (1,-2) {};
      \node[gnode] (d1) at (1,-3) {};
      \node[gnode] (a2) at (2,0) {};
      \node[gnode] (c2) at (2,-2) {};
      \node[gnode] (d2) at (2,-3) {};
      \node[gnode] (a3) at (3,0) {};
      \node[gnode] (b3) at (3,-1) {};
      \node[gnode] (d3) at (3,-3) {};
      \node[fnode] (a4) at (4,0) {};
      \node[fnode] (b4) at (4,-1) {};
      \node[fnode] (c4) at (4,-2) {};
      \node[fnode] (d4) at (4,-3) {};

      \draw[eq-] (a3) -- (c1);
      \draw[eq-] (b4) -- (d2);

      \draw[eq-] (a2) -- (b1);
      \draw[q->] (c2) -- (d1);

      \draw[eq-] (b3) -- (c2);

      \draw[q->] (a4) -- (b3);
      \draw[eq-] (c4) -- (d3);

      \draw[q->] (a3) -- (a4);

      \draw[q->] (d1) -- (d2);
      \draw[q->] (d2) -- (d3);
      \draw[q->] (d3) -- (d4);

      \draw[q->] (d1) -- (c1);

      \draw[eq-] ($(a1) + (0.5,0.5)$) -- ($(a1) + (-0.5,-0.5)$);

      \draw[q->] ($(a2) + (0.5,0.5)$) -- (a2);
      \draw[q->] ($(a3) + (0.5,0.5)$) -- (a3);

      \draw[q-] (a3) -- ($(a3) + (0,0.5)$);

      \draw[q->] ($(d1) + (0,-0.5)$) -- (d1);
      \draw[q->] ($(d2) + (0,-0.5)$) -- (d2);
      \draw[q->] ($(d3) + (0,-0.5)$) -- (d3);

      \draw[q->] ($(c1) + (-0.5,0)$) -- (c1);
      \draw[q->] ($(d1) + (-0.5,0)$) -- (d1);

      \draw[q-] (b1) -- ($(b1) + (-0.5,-0.5)$);
      \draw[q-] (c1) -- ($(c1) + (-0.5,-0.5)$);
      \draw[q-] (d1) -- ($(d1) + (-0.5,-0.5)$);

      \draw[q-] (d2) -- ($(d2) + (-0.5,-0.5)$);
      \draw[q-] (d3) -- ($(d3) + (-0.5,-0.5)$);
      \draw[q-] (d4) -- ($(d4) + (-0.5,-0.5)$);
    \end{tikzpicture}
  }
  \caption{In figure~\ref{fig:closing-all-a} the vevs are given to the baryons at the arrows on the diagonal. We apply \eqref{eq:massterm} and \eqref{eq:NfN} repeatedly to get figures \ref{fig:closing-all-b} and \ref{fig:closing-all-c}. This corresponds to closing the $k$ minimal punctures.}
    \label{fig:closing-all}
\end{figure}

  An important point here is that since the $\U(1)_{\beta_i}$ symmetries come from the $\SU(k)_\beta$ symmetry in six dimensions, adding one unit of $\U(1)_{\beta_i}$ flux for every $i$ is equivalent to doing nothing (and similarly for the $\U(1)_{\gamma_i}$ flux).
  To add one unit of flux for every $\U(1)_{\beta_i}$ in a given theory, say $\CT_{\rm IR}$, we consider a sphere with two maximal punctures and $k$ minimal punctures, and glue one of its maximal punctures to that of the given theory. 
  We refer to the resulting theory as $\CT_{\rm UV}$.
  We then give vevs to $k$ baryons charged under distinct $\U(1)_{\beta_i}$.  (Let us suppose that the maximal puncture to which we attach the sphere has $\po = -1$.)
  We can perform this at the level of the (lens) index.
  For simplicity we assume $N=2$.
  Giving a constant vev to a bi-fundamental multiplet $\begin{tikzpicture}
      \node[gnode] (x) at (0,0) {$x$};
      \node[fnode] (z) at (1,0) {$z$};
      \draw[q->] (x) -- node[above] {$ \frac{\sqrt{t} \beta}{\alpha} $} (z);
      \end{tikzpicture}$
  corresponds to computing the residue at $\alpha = \sqrt{t} \beta$, as in section \ref{sec:class-S}. 
  In the quiver we denote this operation as follows, and it turns out that 
    \begin{align}
      \begin{tikzpicture}
      \node[gnode] (x) at (0,0) {$x$};
      \node[fnode] (z) at (1,0) {$z$};
      \draw[q->, vev] (x) -- node[above] {$\frac{\sqrt{t}\beta}{\alpha}$} (z);
      \end{tikzpicture}
      ~:=
      2 \Xi \Res_{\alpha = \sqrt{t} \beta} \left[ \frac{1}{\alpha} \left(
      \begin{tikzpicture}
      \node[gnode] (x) at (0,0) {$x$};
      \node[fnode] (z) at (1,0) {$z$};
      \draw[q->] (x) -- node[above] {$\frac{\sqrt{t}\beta}{\alpha}$} (z);
      \end{tikzpicture}
      \right) \right]
       =     \begin{tikzpicture}[baseline=(z.base)]
      \node[gnode] (x) at (0,0) {$x$};
      \node[fnode] (z) at (1,0) {$z$};
      \draw[eq-] (x) -- (z);
      \end{tikzpicture}
      ~\;.
      \label{eq:giving-vev}
    \end{align}
  Closing the $k$ minimal punctures corresponds to giving vevs to the baryons built out of bi-fundamentals in $\CT_{\rm UV}$, as in figure~\ref{fig:closing-all-a}.
  By \eqref{eq:giving-vev}, the vevs turn the cubic superpotential terms involving the stuck-out arrows into mass terms,
  so that the corresponding bi-fundamentals are integrated out by \eqref{eq:massterm}, and the quiver becomes that of figure~\ref{fig:closing-all-b}.
  Figure~\ref{fig:closing-all-b} contains a number of gauge nodes with two arrows attached, which exhibit confinement and chiral symmetry breaking.
  By the identity \eqref{eq:NfN}, further pairs of nodes are identified, as in figure~\ref{fig:closing-all-c}.
  Repeating this process until all the flavor nodes at the end are identified with the gauge nodes, we recover $\CT_{\rm IR}$.

\subsection{Construction of the surface defects}
\label{subsec:surfaceSk}
  Surface defects in class $\CS_k$ theories are realized via RG flows as discussed in \cite{Gaiotto:2015usa,Maruyoshi:2016caf}.
  A surface defect in $\CT_{\rm IR}$ is realized in a similar way to the closing of the minimal punctures in section~\ref{sec:Sk-review}.
  We focus on $\CT_{\rm IR}$ with a maximal puncture with $(\pc, \po) = (0, -1)$. 
  We first construct $\CT_{\rm UV}$ by gluing to $\CT_{\rm IR}$ a sphere with two maximal and $k$ minimal punctures.
  Next, we give constant vevs to the baryons with fugacities $(\sqrt{t} \beta_i/\alpha_i)^N$ with $i \neq *$.
  Finally, we give the position-dependent vev to the baryon with fugacity $(\sqrt{t} \beta_*/\alpha_*)^N$.
  There are $k$ choices for the index $*$, and each choice leads to a different surface defect.
  Of course, we could follow the same procedure with the roles of $\beta_i$ and $\gamma_i$ exchanged, so in total there are $2k$ inequivalent surface defects.

  There are further labels of the surface defect coming from the fact that the theory is placed on the lens space.
  They are determined by the pinching of the poles of the bi-fundamental with fugacity $\sqrt{t}\beta_*/\alpha$ as in section~\ref{subsec:poles}:
  the poles are at $\alpha^2 = t\beta_*^2 p^aq^b$ with $\U(1)_\alpha$ holonomy $\mu$ satisfying \eqref{eq:pole-condition},
  and for $\mu=0$ we label them as
    \begin{align}
      \alpha
       =     \sqrt{t} \beta_* (p^r)^{\frac{n_1}{2}} (q^r)^{\frac{n_2}{2}} (pq)^{\frac{n_3}{2}}
    \end{align}
  with nonnegative integers $n_1$, $n_2$ and $n_3$.
  Therefore, the surface defects in this construction are labeled by $(n_1, n_2, n_3)$ (and $\mu$),
  in addition to the $2k$ labels by $\beta_i$ and $\gamma_i$.
  The pinching in the $(p,p)$, $(q,q)$, $(p,q)$ and $(q,p)$ sectors proceeds as in section~\ref{subsec:poles}.

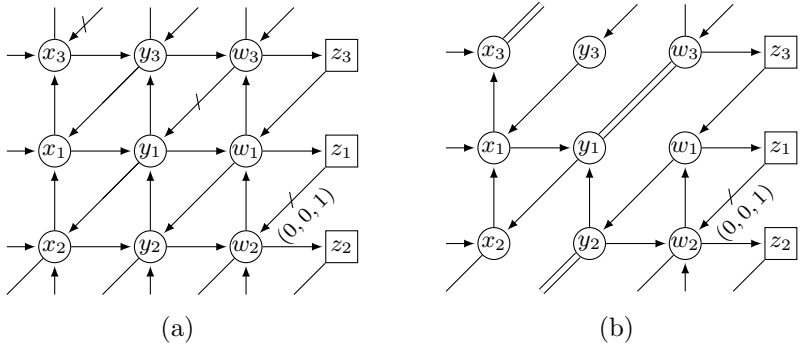
\begin{figure}
  \centering
  \subcaptionbox{\label{fig:surface-defect-Ska}}{
    \begin{tikzpicture}
      \node[gnode] (a1) at (1,0) {$x_3$};
      \node[gnode] (b1) at (1,-1) {$x_1$};
      \node[gnode] (c1) at (1,-2) {$x_2$};
      \node[gnode] (a2) at (2,0) {$y_3$};
      \node[gnode] (b2) at (2,-1) {$y_1$};
      \node[gnode] (c2) at (2,-2) {$y_2$};
      \node[gnode] (a3) at (3,0) {$w_3$};
      \node[gnode] (b3) at (3,-1) {$w_1$};
      \node[gnode] (c3) at (3,-2) {$w_2$};
      \node[fnode] (a4) at (4,0) {$z_3$};
      \node[fnode] (b4) at (4,-1) {$z_1$};
      \node[fnode] (c4) at (4,-2) {$z_2$};

      \draw[q->,vev] (a3) -- (b2);
      \draw[q->] (b3) -- (c2);

      \draw[q->] (a4) -- (b3);
      \draw[q->,vev] (b4) -- node[below, sloped] {$(0,0,1)$} (c3);

      \draw[q->] (a2) -- (a3);
      \draw[q->] (a3) -- (a4);

      \draw[q->] (b2) -- (b3);
      \draw[q->] (b3) -- (b4);

      \draw[q->] (c2) -- (c3);
      \draw[q->] (c3) -- (c4);

      \draw[q->] (b2) -- (a2);
      \draw[q->] (c2) -- (b2);

      \draw[q->] (b3) -- (a3);
      \draw[q->] (c3) -- (b3);

      \draw[q->] (a2) -- (b1);
      \draw[q->] (b2) -- (c1);

      \draw[q->] (b1) -- (a1);
      \draw[q->] (c1) -- (b1);

      \draw[q->, vev] ($(a1) + (0.5,0.5)$) -- (a1);
      \draw[q->] ($(a2) + (0.5,0.5)$) -- (a2);
      \draw[q->] ($(a3) + (0.5,0.5)$) -- (a3);

      \draw[q-] (a1) -- ($(a1) + (0,0.5)$);
      \draw[q-] (a2) -- ($(a2) + (0,0.5)$);
      \draw[q-] (a3) -- ($(a3) + (0,0.5)$);

      \draw[q->] ($(c1) + (0,-0.5)$) -- (c1);
      \draw[q->] ($(c2) + (0,-0.5)$) -- (c2);
      \draw[q->] ($(c3) + (0,-0.5)$) -- (c3);

      \draw[q->] (a1) -- (a2);
      \draw[q->] (b1)  -- (b2);
      \draw[q->] (c1)  -- (c2);

      \draw[q->] ($(a1) + (-0.5,0)$) -- (a1);
      \draw[q->] ($(b1) + (-0.5,0)$) -- (b1);
      \draw[q->] ($(c1) + (-0.5,0)$) -- (c1);

      \draw[q-] (a2) -- ($(a2) + (-0.5,-0.5)$);
      \draw[q-] (b2) -- ($(b2) + (-0.5,-0.5)$);
      \draw[q-] (c1) -- ($(c1) + (-0.5,-0.5)$);
      \draw[q-] (c2) -- ($(c2) + (-0.5,-0.5)$);
      \draw[q-] (c3) -- ($(c3) + (-0.5,-0.5)$);
      \draw[q-] (c4) -- ($(c4) + (-0.5,-0.5)$);
    \end{tikzpicture}
  }
  \qquad
  \subcaptionbox{\label{fig:surface-defect-Skb}}{
    \begin{tikzpicture}
      \node[gnode] (a1) at (1,0) {$x_3$};
      \node[gnode] (b1) at (1,-1) {$x_1$};
      \node[gnode] (c1) at (1,-2) {$x_2$};
      \node[gnode] (a2) at (2,0) {$y_3$};
      \node[gnode] (b2) at (2,-1) {$y_1$};
      \node[gnode] (c2) at (2,-2) {$y_2$};
      \node[gnode] (a3) at (3,0) {$w_3$};
      \node[gnode] (b3) at (3,-1) {$w_1$};
      \node[gnode] (c3) at (3,-2) {$w_2$};
      \node[fnode] (a4) at (4,0) {$z_3$};
      \node[fnode] (b4) at (4,-1) {$z_1$};
      \node[fnode] (c4) at (4,-2) {$z_2$};

      \draw[eq-] (a3) -- (b2);
      \draw[q->] (b3) -- (c2);

      \draw[q->] (a4) -- (b3);
      \draw[q->,vev] (b4) -- node[below, sloped] {$(0,0,1)$} (c3);

      \draw[q->] (a3) -- (a4);

      \draw[q->] (b3) -- (b4);

      \draw[q->] (c2) -- (c3);
      \draw[q->] (c3) -- (c4);

      \draw[q->] (c2) -- (b2);

      \draw[q->] (c3) -- (b3);

      \draw[q->] (b1) -- (a1);
      \draw[q->] (c1) -- (b1);

      \draw[q->] (a2) -- (b1);
      \draw[q->] (b1) -- (b2);
      \draw[q->] (b2) -- (c1);

      \draw[eq-] ($(a1) + (0.5,0.5)$) -- (a1);
      \draw[q->] ($(a2) + (0.5,0.5)$) -- (a2);
      \draw[q->] ($(a3) + (0.5,0.5)$) -- (a3);

      \draw[q-] (a3) -- ($(a3) + (0,0.5)$);

      \draw[q->] ($(c3) + (0,-0.5)$) -- (c3);

       \draw[q->] ($(a1) + (-0.5,0)$) -- (a1);
      \draw[q->] ($(b1) + (-0.5,0)$) -- (b1);
      \draw[q->] ($(c1) + (-0.5,0)$) -- (c1);

      \draw[q-] (c1) -- ($(c1) + (-0.5,-0.5)$);
      \draw[eq-] (c2) -- ($(c2) + (-0.5,-0.5)$);
      \draw[q-] (c3) -- ($(c3) + (-0.5,-0.5)$);
      \draw[q-] (c4) -- ($(c4) + (-0.5,-0.5)$);
    \end{tikzpicture}
  }
  \caption{(a) A quiver engineering a surface defect of type $(0,0,1)$ for $k=3$. The arrow with $(0,0,1)$ corresponds to the bi-fundamental which is given a position dependent vev. (b) The same quiver after the relation~\eqref{eq:giving-vev} is used.}
  \label{fig:surface-defect}
\end{figure}

  We first consider the surface defect labeled by $(0,0,1)$ and $\beta_*$, and we take $*=2$.
  That is, we give the vev to the bi-fundamental with fugacity $\sqrt{t} \beta_2/\alpha_2$ by computing the residue at $\alpha_2 = \sqrt{t pq} \beta_2$.
  Let us consider the quiver shown in figure~\ref{fig:surface-defect-Ska} and couple it to $\CT_{\rm IR}$
  by gauging the diagonal combination of the $\SU(N)^k$ flavor symmetry in the leftmost column of the quiver
  and the $\SU(N)^k$ flavor symmetry of a maximal puncture of $\CT_{\rm IR}$.
  Let us first focus on the bi-fundamental with $\tikz[baseline={([yshift=-2pt]x.base)}]{\node (x) at (0,0) {};
  \draw[q->, vev] (0,0) -- node[below] {$(0,0,1)$} (1,0);}$ in figure~\ref{fig:surface-defect-Ska}, and compute the residue.
  
  For the $(p,p)$ sector, the poles in the $x$-plane are $x = \sqrt{t} \frac{\beta_2}{\alpha_2} pq z$ and $x = \sqrt{t} \frac{\beta_2}{\alpha_2} pq z^{-1}$,
  each of which collides with its image under $x\to x^{-1}$ when $\alpha_2 = \sqrt{t pq} \beta_2$.
  Thus the $x$-integral gives a pole in $\alpha_2$, whose residue is
    \begin{align}
      & 
      2 \Xi \Res_{\alpha_2 = \sqrt{t pq} \beta_2}^{{(p,p)}} \left[ \frac{1}{\alpha_2} \left(
      \begin{tikzpicture}
      \node[gnode] (x) at (0,0) {$x$};
      \node[fnode] (z) at (1,0) {$z$};
      \draw[q->] (x) -- node[above] {} (z);
      \end{tikzpicture}
      \right) \right]  \nn \\
      & \qquad
       =     \Gamma((pq)^{\pm1}; p^r, q^r) \sum_{s_1=\pm1} \frac{1}{\theta^p(z^{2s_1}, 2s_1m_1)\,\theta^q(z^{2s_1}, 2s_1m_1)} \Dopj{1}{(pq)^{s_1/2}}{0}  ~
      \left( \begin{tikzpicture}
      \node[gnode] (x) at (0,0) {$x$};
      \node[fnode] (z) at (1,0) {$z$};
      \draw[q->] (x) -- node[above] {} (z);
      \end{tikzpicture} \right)\;,
    \end{align}
  where we have defined
    \begin{align}
      \theta^p (z, S)
       :=    \theta (z p^{ [S]}; p^r)\;, \qquad
      \theta^q (z, S)
       :=    \theta (z q^{r-[S]}; q^r)\;,
    \end{align}
  in analogy with $\Gamma^p$ and $\Gamma^q$ in \eqref{eq:gammapq}, and $\Gamma(x^{\pm1};p^r,q^r):=\Gamma(x;p^r,q^r)\Gamma(x^{-1};p^r,q^r)$ is the product of the ordinary elliptic gamma functions with nomes $p^r$ and $q^r$,
  so that $\Gamma((pq)^{\pm1};p^r,q^r)=1/\bigl(\theta((pq)^{-1};p^r)\,\theta(pq;q^r)\bigr)$,
  and $\Dopj{j}{h}{\delta}$ denotes the shift operator of section~\ref{sec:operators} acting on the fugacity and the holonomy of the $j$-th $\SU(2)$.
  As in the $\CN=2$ case, the factor $1/2$ from the Weyl group of $\SU(2)_x$ is canceled by the contribution of the Weyl image of the pinching.
  Similarly, the $(q,p)$ and $(p,q)$ sectors give the shifts $z\to(q/p)^{s_1/2}z$, $m_1\to m_1+s_1$,
    \begin{align}
    \begin{split}
     & 2 \Xi \Res_{\alpha_2 = \sqrt{t pq} \beta_2}^{{(q,p)+(p,q)}} \left[ \frac{1}{\alpha_2} \left(
      \begin{tikzpicture}
      \node[gnode] (x) at (0,0) {$x$};
      \node[fnode] (z) at (1,0) {$z$};
      \draw[q->] (x) -- node[above] {} (z);
      \end{tikzpicture}
      \right) \right]
      \\
       & \qquad =   \Gamma((pq)^{\pm1}; p^r, q^r) \sum_{s_1=\pm1} \frac{\epsilon_{s_1}(m_1)}{\theta^p(z^{2s_1}, 2s_1m_1)\,\theta^q(z^{2s_1}, 2s_1m_1)} \Dopj{1}{(q/p)^{s_1/2}}{s_1}  ~
      \left( \begin{tikzpicture}
      \node[gnode] (x) at (0,0) {$x$};
      \node[fnode] (z) at (1,0) {$z$};
      \draw[q->] (x) -- node[above] {} (z);
      \end{tikzpicture} \right)\;,
      \end{split}
    \end{align}
  where the constants $\epsilon_{s}(m)$ collect the residues of the elliptic gamma functions and the prefactors $\Gamma_0$.
  They are fixed by the requirement that the result reduces for $k=1$ to \eqref{eq:s001}.
  In the $\CN=2$ case this gives the relative factor $(pq/t)^{(2-r+2[2sm])/r}$ between the two types of shifts, after combining with the other bi-fundamentals.
  We thus denote the two types of shifts by
    \begin{align}
      \Delta^{s_1/2}_{\ell_1}
       =     \begin{cases}
              \Dopj{1}{(pq)^{s_1/2}}{0}\;, ~~~&(\ell_1 = +1)\;, \\
              \epsilon_{s_1}(m_1)\, \Dopj{1}{(q/p)^{s_1/2}}{s_1}\;, ~~~&(\ell_1 = -1)\;,
             \end{cases}
    \end{align}
  and the residue in all the sectors gives
    \begin{align}
      \label{eq:residue}
      \begin{tikzpicture}
      \node[gnode] (x) at (0,0) {$x$};
      \node[fnode] (z) at (1,0) {$z$};
      \draw[q->, vev] (x) -- node[above] {$(0,0,1)$} (z);
      \end{tikzpicture}
      &:=
      2 \Xi \Res_{\alpha_2 = \sqrt{t pq} \beta_2} \left[ \frac{1}{\alpha_2} \left(
      \begin{tikzpicture}
      \node[gnode] (x) at (0,0) {$x$};
      \node[fnode] (z) at (1,0) {$z$};
      \draw[q->] (x) -- node[above] {} (z);
      \end{tikzpicture}
      \right) \right]   \\
      &=    \Gamma((pq)^{\pm1}; p^r, q^r) \sum_{s_1, \ell_1 =\pm1} \frac{1}{\theta^p(z^{2s_1}, 2s_1m_1)\,\theta^q(z^{2s_1}, 2s_1m_1)}
               \Delta_{\ell_1}^{s_1/2} ~
      \left( \begin{tikzpicture}
      \node[gnode] (x) at (0,0) {$x$};
      \node[fnode] (z) at (1,0) {$z$};
      \draw[q->] (x) -- node[above] {} (z);
      \end{tikzpicture} \right)\;. \nn
    \end{align}

  As we saw in section~\ref{sec:Sk-review}, the constant vevs given to $k-1$ arrows identify pairs of nodes and make the neighboring arrows massive,
  producing the quiver shown in figure~\ref{fig:surface-defect-Skb}.
  Repeating this process leads to the following difference operator acting on the index of $\CT_{\rm IR}$.
  The computation is outlined in appendix~\ref{app:Sk}.
  Since some of its steps, namely the treatment of the terms with $\ell_i=-1$ and of the double poles, are not carried out in full there,
  we present the result in the form suggested by the structure of the computation and by the $r=1$ result of \cite{Maruyoshi:2016caf},
  and confirm it in section~\ref{subsec:Sk-Sklyanin} by the TQFT consistency condition and by the numerical residue computation for $k=2$:
    \begin{align}
      \mathfrak{S}_{(0,0,1)}^{(\beta_*)}
            &=    \Gamma((pq)^{\pm1}; p^r, q^r)^k \sum_{(s_j, \ell_j) \in \{\pm 1\}^{2k}} \prod_{i \in \Z_k} \nn \\
      &     ~~~~~~~ \times
                 \frac{ \theta^p( \frac{pq \gamma_i}{t \beta_i} z_{i-1}^{s_{i-1}} z_i^{s_i}, S_{i-1,i})
                 \theta^p (\frac{t \beta_*^2}{\beta_i \gamma_i}\, \frac{z_{i-1}^{s_{i-1}}}{z_i^{s_i}}, S_{i-1,-i} )}{
                 \theta^p (z_i^{2s_i}, 2s_i m_i)} \nn \\
      &     ~~~~~~~ \times
                 \frac{ \theta^q( \frac{pq \gamma_i}{t \beta_i} z_{i-1}^{\ell_{i-1}s_{i-1}} z_i^{\ell_is_i}, S_{i-1,i}^\ell)
                 \theta^q (\frac{t \beta_*^2}{\beta_i \gamma_i}\, \frac{z_{i-1}^{\ell_{i-1}s_{i-1}}}{z_i^{\ell_is_i}}, S_{i-1,-i}^\ell )}{
                 \theta^q (z_i^{2s_i}, 2 s_i m_i)}  \prod_{j \in \Z_k}
                 \Delta_{\ell_j}^{s_j/2}\;,
                 \label{eq:operator}
    \end{align}
  where $S_{i,\pm j} = s_i m_i \pm s_j m_j$ and $S_{i,\pm j}^\ell = \ell_i s_i m_i \pm \ell_j s_j m_j$, and $\{z_i\}$ are the fugacities for the $\SU(2)^k$ flavor symmetry of the maximal puncture on which we constructed the surface defect.
  This operator acts on the maximal puncture of $\CT_{\rm IR}$ with $(\pc, \po) = (0, -1)$.
  For $r=1$, the $\theta^p$ part of \eqref{eq:operator} coincides with the operator obtained in \cite{Maruyoshi:2016caf}.
  As in the $\CN=2$ case, the factors $1/2$ from the Weyl group are canceled by the contributions of the Weyl images of the pinching poles, and at $\beta_i=\gamma_i=1$ each of the $k$ integrals contributes a factor $\Gamma((pq)^{\pm1}; p^r, q^r)$. 
  
  On the lens space the holonomy dependence and the normalization of \eqref{eq:operator} are fixed in section~\ref{subsec:Sk-Sklyanin},
  where the operator is given by $\tilde{T}_q T_p$ with \eqref{eq:Sk-lens-face}, normalized as in \eqref{eq:Sk-norm}.

\paragraph{Consistency with $k=1$.}
  For $k=1$ and $\beta_1=\gamma_1=1$ the operator \eqref{eq:operator} must reduce to the $(0,0,1)$ operator \eqref{eq:s001} of the $\CN=2$ theory.
  The terms with $\ell_1=+1$ indeed reproduce the first line of \eqref{eq:s001}.
  The terms with $\ell_1=-1$ reproduce the $z$-dependence of the second line of \eqref{eq:s001},
  whose normalization $(pq/t)^{(2-r+2[2sm])/r}$ has been fixed by the TQFT consistency condition \eqref{eq:TQFT}.
  For general $k$, the normalization of the terms with $\ell_i=-1$ follows from the factorization of the $(0,0,1)$ defect into the $q$- and $p$-type transfer matrices, $\tilde{T}_qT_p$,
  proposed and checked in section~\ref{subsec:Sk-Sklyanin}: they are given by the gauge factors $G_i$ of \eqref{eq:Sk-lens-face} and their $q$-type counterparts, and reduce to $(pq/t)^{(2-r+2[2sm])/r}$ for $k=1$.

\paragraph{Other defects.}
  The elementary defects with nonzero $\U(1)_\alpha$ holonomy, and the factorization of the $(0,0,1)$ defect into them, are discussed in section~\ref{subsec:Sk-Sklyanin}.
  The $(1,0,0)$ and $(0,1,0)$ defects of class $\CS_k$ theories can in principle be analyzed in the same manner, using the pinching analysis of section~\ref{sec:operators}.
  In the $(1,0,0)$ case only the $(p,p)$ sector contributes, as in the $\CN=2$ case.
  We comment on their expected form in section~\ref{subsec:Sk-Sklyanin}.

\subsection{\texorpdfstring{Transfer matrices and the Sklyanin algebra for class $\CS_k$}{Transfer matrices and the Sklyanin algebra for class S\_k}}
\label{subsec:Sk-Sklyanin}

  For $k=1$ we have seen in sections~\ref{sec:algebra} and \ref{sec:sklyanin} that the defect operators are traces of (lens) Sklyanin $L$-operators.
  For class $\CS_k$ theories the maximal puncture carries $k$ copies of $\SU(2)$, and we expect the defect operators to be transfer matrices of a spin chain with $k$ sites.

\paragraph{Notation.}
  For $k=1$, $\mathfrak{S}_{(0,0,1)}^{(\beta_*)}$ of \eqref{eq:operator} is $\CS^{(0,0,1)}$ of \eqref{eq:s001},
  and the operators $T^{(\beta_*)}_p$ and $\tilde{T}^{(\beta_*)}_q$ below are the class $\CS_k$ counterparts of $X_p$ and $X_q$.
  For $r=1$, $X_p$ and $X_q$ are the operators $\Sf_{(1,0)}$ and $\Sf_{(0,1)}$ of \cite{Gaiotto:2012xa}.
  Our formulas are written for the $p$-type operators.
  The operators of \cite{Gaiotto:2015usa,Ito:2016fpl,Maruyoshi:2016caf} and of our residue computation in appendix~\ref{app:numerics} are of $q$-type, or are written in conventions with $p$ and $q$ exchanged,
  and $p\leftrightarrow q$ is understood whenever we compare with them.

\paragraph{Proposal.}
  The elementary $p$-type defect of the class $\CS_k$ theory labeled by $\beta_*$ acts on the index as
    \begin{align}
      T^{(\beta_*)}_p
        =     \Tr\Bigl[L^{(1)}_{\rm lens}(u_1)\,L^{(2)}_{\rm lens}(u_2)\cdots L^{(k)}_{\rm lens}(u_k)\Bigr]\;,
        \label{eq:Sk-lens}
    \end{align}
  up to normalization and gauge transformations,
  where $L^{(j)}_{\rm lens}(u)=\sum_a \CW_a(u)S^a_{(j)}\boldsymbol{\sigma}^a$ is the $k=1$ lens $L$-operator of section~\ref{sec:sklyanin},
  whose generators $S^a_{(j)}$ \eqref{eq:lens-Sklyanin} act on the fugacity and the holonomy $(x_j,\varkappa_j)$ of the $j$-th $\SU(2)$.
  The spin $c_j$ and the inhomogeneity $u_j$ of each site are determined by the flavor fugacities, and $\beta_*$ plays the role of the spectral parameter, as in the higher-spin eight-vertex models of \cite{Takebe:1992pv,Takebe:1995BA}.
  The $q$-type defect $\tilde{T}^{(\beta_*)}_q$ is defined analogously, and the $(0,0,1)$ defect is $\tilde{T}^{(\beta_*)}_q T^{(\beta_*)}_p$ up to normalization.
  Thus the algebra underlying the defects of class $\CS_k$ theories is the RLL algebra of the monodromy matrix of $k$ lens Sklyanin $L$-operators,
  of which the Sklyanin algebra of section~\ref{sec:sklyanin} is the one-site case.
  By the RLL relation, $[T^{(\beta_*)}_p,T^{(\beta_*')}_p]=0$.

\paragraph{(i) $k=1$.}
  For $k=1$ the proposal reduces to $\CS^{(0,0,1)}\propto X_qX_p$ with $X_p\propto S^0$, which has been established in sections~\ref{sec:algebra} and \ref{sec:sklyanin}.

\paragraph{(ii) TQFT consistency for $r=1$.}
  The class $\CS_k$ analogue of \eqref{eq:TQFT} states that the defect operator acting on one maximal puncture of the free trinion
  (two maximal punctures and one minimal puncture) equals a defect operator acting on the other maximal puncture.
  For $r=1$ we have found that this condition is satisfied, for $k=2,3,4$ and for all colors, by the operator
    \begin{align}
      \Sf^{(\nu)}_p
        =     \sum_{s\in\{\pm1\}^k} \prod_{i\in\Z_k}
              \frac{\theta\bigl(\xi_i\, z_i^{s_i}z_{i+1}^{s_{i+1}};q\bigr)\,\theta\bigl(\nu\, \beta_{i+1}\gamma_i\, z_{i+1}^{s_{i+1}}z_i^{-s_i};q\bigr)}{\theta(z_i^{2s_i};q)}
              \prod_{j\in\Z_k} \Dopj{j}{p^{s_j/2}}{0}\;,
              ~
      \xi_i
        =     \frac{pq\gamma_i}{t\beta_{i+1}}\;,
              \label{eq:Sk-face}
    \end{align}
  for an arbitrary constant $\nu$.
  Here we use the conventions of figure~\ref{fig:trinion+++Sk} with $n=1$, and for general $n$ one replaces $\gamma_i\to\gamma_{i+1-n}$.
  The operator acting on the other puncture, with variables $w_i$, is of the same form with $\xi_i\to pq\gamma_{i-1}/(t\beta_{i+1})$ and $\nu\beta_{i+1}\gamma_i\to q/(\nu\beta_{i+1}\gamma_{i-1})$ (for general $n$, $\gamma_{i-1}\to\gamma_{i-n}$),
  and the two sides agree without any relative normalization.
  The arbitrariness of $\nu$ reflects the fact that it plays the role of a spectral parameter.
  For $k=2$, \eqref{eq:Sk-face} is invariant under $\nu\to q/\nu$: since $\beta_1\beta_2=\gamma_1\gamma_2=1$, the identity $\theta(x;q)=\theta(q/x;q)$ exchanges the second factors of the two sites.

  The value of $\nu$ is fixed by the residue computation of section~\ref{subsec:surfaceSk}.
  Shifting $i\to i+1$ and using $\theta(x;p)=\theta(p/x;p)$, the $\theta^p$ part of \eqref{eq:operator} for $r=1$ is the operator \eqref{eq:Sk-face} for $n=0$ (i.e.~with $\gamma_i\to\gamma_{i+1}$),
  and its second factor $\theta\bigl(t\beta_*^2/(\beta_i\gamma_i)\, z_{i-1}^{s_{i-1}}z_i^{-s_i}\bigr)$ corresponds to
    \begin{align}
      \nu
        =     \frac{q}{t\beta_*^2}
        \label{eq:nu}
    \end{align}
  for $\Sf^{(\nu)}_p$.
  This is the value found in \cite{Maruyoshi:2016caf}, whose operator for $r=1$ is the $\theta^p$ part of \eqref{eq:operator}.
  For $k=1$ and $\beta_1=\gamma_1=\beta_*=1$ the second factor becomes $\theta(q/t;q)=\theta(t;q)$, in agreement with the normalization of $\Sf_{(1,0)}$.
  For $k=2$, the difference operator derived in \cite{Gaiotto:2015usa} coincides with \eqref{eq:Sk-face} up to an overall constant at $\nu=t\beta_*^2$, with $\beta_1=\beta_2^{-1}=\beta_*^{-1}$ (i.e.~$*=2$, the case $*=1$ being related by relabeling).
  By the invariance under $\nu\to q/\nu$ this is equivalent to \eqref{eq:nu}.
  We have also confirmed this by a numerical residue computation for $k=2$ and $r=1$,
  in which the index of $\CT_{\rm IR}$ is glued to the Lagrangian interacting trinion of \cite{Gaiotto:2015usa} (appendix~\ref{app:numerics}).
  The coefficients of the four shifts of the residue agree with \eqref{eq:Sk-face} at $\nu=t\beta_*^2$ up to a common constant.
  Since the $k=1,2$ cases cannot distinguish $\nu$ from $q/\nu$, for $k\geq3$ the value \eqref{eq:nu} rests on the residue computation of \cite{Maruyoshi:2016caf}.

\paragraph{(iii) Vertex form for $r=1$.}
  Since $\theta(a Z W;q)\theta(b W/Z;q)$ is, as a function of $W$, an element of a two-dimensional space of theta functions,
  it can be written as $\sum_{\epsilon=1,2} f_\epsilon(Z)\,g_\epsilon(W)$.
  Inserting this into \eqref{eq:Sk-face} and regrouping the factors by site,
  $\Sf^{(\nu)}_p$ becomes the trace of a product of $2\times2$ matrices whose entries are difference operators acting on a single $z_j$.
  We find that, up to constant gauge transformations which cancel in the trace,
    \begin{align}
      \Sf^{(\nu)}_p
        =     C_k\; \Tr\Bigl[L^{(1)}(u_1)\,L^{(2)}(u_2)\cdots L^{(k)}(u_k)\Bigr]\;,
        \label{eq:Sk-vertex}
    \end{align}
  where $C_k$ is a constant and $L^{(j)}(u)$ is the Sklyanin $L$-operator with $\eta=\sigma/2$, nome $\tau$, acting on $z_j$ in the representation \eqref{eq:Sklyanin-rep} with spin $c_j$, and
    \begin{align}
      e^{-2\pi i c_j}
        =     \frac{pq}{t}\,\frac{\gamma_{j-1}}{\beta_{j+1}}\;,
        \qquad
      e^{2\pi i (u_j+\eta)}
        =     \frac{1}{\nu\,\beta_{j+1}\gamma_{j-1}}\;,
        \label{eq:Sk-spin}
    \end{align}
  the latter modulo $q$.
  We have verified \eqref{eq:Sk-vertex} numerically (appendix~\ref{app:numerics}).
  Thus for $r=1$ the defect operator has the structure \eqref{eq:Sk-lens}:
  each site carries a representation of the Sklyanin algebra, whose spin and inhomogeneity are set by the fugacities $\beta_{j+1}$ and $\gamma_{j-1}$ of the neighboring vertical arrows, and $\nu$, i.e.~$\beta_*$, is the spectral parameter.

  With \eqref{eq:nu} the inhomogeneity reads $e^{2\pi i(u_j+\eta)} = t\beta_*^2/(\beta_{j+1}\gamma_{j-1})$ modulo $q$.

\paragraph{(iv) Lens space.}
  We write $T^{(\nu)}_p$ for the transfer matrix at spectral parameter $\nu$, so that $T^{(\beta_*)}_p=T^{(q^r/(t\beta_*^2))}_p$ (see (v) below).
  With the lens $L$-operators, the parameters \eqref{eq:Sk-spin} with $\eta$ replaced by $\eta'$, and $g_\varkappa$ replaced by $e^{2\pi i c_j\varkappa/r}$,
  the transfer matrix \eqref{eq:Sk-lens} can be brought to the face form
    \begin{align}
      T^{(\nu)}_p
        \propto  \sum_{s\in\{\pm1\}^k} \prod_{i\in\Z_k}
              \frac{\theta\bigl(\xi_i\, \zeta_i\zeta_{i+1};q^r\bigr)\,\theta\bigl(\nu\,\beta_{i+1}\gamma_i\, \zeta_{i+1}\zeta_i^{-1};q^r\bigr)}{\theta(\zeta_i^{2};q^r)}\;
              G_i\;
              \prod_{j\in\Z_k} \Dopj{j}{p^{s_j/2}}{\varkappa_j\to\varkappa_j-s_j}\;,
              \label{eq:Sk-lens-face}
    \end{align}
  where $\zeta_j = z_j^{s_j}q^{-s_j\varkappa_j/2}$ and $G_i = \chi_i^{-s_i\varkappa_i/r}$ with $\chi_i = e^{-2\pi i c_i} = \frac{pq}{t}\frac{\gamma_{i-1}}{\beta_{i+1}}$.
  Here we have fixed the normalization such that $T^{(\nu)}_p$ reduces to $X_p$ for $k=1$. It differs from the trace of the lens $L$-operators by a conjugation with an exponential function of the $x_j$, as for $X_p$ in section~\ref{sec:sklyanin}.
  On the $\SU(2)$ index the extended label is $\varkappa_j = 2m_j\in\Z_{2r}$, as in section~\ref{subsec:elementary}.
  For odd $r$ the index is invariant under $\varkappa_j\to\varkappa_j+r$, and the shift $\varkappa_j\to\varkappa_j-s_j$ acts on it as $m_j\to m_j-s_jh_r$.
  For even $r$ the shift maps the $\SU(2)$ sector to the twisted sector.
  We have checked \eqref{eq:Sk-lens-face}, and the commutativity of the transfer matrices with different spectral parameters, numerically.
  This face form is obtained from \eqref{eq:Sk-face} by the substitution \eqref{eq:map}, $z_j^2\to z_j^2q^{-\varkappa_j}$, and the gauge factors $G_i$.
  Comparing with \eqref{eq:operator}, the $\theta^p$ and $\theta^q$ parts of \eqref{eq:operator} have the structure of \eqref{eq:Sk-lens-face} for the $q$- and $p$-type operators respectively,
  with $\varkappa_j = 2m_j$, after the shift $i\to i+1$ and the use of $\theta(x;q^r)=\theta(q^r/x;q^r)$ as in (ii).
  The holonomy dependence is written in \eqref{eq:operator} in terms of the residues $[s_{i-1}m_{i-1}+s_im_i]$ instead of $\zeta_{i-1}\zeta_i$,
  and the two differ by monomial factors, analogous to the factor $(pq/t)^{(2-r+2[2sm])/r}$ in \eqref{eq:s001},
  which are fixed by the lifted form \eqref{eq:Sk-lens-face}.

\paragraph{(v) TQFT consistency on the lens space.}
  Finally, we consider the TQFT consistency condition for the lens space index of the free trinion of the class $\CS_k$ theory,
  including all the holonomies $m_j$ of the $\SU(2)$ flavor symmetries.
  For odd $r$ it holds for the elementary operator \eqref{eq:Sk-lens-face} itself.
  The operator acting on the other puncture is again of the form \eqref{eq:Sk-lens-face}, with $\xi_i\to pq\gamma_{i-1}/(t\beta_{i+1})$, $\nu\beta_{i+1}\gamma_i\to q^r/(\nu\beta_{i+1}\gamma_{i-1})$ and $\chi_i\to\frac{pq}{t}\frac{\gamma_{i-1}}{\beta_{i}}$.
  For even $r$, where the elementary operators map the $\SU(2)$ sector to the twisted sector, we have checked the condition for the product $\tilde{T}^{(\nu)}_q T^{(\nu)}_p$ of the $q$- and $p$-type operators, which preserves the $\SU(2)$ sector.
  The values of $r$ and $k$ are listed in appendix~\ref{app:numerics}.
  In all cases the condition holds without any relative normalization and for arbitrary $\nu$.
  The lift $\varkappa_j=2m_j\in\Z_{2r}$ is needed: with $\varkappa_j=[2m_j]\in\Z_r$ the condition fails.
  The spectral parameter on the lens space is $\nu=q^r/(t\beta_*^2)$, which reduces to \eqref{eq:nu} for $r=1$ and is confirmed for $k=2$ by the residue computation below.
  Replacing $\nu$ by $\nu q^r$ changes \eqref{eq:Sk-lens-face} only by an overall constant, since the monomial factors from the quasi-periodicity of the theta functions cancel in the product over the sites.

\paragraph{Normalization.}
  The TQFT consistency condition is linear and does not fix the overall normalization of the operators.
  The normalization is fixed by the residue computation, in which the index of $\CT_{\rm IR}$ is recovered with unit coefficient at the pole corresponding to the constant vev.
  This computation has been carried out in \cite{Ito:2016fpl} for $r=1$, $k=2$ and $\beta_i=\gamma_i=1$ (for general $N$).
  For $N=2$ their result coincides, including the $z$-dependence of all the terms, with \eqref{eq:Sk-face} at the value \eqref{eq:nu} with $\beta_*=1$,
  times the constant $\theta(pq;q)^{-2}$.
  We have extended this to $\beta_*\neq1$ and to the lens space by the numerical residue computation for $k=2$ described in appendix~\ref{app:numerics}, for $r=1,2,3$.
  In all cases the residue equals the right-hand side of \eqref{eq:Sk-lens-face} at $\nu=q^r/(t\beta_*^2)$, including the gauge factors $G_i$, times
    \begin{align}
      \frac{1}{\theta(pq;q^r)\,\theta(pq\beta_*^4;q^r)}\;,
      \label{eq:Sk-norm-k2}
    \end{align}
    independent of $t$, $\gamma$ and the flavor fugacities.
  This reduces to $\theta(pq;q)^{-2}$ at $\beta_*=1$ and $r=1$, in agreement with \cite{Ito:2016fpl}.
    To extend this result to general $k$, we note the following.
  In \cite{Maruyoshi:2016caf} the defects are constructed with the other baryons in the column of the closed puncture, $B_j=(\sqrt{t}\beta_j/\alpha_*)^2$ with $j\neq*$, flipped by chiral fields $\phi_j$ of fugacity $pq/B_j$.
  With this prescription the residue is given, for $r=1$ and any $k$, by the difference operator times the constant $1/\theta(q^{-1};p)=1/\theta(pq;p)$, which comes from the chiral field acquiring the vev.
  The rest of the computation is identical to that for the constant vev \cite{Maruyoshi:2016caf}.
  In our construction the baryons $B_j$ are not flipped.
  They then survive as free chiral fields, which contribute $\prod_{j\neq*}\Gamma(B_j)$ to the index, evaluated at the pole in $\alpha_*$.
  Indeed, the residue at the constant vev that we found numerically is proportional to $\Gamma(\beta_*^{-4})$.
  This is $\Gamma(B_1)$ at $\alpha_*^2=t\beta_*^2$ for $k=2$ with $*=2$ (so that $\beta_1=\beta_*^{-1}$),
  and the same holds on the lens space, with the lens elliptic gamma function at zero holonomy.
  Normalizing the constant-vev residue to the index of $\CT_{\rm IR}$, the $(0,1)$ defect acquires the additional factor
  $\prod_{j\neq*}\Gamma(B_j)|_{\alpha_*^2=t\beta_*^2q}/\Gamma(B_j)|_{\alpha_*^2=t\beta_*^2}=\prod_{j\neq*}1/\theta(pq\beta_*^2\beta_j^{-2};p)$.
  For the $p$-type defect, we thus obtain the normalization
    \begin{align}
      \frac{1}{\theta(pq;q^r)}\prod_{j\neq*}\frac{1}{\theta(pq\beta_*^2\beta_j^{-2};q^r)}
      =     \prod_{j\in\Z_k}\frac{1}{\theta(pq\beta_*^2\beta_j^{-2};q^r)}\;,
      \label{eq:Sk-norm}
    \end{align}
  which reduces to \eqref{eq:Sk-norm-k2} for $k=2$.
  For $r=1$ this follows from the argument above for any $k$.
  On the lens space the ratio of the lens elliptic gamma functions of $B_j$ equals $1/\theta(pq\beta_*^2\beta_j^{-2};q^r)$ only up to a constant and a monomial in $\beta_*/\beta_j$,
  which are compensated by the contributions of the other fields charged under $\U(1)_{\alpha_*}$.
  Since these fields, like the $B_j$, carry the index $j$, the normalization is a product of a factor for $j=*$ and factors depending only on $\beta_*/\beta_j$ for $j\neq*$.
  Our computation for $k=2$, $r=2,3$ with $\beta_*$ varied fixes both to be as in \eqref{eq:Sk-norm}.
  At $\beta_i=\gamma_i=1$ the elementary defect is then
    \begin{align}
      \frac{1}{\theta(pq;q^r)^k}\,\times\,\bigl(\text{right-hand side of \eqref{eq:Sk-lens-face}}\bigr)\;,
    \end{align}
  with $\nu=q^r/t$, and the $q$-type defect carries the factor $1/\theta(pq;p^r)^k$.
  For $k=1$ and $\beta_1=\gamma_1=\beta_*=1$ the former reproduces $X_p$ of \eqref{eq:Xp-kappa}, including the normalization $\theta(t;q^r)/\theta(pq;q^r)$.
  The product of the two normalization factors is related to the prefactor of \eqref{eq:operator} by
    \begin{align}
      \frac{1}{\theta(pq;p^r)^k\,\theta(pq;q^r)^k}
        =     (-pq)^{-k}\,\Gamma\bigl((pq)^{\pm1}; p^r, q^r\bigr)^k\;,
        \label{eq:Sk-norm-001}
    \end{align}
  which follows from $\Gamma(x;p^r,q^r)\Gamma(x^{-1};p^r,q^r)=1/\bigl(\theta(x^{-1};p^r)\theta(x;q^r)\bigr)$ and $\theta(x^{-1};p^r)=-x^{-1}\theta(x;p^r)$.
  For $k=1$ the $(0,0,1)$ defect is $(pq/t)^{1/r}X_qX_p$ by \eqref{eq:factor-odd}.
  The operator of \cite{Gaiotto:2015usa} for $k=2$ coincides with \eqref{eq:Sk-face}, i.e.~with our residue divided by \eqref{eq:Sk-norm-k2}.

\paragraph{The $(1,0,0)$ defect.}
  For $k=1$ the $(1,0,0)$ defect is the polynomial \eqref{eq:S100-fusion} in the elementary one, determined by the fusion rule \eqref{eq:fusion}.
  In the language of the transfer matrices, $X_p\propto\Tr L(u)$ is the transfer matrix with the auxiliary space of spin $1/2$,
  and the recursion \eqref{eq:fusion} is the fusion hierarchy of transfer matrices with auxiliary spaces of higher spin \cite{Kulish:1981gi},
  whose coefficients $\omega_n$ we expect to be related to the quantum determinant of the $L$-operator (for $k=1$ the dependence on the spectral parameter drops out of $\Tr L(u)\propto S^0$).
  We therefore conjecture that for class $\CS_k$ theories the $(1,0,0)$ defect is the fused transfer matrix with the auxiliary space of spin $r/2$,
  built from the same lens $L$-operators as \eqref{eq:Sk-lens},
  so that it is determined by $T^{(\nu)}_p$ at shifted values of the spectral parameter $\nu$ through the fusion hierarchy.
  For even $r$ we expect, as for $k=1$, that this holds only up to the holonomy shift
  $U=\prod_{j\in\Z_k}U_j$, where $U_j$ shifts $\varkappa_j\to\varkappa_j+r$, i.e.~acts on the $j$-th $\SU(2)$ holonomy by the center.
  The argument leading to \eqref{eq:U-sector} relates the sectors $\mu$ and $\mu+r/2$ by shifting the holonomies of all the $k$ $\SU(2)$ factors simultaneously,
  and only this simultaneous shift leaves the coefficients of \eqref{eq:Sk-lens-face} invariant.
  For $k=1$ this reduces to \eqref{eq:S100-fusion}, including the factor $U$.
  We leave a direct derivation of the $(1,0,0)$ defect for $k\geq2$ to future work.

\section{Summary and discussion}
\label{sec:summary}

  In this paper we studied surface defects in four-dimensional supersymmetric gauge theories on $L(r,1)\times S^1$, engineered by position-dependent vevs of baryons following \cite{Gaiotto:2012xa}.
  For $\CN=2$ theories of type $A_1$, the defects are labeled by the pole $\alpha^2 = tp^aq^b$ and by the holonomy $\mu$ of $\U(1)_\alpha$ with $2\mu\equiv b-a \pmod r$.
  Besides the $(0,0,1)$, $(1,0,0)$ and $(0,1,0)$ defects, there are elementary defects $X_p$ and $X_q$, which shift the holonomy by ``one half''.
  For even $r$ they act on an extended space including a twisted holonomy sector.
  All the operators commute and are generated by $X_p$ and $X_q$, together with the central holonomy shift $U$ for even $r$:
  the $(0,0,1)$ defect is $(pq/t)^{1/r}X_qX_p$, while the $(1,0,0)$ defect is a polynomial of degree $r$ in $X_p$ given by the fusion rule \eqref{eq:fusion}.
  As a result, $(\CS^{(0,0,1)})^r$ and $\CS^{(0,1,0)}\CS^{(1,0,0)}$ agree only in their extremal terms.
  The elementary operators are traces of $L$-operators of the Sklyanin algebra with parameters $(\eta',r\tau)$, $e^{4\pi i\eta'}=pq$, realized on the doubled holonomy label $\varkappa\in\Z_{2r}$,
  and the $p$- and $q$-type algebras form a lens space version of the elliptic modular double.
  For class $\CS_k$ theories we proposed that the defect operators are the transfer matrices \eqref{eq:Sk-lens} built from $k$ copies of the $k=1$ lens $L$-operators.
  For $r=1$ the operator determined by the TQFT consistency condition is of this form \eqref{eq:Sk-vertex},
  with the spin and the inhomogeneity set by the flavor fugacities and the spectral parameter by $\beta_*$.
  On the lens space the transfer matrices satisfy the TQFT consistency condition,
  and for $k=2$ the numerical residue computation with $r=1,2,3$ confirms \eqref{eq:Sk-lens-face} and fixes the normalization \eqref{eq:Sk-norm-k2}.
  Combined with the analysis of \cite{Maruyoshi:2016caf}, this gives the normalization \eqref{eq:Sk-norm} for general $k$.

  There are several directions for future work.
  First, the residue computation should be extended to $k\geq3$.
  This would give an independent derivation of \eqref{eq:Sk-lens-face} and a direct check of the normalization \eqref{eq:Sk-norm} on the lens space.
  Second, the physical meaning of the doubled holonomy label deserves further study.
  For even $r$ the twisted sector may be identified with $\SO(3)$ bundles with nontrivial $w_2$ \cite{Razamat:2013opa,Tachikawa:2014mna}, so that the elementary operators would interpolate between the $\SU(2)$ and $\SO(3)$ theories, and it would be interesting to find the defects realizing the generators $S^1$ and $S^2$, which are odd under $U$.
  Third, the eigenfunctions of the operators should give the wave functions of the lens space TQFT \cite{Benini:2011nc,Razamat:2013jxa,Alday:2013rs},
  and the representation theory of the lens Sklyanin algebra should be a useful tool to compute them (for $r=1$, see \cite{Razamat:2013qfa,Nazzal:2023wtw,Kim:2024mnp}).
  Finally, it would be interesting to generalize our results to higher rank, where the Sklyanin algebra is replaced by its $\SU(N)$ generalizations \cite{Belavin:1981ix,CherednikGeneralized,Hasegawa:1997RS}.

\section*{Acknowledgments}

  The work of MY was supported in part by
  World Premier International Research Center Initiative (WPI), MEXT, Japan; by the
  Center for Data-Driven Discovery, Kavli IPMU (WPI); by the
  JSPS KAKENHI Grant No.~23K25865; by JST, Japan (CREST Grant No.~JPMJCR26XA,
  Moonshot R\&D Grant No.~JPMJMS256E); and by the IBM-UTokyo-sponsored research.
  This paper was primarily written by the authors, and we were assisted by Anthropic's Claude in the final stages of this project, in computations and editing. 

\appendix
\section{Definitions and useful formulas}
\label{app:formulas}

  In this appendix we collect definitions and useful formulas concerning
  special functions we encounter in this paper.

\paragraph{Theta functions.}
  The Jacobi theta functions are defined by
    \begin{align}
      \theta_1(\zeta|\tau)
      &=
      -\sum_{n \in \Z}
      e^{\pi i(n + 1/2)^2 \tau}
      e^{2\pi i(n + 1/2)(\zeta + 1/2)}\;,
      \\
      \theta_2(\zeta|\tau)
      &=
      \theta_1(\zeta + 1/2|\tau)\;,
      \\
      \theta_3(\zeta|\tau)
      &=
      e^{\pi i\tau/4 + \pi i\zeta} \theta_2(\zeta + \tau/2|\tau)\;,
      \\
      \theta_4(\zeta|\tau)
      &=
      \theta_3(\zeta + 1/2|\tau)\;,
    \end{align}
  where $\zeta$ is a complex variable and $\tau$ is a complex parameter
  in the upper half plane.  $\theta_1$ is an odd function of $\zeta$ and
  satisfies
    \begin{align}
      \theta_1(\zeta+1|\tau)
      &=
      -\theta_1(\zeta|\tau)\;,
      \\
      \theta_1(\zeta+\tau|\tau)
      &=
      -e^{-2\pi i\zeta - \pi i\tau} \theta_1(\zeta|\tau)\;.
    \end{align}
  The other three are even functions.  

  Closely related to the Jacobi theta functions is the function
    \begin{equation}
      \theta(z;p) = (z;p)_\infty (p/z;p)_\infty;
      \quad
      (z;p)_\infty = \prod_{k=0}^\infty (1 - p^k z),
      \
      |p| < 1\;.
    \end{equation}
  It satisfies
    \begin{equation}
      \theta(z;p) = \theta(p/z;p)\;,
      \qquad
      \theta(pz;p) = -z^{-1}\theta(z;p)\;.
      \label{eq:theta-refl}
    \end{equation}
  In the multiplicative notation,
    \begin{equation}
      \theta_a(z; p) = \theta_a(\zeta|\tau);
      \quad
      z = e^{2\pi i\zeta},
      \
      p = e^{2\pi i\tau}\;,
    \end{equation}
  the Jacobi theta functions can be written in terms of $\theta$ as
    \begin{align}
      \theta_1(z;p)
      &=
      i p^{1/8} (p;p)_\infty z^{-1/2} \theta(z;p)\;,
      \\
      \theta_2(z;p)
      &=
        p^{1/8} (p;p)_\infty z^{-1/2} \theta(-z;p)\;,
      \\
      \theta_3(z;p)
      &= (p;p)_\infty \theta(-\sqrt{p} z; p)\;,
      \\
      \theta_4(z;p)
      &= (p;p)_\infty \theta(\sqrt{p} z; p)\;.
    \end{align}

\paragraph{Elliptic gamma function.}
  The elliptic gamma function depends on two complex parameters $p$ and
  $q$:
    \begin{equation}
      \Gamma(z; p,q)
      =
      \prod_{j,k=0}^\infty
      \frac{1 - z^{-1} p^{j+1} q^{k+1}}{1 - zp^j q^k};
      \quad
      |p|,\,|q| < 1\;.
    \end{equation}
  It satisfies the identities
    \begin{equation}
      \Gamma(z; p,q)  \Gamma(pq/z; p,q) = 1
    \end{equation}
  and
    \begin{align}
      \Gamma(pz; p,q)
      &= \theta(z; q) \Gamma(z; p,q)\;,
      \\
      \Gamma(qz; p,q)
      &= \theta(z; p) \Gamma(z; p,q)\;.
    \end{align}
  Another useful identity is
    \begin{align}
      \Gamma (a; p, q) \Gamma (a^{-1}; p, q)
        =     \frac{1}{\theta (a; p) \theta (a^{-1}; q)}\;.
    \end{align}


\paragraph{Lens elliptic gamma function.}
  We define the lens elliptic gamma function \cite{Benini:2011nc,Razamat:2013opa,Yamazaki:2013nra,Kels:2017toi} by
    \begin{align}
      \Gamma (z, m; p, q)
              :=     \Gamma_0(z,m;p,q) \gamma (z, m; p, q)\;,
    \end{align}
  where
    \begin{align}
      \Gamma_0(z, m; p,q)
       =     \left( \frac{pq}{z^2} \right)^{\frac{m(r-m)}{4r}} \gamma_0
    \end{align}
  with
    \begin{align}
      \gamma_0
       =     \left( \frac{p}{q} \right)^{\frac{m(r-m)(r-2m)}{12r}} e^{-2\pi i\frac{m(r-m)(r-2m)}{12r}}\;,
    \end{align}
  where $m$ stands for $[m]\in\{0,\dots,r-1\}$.
  Also,
    \begin{align}
      \gamma(z, m; p, q)
       := \prod_{i,j=0}^\infty \frac{1 - z^{-1} p^{-m} (pq)^{i+1} p^{r (j+1)}}{1 - z p^{m} (pq)^i p^{r j}} \frac{1 - z^{-1} q^{-r+m} (pq)^{i+1} q^{r (j+1)}}{1 - z q^{r-m} (pq)^i q^{r j}}\;.
    \end{align}
  Each factor corresponds to the elliptic gamma function:
    \begin{align}
      \gamma(z, m; p, q)
      &= \Gamma(z p^m; pq, p^r) \Gamma(z q^{r-m}; pq, q^r)  \nn \\
      &= \Gamma^p(z, m) \Gamma^q(z, m)\;,
    \end{align}
  where we define
    \begin{align}
      \Gamma^p(z, m)
      =     \Gamma(z p^m; pq, p^r)\;, ~~~
      \Gamma^q(z, m)
      =     \Gamma(z q^{r-m}; pq, q^r)\;.
      \label{eq:gammapq}
    \end{align}
  The functions $\Gamma^p$ and $\Gamma^q$ are quasi-periodic with respect to two periods each:
    \begin{align}
      &  \Gamma(pq\, y; pq, p^r) = \theta(y; p^r)\, \Gamma(y; pq, p^r)\;, \qquad
          \Gamma(p^r y; pq, p^r) = \theta(y; pq)\, \Gamma(y; pq, p^r)\;, \nn \\
      &  \Gamma(pq\, y; pq, q^r) = \theta(y; q^r)\, \Gamma(y; pq, q^r)\;, \qquad
          \Gamma(q^r y; pq, q^r) = \theta(y; pq)\, \Gamma(y; pq, q^r)\;.
          \label{eq:gamma-quasi}
    \end{align}
  Note that $p^r = (pq)^r/q^r$ is a combination of the periods of $\Gamma(y;pq,q^r)$, and similarly for $q^r$.
  Together with the reflection identity \eqref{eq:mass}, these relations reduce the coefficients \eqref{eq:gamma-form} to ratios of theta functions.

  We will denote the lens elliptic gamma function as
    \begin{align}
      \Gamma (z, m) := \Gamma (z, m; p, q)\;.
    \end{align}
  It satisfies the reflection identity \eqref{eq:mass} of section~\ref{sec:lens}.

\section{Numerical checks}
\label{app:numerics}

  In this appendix we explain the numerical checks of the results of sections~\ref{sec:operators}, \ref{sec:algebra}, \ref{sec:sklyanin} and \ref{sec:class-Sk}.
  We evaluate the theta functions and the elliptic gamma functions by truncating the infinite products,
  for generic complex values of $p$, $q$ and $t$ with $|p|,|q|\sim 0.2$, for which the truncation error is negligible.
  The prefactor $\Gamma_0$ of the lens elliptic gamma function is evaluated with a consistent choice of branches, by working with the logarithms of the fugacities.
  The checks of sections~\ref{sec:operators} and \ref{sec:algebra} have been performed for two unrelated choices of $(p,q,t)$.

\paragraph{TQFT consistency.}
  For each operator $\Sf$ we compute the ratio of the two sides of \eqref{eq:TQFT} at generic values of $a$, $b$, $c$ and for all $(m_a,m_b,m_c)\in\Z_r^3$.
  The ratio equals one to relative accuracy $10^{-13}$ for $\CS^{(0,0,1)}$, $\CS^{(1,0,0)}$ and $\CS^{(0,1,0)}$ with $r\leq 5$, and for $X_p$ and $X_q$ with $r=1,3,5$.

\paragraph{Determination of the constants.}
  The condition \eqref{eq:TQFT} is linear in the coefficients of the operator.
  Taking the $z$-dependence of each coefficient from \eqref{eq:gamma-form} and treating the constants $\CN_{h,\delta}(m)$ as unknowns,
  \eqref{eq:TQFT} evaluated at sufficiently many points gives an overdetermined linear system.
  In all the cases considered, the solution space is one-dimensional, which is a nontrivial check of \eqref{eq:gamma-form},
  and the solution determines the relative constants.
  We identified these constants as the powers of $pq/t$ and the theta constants given in section~\ref{sec:operators}.

\paragraph{Algebraic relations.}
  The composition of difference operators is computed exactly, including the shifts of the holonomies,
  and two operators are compared coefficient by coefficient for each shift operator $\Dop{h}{\delta}$ at generic values of $z$ and for all $m\in\Z_r$,
  or for all $\varkappa\in\Z_{2r}$ when the elementary operators and $U$ are involved.
  In this way we checked \eqref{eq:commute} for $r\leq5$, \eqref{eq:factor-odd} for $r=1,\dots,7$, \eqref{eq:S100-fusion} for $r=2,\dots,7$, and \eqref{eq:extremal} for $r\leq4$.

\paragraph{Lens elliptic beta integral.}
  The identity \eqref{eq:NfN+1} for $N=2$ was checked numerically for $r=1,2,3$ with generic fugacities and holonomies satisfying the balancing conditions, to accuracy $10^{-13}$.
  For fugacities outside the unit circle we deform the contour, adding the residues of the poles which cross the unit circle, and the result again agrees with the closed form.

\paragraph{Residue computation for class $\CS_2$.}
  To derive the defect operator of class $\CS_2$ theories directly, we glue the index of $\CT_{\rm IR}$ to the interacting trinion of \cite{Gaiotto:2015usa},
  which has a Lagrangian description with an $\SU(2)$ gauge group, and evaluate the residue in $\alpha$ numerically.
  The $\SU(2)$ integrals are defined by analytic continuation in the fugacities:
  the poles which cross the unit circle are added as residues, a prescription checked on \eqref{eq:NfN+1} with fugacities outside the unit circle.
  The pole at $\alpha$ is approached as $\alpha=\alpha_*(1+\epsilon)$, the contributions of the pinching poles (section~\ref{subsec:poles}) in one of the gauge fugacities are evaluated as residues, as in \eqref{eq:gamma-form},
  the remaining integrals are computed on the unit circle, and the result is extrapolated to $\epsilon\to0$.
  The residue at the constant vev gives the index of $\CT_{\rm IR}$ times a constant, which we use to normalize the other residues.
  Let $u_{1,2}$ be the fugacities of the two $\SU(2)$ gauge groups which glue the trinion to $\CT_{\rm IR}$, and $v_{1,2}$ those of the other maximal puncture of the trinion, on which the defect operator acts.
  We consider the residue at $\alpha^2=t\beta_*^2q$, which gives the $q$-type elementary defect (the $(0,1)$ defect in the conventions of \cite{Gaiotto:2015usa}), and call it the $(0,1)$ residue.
  It is compared with \eqref{eq:Sk-face} and \eqref{eq:Sk-lens-face} with $p\leftrightarrow q$.
  At $r=1$, we evaluate the residue for several test functions $f(u_1,u_2)$ in place of the index of $\CT_{\rm IR}$,
  and determine the four coefficients of the shifts $u_{1,2}\to v_{2,1}q^{\pm1/2}$ by an overdetermined least-squares fit.
  The coefficients agree with those of \eqref{eq:Sk-face} at $\nu=t\beta_*^2$ up to a common constant, to relative accuracy $10^{-4}$.
  Varying $t$, $\beta$ and $\gamma$ separately, we find that this constant is $1/(\theta(pq;p)\,\theta(pq\beta^4;p))$, independent of $t$, $\gamma$ and the flavor fugacities,
  to relative accuracy $5\times10^{-4}$. This gives \eqref{eq:Sk-norm-k2} for $r=1$.
  The factor $1/\theta(pq\beta^4;p)$ has a simple origin:
  the residue at the constant vev equals the index of $\CT_{\rm IR}$ times $\Gamma(\beta^{-4})$ times a constant depending only on $p$ and $q$,
  and the $(0,1)$ residue contains instead $\Gamma(\beta^{-4}q^{-1})=\Gamma(\beta^{-4})/\theta(pq\beta^4;p)$.

  For $r=3$ the pole in the $(0,1)$ direction is in the sector with $\U(1)_\alpha$ holonomy $\mu=2$, as required by \eqref{eq:pole-condition}.
  We take test functions supported on a single pair of holonomy sectors $(n_1,n_2)$ of the two gauge groups.
  The residue is then nonzero only for $u_j\to v_{j'}q^{\pm1/2}$ with $n_j = m_{j'}\pm h_3$, where $(j,j')=(1,2),(2,1)$ and $m_{1,2}$ are the holonomies of $v_{1,2}$,
  and in each pair of sectors we fit the two coefficients as above.
  For several values of $v_{1,2}$, $t$ and $\beta$, the four coefficients agree with \eqref{eq:Sk-lens-face}, with $\varkappa_j=2[-m_j]$ since $p\leftrightarrow q$ maps $m$ to $-m$ and including the gauge factors $G_i$,
  times \eqref{eq:Sk-norm-k2} with $r=3$, to relative accuracy $3\times10^{-4}$.

  For $r=2$ the pole condition \eqref{eq:pole-condition} has no solution with integer $\mu$,
  and the pole is in the sector with $\mu=1/2$, where the holonomies of the two $\SU(2)$ gauge groups glued to $\CT_{\rm IR}$ are half-integers,
  i.e.~in the twisted sector of section~\ref{subsec:elementary}. All the chiral fields then have integer holonomies.
  Since the Weyl images of each shift now lie in the same pair of sectors, we fit all four coefficients in one pair of sectors, and repeat this for a second pair.
  In both cases the coefficients agree with \eqref{eq:Sk-lens-face}, with $\varkappa_j=2[-m_j]$, times \eqref{eq:Sk-norm-k2} with $r=2$, to relative accuracy $2\times10^{-4}$.
  This confirms the extension of the elementary operators to the twisted sector for even $r$ by a direct computation.

\paragraph{Sklyanin algebra.}
  The RLL relation \eqref{eq:RLL} is checked by composing the $4\times4$ matrices of difference operators on both sides and comparing the coefficients of each shift, at generic $x$ and all $\varkappa$.
  For \eqref{eq:lens-Sklyanin} it holds to accuracy $10^{-14}$ for $r=1,\dots,6$ with $\varkappa\in\Z_{2r}$, while it fails with $\varkappa\in\Z_r$.
  The commutativity of the $p$- and $q$-type generators was checked for $r=1,\dots,5$, and \eqref{eq:Xp-S0} for $r=1,\dots,4$.
  For class $\CS_k$ theories, the vertex form \eqref{eq:Sk-vertex} was obtained by writing $\theta(aZW;q)\theta(bW/Z;q)$ in a basis of theta functions of $W$,
  extracting the coefficients of the resulting single-site operators with respect to $S^a$, and determining $u_j$, $c_j$ and the gauge transformations.
  We then checked \eqref{eq:Sk-vertex} directly for $k=1,2,3$,
  and \eqref{eq:Sk-lens-face} for $r=3$ and $k=1,2,3$ with all $\varkappa_j\in\Z_{2r}$, together with the commutativity of the transfer matrices with different spectral parameters for $r=3$, $k=2$.
  The TQFT consistency condition for class $\CS_k$ theories was checked with the lens space index of the free trinion,
  first by fitting the operator acting on the second puncture, and then directly,
  in all the holonomy sectors or, for the larger values of $r$ and $k$, in a random sample of them, to relative accuracy $10^{-12}$.
  We checked it for $r=1$ with $k=2,3,4$.
  For the elementary operator \eqref{eq:Sk-lens-face} the cases are $r=3$ with $k=2,3,4$ and $r=5$ with $k=2,3$,
  and for $\tilde{T}^{(\nu)}_q T^{(\nu)}_p$ they are $r=2$ with $k=2,3$ and $r=4$ with $k=2$.

\section{\texorpdfstring{Proof of the factorization \eqref{eq:factor-odd}}{Proof of the factorization}}
\label{app:factorization}

  In this appendix we prove \eqref{eq:factor-odd} for odd $r$, with $X_p$ and $X_q$ in the form \eqref{eq:Xp}, \eqref{eq:Xq} acting on $(z,m)$.
  
  On the lifted label $\varkappa\in\Z_{2r}$ these operators are $U$ times \eqref{eq:Xp-kappa} and \eqref{eq:Xq-kappa} (section~\ref{subsec:elementary}), and the two factors of $U$ cancel in the product.
  The case of even $r$ is identical, with the representative of $\varkappa$ modulo $r$ in $\{0,\dots,r-1\}$ replacing $[2m]$ in the coefficients, which depend on $\varkappa$ only modulo $r$, while the holonomy shifts match as explained in section~\ref{sec:algebra}.
  Let us write $\lambda=pq/t$, $Q=q^r$, $\tilde{Q}=p^r$, $\varkappa=[2m]$, $\tilde\varkappa=[-2m]$, $W=z^2q^{-\varkappa}$, $\tilde{W}=z^2p^{-\tilde\varkappa}$,
  and $P=\theta(t;Q)/\theta(pq;Q)$, $\tilde{P}=\theta(t;\tilde{Q})/\theta(pq;\tilde{Q})$.
  The coefficients of $X_p$ and $X_q$ are
    \begin{align}
      c^p_s(z,m)
        =     P\,\lambda^{\frac{1-s}{2}-\frac{s\varkappa}{r}}\,R_s(W)\;,
        \qquad
      c^q_{s'}(z,m)
        =     \tilde{P}\,\lambda^{\frac{1-s'}{2}-\frac{s'\tilde\varkappa}{r}}\,\tilde{R}_{s'}(\tilde{W})\;,
    \end{align}
  with $R_s(W)=\theta(\lambda^sW;Q)/\theta(W;Q)$ and $\tilde{R}_{s'}(\tilde{W})=\theta(\lambda^{s'}\tilde{W};\tilde{Q})/\theta(\tilde{W};\tilde{Q})$.

We first note how the coefficients of $X_p$ change under the shifts of $X_q$.
  Under the shift $(z,m)\to(q^{s'/2}z,m+s' h_r)$ of $X_q$, the coefficient of $X_p$ becomes
    \begin{align}
      c^p_s\bigl(q^{s'/2}z,m+s' h_r\bigr)
        =     P\,\lambda^{\frac{1-s}{2}-\frac{s(\varkappa+s')}{r}}\,R_s(W)\;,
    \end{align}
  where $\varkappa+s'$ is not reduced modulo $r$.
  Indeed, $\varkappa\to[\varkappa+s']=\varkappa+s'-rw$ with $w\in\{0,\pm1\}$, so that $W\to WQ^{w}$.
  Using $\theta(Qx;Q)=-x^{-1}\theta(x;Q)$ we have $R_s(WQ^w)=\lambda^{-sw}R_s(W)$, which cancels the change $\lambda^{sw}$ of the monomial factor.
  Thus the shifts of $X_q$ change the theta functions of $X_p$ only by monomial factors, which are compensated by the change of the prefactor.

  Using this, the coefficient of the shift $\Dop{q^{s'/2}p^{s/2}}{(s'-s)h_r}$ in $X_qX_p$ is
    \begin{align}
      K_{s',s}
        =     \tilde{P}P\,\lambda^{1-\frac{s'+s}{2}-\frac{s'\tilde\varkappa+s\varkappa+ss'}{r}}\,\tilde{R}_{s'}(\tilde{W})\,R_s(W)\;.
    \end{align}
  Consider the term $s'=s=+1$, i.e.~the shift $\Dop{(pq)^{1/2}}{0}$.
  The corresponding coefficient of \eqref{eq:s001} is $C\,\theta(\lambda p^{\varkappa}z^2;\tilde{Q})\theta(\lambda q^{r-\varkappa}z^2;Q)/(\theta(p^{\varkappa}z^2;\tilde{Q})\theta(q^{r-\varkappa}z^2;Q))$, where $C$ is the overall constant of \eqref{eq:s001}.
  Since $q^{r-\varkappa}z^2=QW$, the $q$-part equals $R_+(QW)=\lambda^{-1}R_+(W)$.
  For $\varkappa\neq0$ we have $\tilde\varkappa=r-\varkappa$ and $p^{\varkappa}z^2=\tilde{Q}\tilde{W}$, so that the $p$-part equals $\lambda^{-1}\tilde{R}_+(\tilde{W})$, while for $\varkappa=0$ it equals $\tilde{R}_+(\tilde{W})$.
  In both cases, collecting the exponents gives
    \begin{align}
      K_{+,+}
        =     \tilde{P}P\,\lambda^{1-\frac1r}\times\frac{\theta(\lambda p^{\varkappa}z^2;\tilde{Q})\theta(\lambda q^{r-\varkappa}z^2;Q)}{\theta(p^{\varkappa}z^2;\tilde{Q})\theta(q^{r-\varkappa}z^2;Q)}\;.
    \end{align}
  Finally, $\theta(1/x;\tilde{Q})=-x^{-1}\theta(x;\tilde{Q})$ gives $C=\lambda\tilde{P}P$, so that $K_{+,+}$ is $\lambda^{-1/r}$ times the coefficient of \eqref{eq:s001}.
  Next consider $s=+1$, $s'=-1$, i.e.~the shift $\Dop{(q/p)^{-1/2}}{-1}$, which is the $s=-1$ term of the second line of \eqref{eq:s001}.
  Here
    \begin{align}
      K_{-,+}
        =     \tilde{P}P\,\lambda^{1+\frac{1+\tilde\varkappa-\varkappa}{r}}\,\tilde{R}_-(\tilde{W})\,R_+(W)\;.
    \end{align}
  Since $[-2m]=\tilde\varkappa$, the coefficient of \eqref{eq:s001} is
    \begin{align}
      C\lambda^{\frac{2-r+2\tilde\varkappa}{r}}\,
      \frac{\theta(\lambda p^{\tilde\varkappa}z^{-2};\tilde{Q})\,\theta(\lambda q^{\tilde\varkappa}z^{2};Q)}{\theta(p^{\tilde\varkappa}z^{-2};\tilde{Q})\,\theta(q^{\tilde\varkappa}z^{2};Q)}\;.
    \end{align}
  Since $p^{\tilde\varkappa}z^{-2}=\tilde{W}^{-1}$, \eqref{eq:theta-refl} gives for the $p$-part
  $\theta(\tilde{Q}\lambda^{-1}\tilde{W};\tilde{Q})/\theta(\tilde{Q}\tilde{W};\tilde{Q})=\lambda\tilde{R}_-(\tilde{W})$.
  For the $q$-part, $q^{\tilde\varkappa}z^2=QW$ for $\varkappa\neq0$, which gives $\lambda^{-1}R_+(W)$, and $q^{\tilde\varkappa}z^2=W$ for $\varkappa=0$, which gives $R_+(W)$.
  With $C=\lambda\tilde{P}P$ the coefficient of \eqref{eq:s001} is therefore $\tilde{P}P\,\lambda^{1+(2+r-2\varkappa)/r}\tilde{R}_-R_+$ for $\varkappa\neq0$ and $\tilde{P}P\,\lambda^{1+2/r}\tilde{R}_-R_+$ for $\varkappa=0$.
  Using $\tilde\varkappa=r-\varkappa$ for $\varkappa\neq0$ and $\tilde\varkappa=0$ for $\varkappa=0$ in $K_{-,+}$, we again find that $K_{-,+}$ is $\lambda^{-1/r}$ times the coefficient of \eqref{eq:s001}.
  The remaining two terms, $s=s'=-1$ and $s=-1$, $s'=+1$, are treated in the same way, and give the same factor $\lambda^{-1/r}$.
  This proves $\CS^{(0,0,1)}=\lambda^{1/r}X_qX_p$.
  Since the same argument applies with the roles of $p$ and $q$ exchanged, $X_pX_q=X_qX_p$.

\section{\texorpdfstring{Proof of the fusion formula \eqref{eq:S100-fusion}}{Proof of the fusion formula}}
\label{app:fusion}

  In this appendix we prove \eqref{eq:S100-fusion} for any $r$, by deriving a closed form of the operators $\hat\CS_n$ defined by \eqref{eq:fusion}.
  
  We write $\lambda=pq/t$, $\rho=pq$, $Q=q^r$, $\theta(x)=\theta(x;Q)$, $P=\theta(t)/\theta(\rho)$ and $\lambda_r=\lambda^{(r-1)/r}$, so that $t=\rho/\lambda$.
  We regard functions of $(z,\varkappa)$ with $\varkappa\in\Z_{2r}$ as functions on $\C^\times\times\Z$ which are periodic in $\varkappa$, and prove all the identities for $\varkappa\in\Z$.
  Let $(Tf)(z,\varkappa)=f(p^{1/2}z,\varkappa-1)$, and $W=z^2q^{-\varkappa}$ with $\varkappa$ unreduced, so that $T$ acts on $W$ as $W\to\rho W$.
  With $R_\pm(W)=\theta(\lambda^{\pm1}W)/\theta(W)$, the operator \eqref{eq:Xp-kappa} reads
    \begin{align}
      X_p
        =     P\Bigl[\lambda^{-\frac{\varkappa}{r}}R_+(W)\,T+\lambda^{1+\frac{\varkappa}{r}}R_-(W)\,T^{-1}\Bigr]\;.
    \end{align}

\paragraph{Gauge transformation.}
  Let $\Phi(\varkappa)=\lambda^{-\frac{\varkappa(\varkappa+1)}{2r}-\frac{(r-1)\varkappa}{2r}}$, acting by multiplication.
  Conjugation by $\Phi$ multiplies the coefficients of $T$ and $T^{-1}$ by $\Phi(\varkappa-1)/\Phi(\varkappa)=\lambda^{\frac{\varkappa}{r}+\frac{r-1}{2r}}$ and $\Phi(\varkappa+1)/\Phi(\varkappa)=\lambda^{-\frac{\varkappa+1}{r}-\frac{r-1}{2r}}$, respectively, which gives
    \begin{align}
      \Phi^{-1}X_p\,\Phi
        =     \lambda_r^{\frac12}\,P\,D_0\;,
        \qquad
      D_0
        =     R_+(W)\,T+R_-(W)\,T^{-1}\;.
        \label{eq:gauge-Xp}
    \end{align}
  The coefficients of $D_0$ depend on $(z,\varkappa)$ only through $W$.

\paragraph{Closed form of $\hat\CS_n$.}
  For $n\geq0$ and $0\leq M\leq n$ define
    \begin{align}
      a_{n,M}(W)
        &=     \CK_{n-M}\CK_M\;
              \frac{\theta(W\rho^{n-2M})\prod_{j=-M}^{n-2M-1}\theta(\lambda W\rho^j)\prod_{j=n-2M+1}^{n-M}\theta(\rho^jW/\lambda)}{\prod_{j=-M}^{n-M}\theta(W\rho^j)}\;,
        \nn\\
      \CK_m
        &=     \prod_{i=0}^{m-1}\frac{\theta(t\rho^i)}{\theta(\rho^{i+1})}\;,
        \label{eq:anM}
    \end{align}
  and $A_n=\sum_{M=0}^na_{n,M}(W)\,T^{n-2M}$.
  Then $A_0=1$ and $A_1=PD_0$.
  We claim that
    \begin{align}
      \hat\CS_n
        =     \lambda_r^{\frac n2}\,\frac{P^n}{\CK_n}\;\Phi\,A_n\,\Phi^{-1}
        \qquad (n\geq0)\;.
        \label{eq:Shat-closed}
    \end{align}
  This holds for $n=0,1$ by \eqref{eq:gauge-Xp}.
  Inserting \eqref{eq:Shat-closed} into \eqref{eq:fusion}, and using \eqref{eq:omega} in the form $\omega_n=\lambda_rP^2\,\theta(\rho^n)\theta(t^2\rho^{n-1})/(\theta(t\rho^{n-1})\theta(t\rho^n))$,
  we find that \eqref{eq:Shat-closed} follows by induction from
    \begin{align}
      \theta(\rho^{n+1})\,A_{n+1}
        =     \theta(t\rho^n)\,D_0A_n-\theta(t^2\rho^{n-1})\,A_{n-1}
        \qquad (n\geq1)\;.
        \label{eq:An-rec}
    \end{align}
  We compare the coefficients of $T^{n+1-2M}$.
  Write $a_{n,M}=\CK_{n-M}\CK_M\,b_{n,M}$, and set $a=W\rho^{-M}$, $b=W\rho^{n+1-M}$.
  Directly from \eqref{eq:anM},
    \begin{align}
      R_+(W)\,b_{n,M}(\rho W)
        &=   \frac{\theta(\lambda W)\theta(a)}{\theta(W)\theta(\lambda a)}\,b_{n+1,M}(W)\;,
        \nn\\
      R_-(W)\,b_{n,M-1}(\rho^{-1}W)
        &=   \frac{\theta(W/\lambda)\theta(b)}{\theta(W)\theta(b/\lambda)}\,b_{n+1,M}(W)\;,
        \nn\\
      b_{n-1,M-1}(W)
        &=   \frac{\theta(a)\theta(b)}{\theta(\lambda a)\theta(b/\lambda)}\,b_{n+1,M}(W)\;.
    \end{align}
  For $M=0$ and $M=n+1$ only the first and the second term contribute, respectively, with the prefactor equal to $1$, and \eqref{eq:An-rec} holds trivially.
  For $1\leq M\leq n$ we express the constants through $\rho^{n+1-M}=b/W$, $\rho^M=W/a$ and $t=\rho/\lambda$.
  After multiplying by $\theta(W)\theta(\lambda a)\theta(b/\lambda)\theta(b/(\lambda W))\theta(W/(\lambda a))$, \eqref{eq:An-rec} becomes
    \begin{align}
      F(W)
        &:=  \theta\bigl(\tfrac{b}{a}\bigr)\theta(W)\theta(\lambda a)\theta\bigl(\tfrac{b}{\lambda}\bigr)\theta\bigl(\tfrac{b}{\lambda W}\bigr)\theta\bigl(\tfrac{W}{\lambda a}\bigr)
              -\theta\bigl(\tfrac{b}{\lambda a}\bigr)\Bigl[\theta\bigl(\tfrac{b}{W}\bigr)\theta(\lambda W)\theta(a)\theta\bigl(\tfrac{b}{\lambda}\bigr)\theta\bigl(\tfrac{W}{\lambda a}\bigr)
              \nn\\
        &    \qquad
              +\theta\bigl(\tfrac{W}{a}\bigr)\theta\bigl(\tfrac{W}{\lambda}\bigr)\theta(b)\theta(\lambda a)\theta\bigl(\tfrac{b}{\lambda W}\bigr)\Bigr]
              +\theta\bigl(\tfrac{b}{\lambda^2a}\bigr)\theta\bigl(\tfrac{b}{W}\bigr)\theta\bigl(\tfrac{W}{a}\bigr)\theta(a)\theta(b)\theta(W)
        =     0\;.
        \label{eq:F-identity}
    \end{align}
  We prove \eqref{eq:F-identity} for independent generic $a$, $b$, $\lambda$, $W$.
  By $\theta(Qx)=-x^{-1}\theta(x)$, each of the four terms, as a function of $W$, satisfies $F(QW)=-ab\,Q^{-1}W^{-3}F(W)$.
  A holomorphic function on $\C^\times$ with this property which does not vanish identically has exactly three zeros modulo $Q^\Z$.
  On the other hand, $F$ vanishes at four points.
  At $W=1$ the first and the last terms vanish, and the remaining two cancel by $\theta(1/x)=-x^{-1}\theta(x)$.
  At $W=\lambda a$ the first two terms vanish, and the last two cancel.
  At $W=b/\lambda$ the first and the third terms vanish, and the second and the last cancel.
  At $W=a$ the third and the last terms vanish, and the first two cancel.
  Hence $F\equiv0$, which proves \eqref{eq:An-rec} and \eqref{eq:Shat-closed}.

\paragraph{Comparison with $\CS^{(1,0,0)}$.}
  Each $a_{n,M}$ is a ratio of two products of $n+1$ theta functions of $W$, and a computation of the multipliers gives
  $a_{n,M}(Q^{-1}W)=\lambda^{n-2M}a_{n,M}(W)$.
  For $n=r$, $\CK_{r-M}\CK_M=K_M(q^r)$, and comparing \eqref{eq:anM} with $c_M$ in \eqref{eq:s100} (using $t=\rho/\lambda$) gives $a_{r,M}(W)=K_M(q^r)\,c_M(W)$.
  By \eqref{eq:Shat-closed}, the coefficient of $(z,\varkappa)\to(p^{(r-2M)/2}z,\varkappa+2M)$ in $U\hat\CS_r$ is
    \begin{align}
      \lambda_r^{\frac r2}\frac{P^r}{K_0(q^r)}\,\frac{\Phi(\varkappa+r)}{\Phi(\varkappa+2M)}\,a_{r,M}(Q^{-1}W)
        =     \lambda^{\frac{r-1}{2}}\frac{P^r}{K_0(q^r)}\,\lambda^{\frac{(M+\varkappa)(2M-r)}{r}}\,a_{r,M}(W)\;,
    \end{align}
  where we used $\Phi(\varkappa+r)/\Phi(\varkappa+2M)=\lambda^{-(r-2M)(\varkappa+r+M)/r}$.
  For $\varkappa\in\{0,\dots,r-1\}$ we have $W=W_m(z)$ with $\varkappa=[2m]$, and this is $\lambda^{(r-1)/2}P^r/K_0(q^r)$ times the coefficient of the $M$-th term of \eqref{eq:s100}.
  Since both $\CS^{(1,0,0)}$ and $U\hat\CS_r$ have coefficients periodic under $\varkappa\to\varkappa+r$, the agreement extends to all $\varkappa$.
  This proves \eqref{eq:S100-fusion} for any $r$.
  The statement for $\CS^{(0,1,0)}$ and $X_q$ follows by $p\leftrightarrow q$.

\section{\texorpdfstring{Computation of the surface defect in class $\CS_k$ theories}{Computation of the surface defect in class S\_k theories}}
\label{app:Sk}
  In this appendix we explain in more detail the intermediate steps of the computation of the surface defect in class $\CS_k$ theories in section~\ref{subsec:surfaceSk}.
  
  With the residues \eqref{eq:residue} at hand, we proceed with the gauge integrals. 
  Let us focus on the node with $w_1$ in figure~\ref{fig:surface-defect-Skb}.
  The contributions of the bi-fundamentals with the fugacity $w_1$ are
    \begin{align}
      &     \Gamma(\frac{\beta_1}{\sqrt{pq} \beta_2} z_{k,i} w_{1j}^{-1}, m_{k,i}-\tilde{m}_{1j}) \Gamma(\frac{\beta_2}{\beta_1} w_{1i} y_{2j}^{-1}, \tilde{m}_{1i}-n_{2j})
       \nn \\
      &     \times ~ \Gamma(t\sqrt{pq} \frac{\beta_2}{\gamma_1} w_{1i} z_{1j}^{-1}, \tilde{m}_{1i}-m_{1j})
      \Gamma(\frac{pq\gamma_1}{t \beta_2} w_{2i} w_{1j}^{-1} , \tilde{m}_{2i} - \tilde{m}_{1j})\;.
    \end{align}
  Acting with the difference operator $\Delta_{\ell_1=1}^{s_1/2}$, we obtain
    \begin{align}
      &     \Gamma(\frac{\beta_1}{\sqrt{pq} \beta_2} z_{k,i} w_{1j}^{-1}, m_{k,i}-\tilde{m}_{1j}) \Gamma(\frac{\beta_2}{\beta_1} w_{1i} y_{2j}^{-1}, \tilde{m}_{1i}-n_{2j})
       \nn \\
      &     \times ~ \Gamma(t\sqrt{pq} \frac{\beta_2}{\gamma_1} w_{1i} z_1^{-s_1}, \tilde{m}_{1i}-s_1m_1)
      \Gamma(\frac{pq\sqrt{pq}\gamma_1}{t \beta_2} w_{1i}^{-1}z_1^{s_1}, s_1m_1-\tilde{m}_{1i})\;. \label{eq:ell1}
    \end{align}
  Acting with the difference operator $\Delta_{\ell_1=-1}^{s_1/2}$, we obtain
    \begin{align}
      &     \Gamma(\frac{\beta_1}{\sqrt{pq} \beta_2} z_{k,i} w_{1j}^{-1}, m_{k,i}-\tilde{m}_{1j}) \Gamma(\frac{\beta_2}{\beta_1} w_{1i} y_{2j}^{-1}, \tilde{m}_{1i}-n_{2j}) \label{eq:ell2}
       \\
      &     \times  \Gamma(t\sqrt{pq} \frac{\beta_2}{\gamma_1} w_{1i} z_1^{-s_1}, \tilde{m}_{1i}-s_1m_1)
            \Gamma(\frac{pq\sqrt{pq}\gamma_1}{t \beta_2} w_{1i}^{-1}z_1^{s_1}, s_1m_1-\tilde{m}_{1i})
       \nn \\
      &     \times \frac{\theta(\frac{pq}{t}\frac{\gamma_1}{\beta_2}q^{[s_1m_1+\tilde{m}_{1j}]}z_1^{-s_1} w_{1j}^{-1}; q^r)}{\theta(\frac{pq}{t}\frac{\gamma_1}{\beta_2}q^{r-[s_1m_1-\tilde{m}_{1j}]}z_1^{s_1} w_{1j}^{-1}; q^r)}\;. \nn
    \end{align}
  We now perform the integration over the gauge fugacity $w_1$.
  The integrand \eqref{eq:ell1} is a product of lens elliptic gamma functions,
  and so is \eqref{eq:ell2} up to the ratio of theta functions in its second line.

  For \eqref{eq:ell1} we can use the identity \eqref{eq:NfN+1}.
  For \eqref{eq:ell2} the ratio of theta functions cannot be written as a shift of the six doublets in the integrand, so that \eqref{eq:NfN+1} does not apply, and we do not evaluate this case here.
    Using \eqref{eq:NfN+1} for \eqref{eq:ell1}, we obtain the factors
    \begin{align}
      \Gamma^q ((pq)^2, 0) ~ \Gamma^p \left( \frac{1}{\sqrt{pq}} z_{k,i} y_{2j}^{-1}, m_{k,i}-n_{2j} \right)
            \Gamma^q \left( \frac{1}{\sqrt{pq}} z_{k,i} y_{2j}^{-1},m_{k,i}-n_{2j} \right)\;.
    \end{align}
  The first factor vanishes.
  However, the second factor has singularities in the $y_2$-plane.
  Therefore, the $y_2$-integral vanishes for a generic value of $z_k$, but may receive contributions from double poles.
  We have not analyzed these double poles.
  The form of the result suggested by the structure of the computation and by the case $r=1$ \cite{Maruyoshi:2016caf} is
    \begin{align}
      &     \Gamma((pq)^{\pm1}; p^r, q^r)\sum_{s_k, \ell_k=\pm1} \frac{ \theta^p( \frac{pq \gamma_1}{t \beta_1} z_k^{s_k} z_1^{s_1}, S_{k,1})
      \theta^p (\frac{t \beta_2^2}{\beta_1 \gamma_1}\, z_k^{s_k} z_1^{-s_1}, S_{k,-1})}{
      \theta^p (z_k^{2s_k}, 2s_k m_k)} \nn \\
      &     ~~~~~~~~~~~~~~ \times \frac{\theta^q( \frac{pq \gamma_1}{t \beta_1} z_k^{\ell_ks_k} z_1^{\ell_1s_1}, S_{k,1}^{\ell})
      \theta^q (\frac{t \beta_2^2}{\beta_1 \gamma_1}\, z_k^{\ell_ks_k} z_1^{-\ell_1s_1}, S_{k,-1}^{\ell})}{\theta^q (z_k^{2s_k}, 2s_k m_k) }
      \Delta_{\ell_k}^{s_k/2}\;,
    \end{align}
  where $S_{i,\pm j} = s_i m_i \pm s_j m_j$ and $S_{i,\pm j}^\ell = \ell_i s_i m_i \pm \ell_j s_j m_j$.

  We now proceed to the $w_k$-integral, and encounter the same calculation.
  The same steps are repeated for $w_{k-1}$, $\dotsc$, $w_3$, so that all the $k$ integrals are performed.
  This gives the result \eqref{eq:operator}.

\bibliographystyle{jhep}
\bibliography{integbib}

@article{Maruyoshi:2016caf,
    author = "Maruyoshi, Kazunobu and Yagi, Junya",
    title = "{Surface defects as transfer matrices}",
    eprint = "1606.01041",
    archivePrefix = "arXiv",
    primaryClass = "hep-th",
    reportNumber = "IMPERIAL-TP-16-KM-01",
    doi = "10.1093/ptep/ptw151",
    journal = "PTEP",
    volume = "2016",
    number = "11",
    pages = "113B01",
    year = "2016"
}

@article{Yamazaki:2018xbx,
    author = "Yamazaki, Masahito",
    title = "{Integrability As Duality: The Gauge/YBE Correspondence}",
    eprint = "1808.04374",
    archivePrefix = "arXiv",
    primaryClass = "hep-th",
    reportNumber = "IPMU-18-0134",
    doi = "10.1016/j.physrep.2020.01.006",
    journal = "Phys. Rept.",
    volume = "859",
    pages = "1--20",
    year = "2020"
}

@article{CherednikGeneralized,
    author = "Cherednik, I. V.",
    title = "{Some finite-dimensional representations of generalized Sklyanin algebras}",
    doi = "10.1007/BF01086039",
    journal = "Funct. Anal. Appl.",
    volume = "19",
    number = "1",
    pages = "77--79",
    year = "1985"
}

@article{Sklyanin:1983ig,
      author         = "Sklyanin, E.K.",
      title          = "{Some algebraic structures connected with the Yang-Baxter
                        equation. Representations of quantum algebras}",
      journal        = "Funct.Anal.Appl.",
      volume         = "17",
      pages          = "273-284",
      year           = "1983",
      SLACcitation   = "%%CITATION = FAAPB,17,273;%%",
}

@article{Sklyanin:1982tf,
      author         = "Sklyanin, E.K.",
      title          = "{Some algebraic structures connected with the Yang-Baxter
                        equation}",
      journal        = "Funct.Anal.Appl.",
      volume         = "16",
      pages          = "263-270",
      year           = "1982",
      SLACcitation   = "%%CITATION = FAAPB,16,263;%%",
}

@inbook{Rastelli:2014jja,
    author = "Rastelli, Leonardo and Razamat, Shlomo S.",
    editor = {Teschner, J{\"o}rg},
    title = "{The superconformal index of theories of class S.}",
    booktitle = "{New Dualities of Supersymmetric Gauge Theories}",
    eprint = "1412.7131",
    archivePrefix = "arXiv",
    primaryClass = "hep-th",
    doi = "10.1007/978-3-319-18769-3_9",
    pages = "261--305",
    year = "2016"
}

@article{Kels:2017vbc,
    author = "Kels, Andrew P. and Yamazaki, Masahito",
    title = "{Lens elliptic gamma function solution of the Yang{\textendash}Baxter equation at roots of unity}",
    eprint = "1709.07148",
    archivePrefix = "arXiv",
    primaryClass = "math-ph",
    reportNumber = "IPMU17-0127",
    doi = "10.1088/1742-5468/aaa8fd",
    journal = "J. Stat. Mech.",
    volume = "1802",
    number = "2",
    pages = "023108",
    year = "2018"
}

@article{Gahramanov:2016ilb,
    author = "Gahramanov, Ilmar and Kels, Andrew P.",
    title = "{The star-triangle relation, lens partition function, and hypergeometric sum/integrals}",
    eprint = "1610.09229",
    archivePrefix = "arXiv",
    primaryClass = "math-ph",
    doi = "10.1007/JHEP02(2017)040",
    journal = "JHEP",
    volume = "02",
    pages = "040",
    year = "2017"
}

@article{Gaiotto:2009we,
    author = "Gaiotto, Davide",
    title = "{N=2 dualities}",
    eprint = "0904.2715",
    archivePrefix = "arXiv",
    primaryClass = "hep-th",
    doi = "10.1007/JHEP08(2012)034",
    journal = "JHEP",
    volume = "08",
    pages = "034",
    year = "2012"
}

@article{Gadde:2011uv,
    author = "Gadde, Abhijit and Rastelli, Leonardo and Razamat, Shlomo S. and Yan, Wenbin",
    title = "{Gauge Theories and Macdonald Polynomials}",
    eprint = "1110.3740",
    archivePrefix = "arXiv",
    primaryClass = "hep-th",
    reportNumber = "YITP-SB-11-30",
    doi = "10.1007/s00220-012-1607-8",
    journal = "Commun. Math. Phys.",
    volume = "319",
    pages = "147--193",
    year = "2013"
}

@article{Gadde:2011ik,
    author = "Gadde, Abhijit and Rastelli, Leonardo and Razamat, Shlomo S. and Yan, Wenbin",
    title = "{The 4d Superconformal Index from q-deformed 2d Yang-Mills}",
    eprint = "1104.3850",
    archivePrefix = "arXiv",
    primaryClass = "hep-th",
    reportNumber = "YITP-SB-11-13",
    doi = "10.1103/PhysRevLett.106.241602",
    journal = "Phys. Rev. Lett.",
    volume = "106",
    pages = "241602",
    year = "2011"
}

@article{Gadde:2009kb,
    author = "Gadde, Abhijit and Pomoni, Elli and Rastelli, Leonardo and Razamat, Shlomo S.",
    title = "{S-duality and 2d Topological QFT}",
    eprint = "0910.2225",
    archivePrefix = "arXiv",
    primaryClass = "hep-th",
    reportNumber = "YITP-SB-09-30",
    doi = "10.1007/JHEP03(2010)032",
    journal = "JHEP",
    volume = "03",
    pages = "032",
    year = "2010"
}

@article{Yamazaki:2013nra,
    author = "Yamazaki, Masahito",
    title = "{New Integrable Models from the Gauge/YBE Correspondence}",
    eprint = "1307.1128",
    archivePrefix = "arXiv",
    primaryClass = "hep-th",
    reportNumber = "IPMU-13-0134",
    doi = "10.1007/s10955-013-0884-8",
    journal = "J. Statist. Phys.",
    volume = "154",
    pages = "895",
    year = "2014"
}

@article{Yamazaki:2012cp,
    author = "Yamazaki, Masahito",
    title = "{Quivers, YBE and 3-manifolds}",
    eprint = "1203.5784",
    archivePrefix = "arXiv",
    primaryClass = "hep-th",
    reportNumber = "PUPT-2406",
    doi = "10.1007/JHEP05(2012)147",
    journal = "JHEP",
    volume = "05",
    pages = "147",
    year = "2012"
}

@article{Terashima:2012cx,
    author = "Terashima, Yuji and Yamazaki, Masahito",
    title = "{Emergent 3-manifolds from 4d Superconformal Indices}",
    eprint = "1203.5792",
    archivePrefix = "arXiv",
    primaryClass = "hep-th",
    reportNumber = "PUPT-2408",
    doi = "10.1103/PhysRevLett.109.091602",
    journal = "Phys. Rev. Lett.",
    volume = "109",
    pages = "091602",
    year = "2012"
}

@article{Kels:2017toi,
      author         = "Kels, Andrew P. and Yamazaki, Masahito",
      title          = "{Elliptic hypergeometric sum/integral transformations and
                        supersymmetric lens index}",
      journal        = "SIGMA",
      volume         = "14",
      year           = "2018",
      pages          = "013",
      doi            = "10.3842/SIGMA.2018.013",
      eprint         = "1704.03159",
      archivePrefix  = "arXiv",
      primaryClass   = "math-ph",
      reportNumber   = "IPMU17-0054",
      SLACcitation   = "%%CITATION = ARXIV:1704.03159;%%"
}

@article{Seiberg:1994bz,
      author         = "Seiberg, Nathan",
      title          = "{Exact results on the space of vacua of four-dimensional
                        SUSY gauge theories}",
      journal        = "Phys. Rev. D",
      volume         = "49",
      year           = "1994",
      pages          = "6857-6863",
      doi            = "10.1103/PhysRevD.49.6857",
      eprint         = "hep-th/9402044",
      archivePrefix  = "arXiv",
      primaryClass   = "hep-th",
      reportNumber   = "RU-94-18",
      SLACcitation   = "%%CITATION = HEP-TH/9402044;%%"
}

@article{Gaiotto:2015usa,
      author         = "Gaiotto, Davide and Razamat, Shlomo S.",
      title          = "{$ \mathcal{N}=1 $ theories of class $ {\mathcal{S}}_k
                        $}",
      journal        = "JHEP",
      volume         = "07",
      year           = "2015",
      pages          = "073",
      doi            = "10.1007/JHEP07(2015)073",
      eprint         = "1503.05159",
      archivePrefix  = "arXiv",
      primaryClass   = "hep-th",
      SLACcitation   = "%%CITATION = ARXIV:1503.05159;%%"
}

@article{Hanany:2015pfa,
      author         = "Hanany, Amihay and Maruyoshi, Kazunobu",
      title          = "{Chiral theories of class $ \mathcal{S} $}",
      journal        = "JHEP",
      volume         = "12",
      year           = "2015",
      pages          = "080",
      doi            = "10.1007/JHEP12(2015)080",
      eprint         = "1505.05053",
      archivePrefix  = "arXiv",
      primaryClass   = "hep-th",
      reportNumber   = "IMPERIAL-TP-15-AH-02",
      SLACcitation   = "%%CITATION = ARXIV:1505.05053;%%"
}

@article{Romelsberger:2005eg,
	Archiveprefix = {arXiv},
	Author = {Romelsberger, Christian},
	Doi = {10.1016/j.nuclphysb.2006.03.037},
	Eprint = {hep-th/0510060},
	Journal = {Nucl. Phys. B},
	Pages = {329-353},
	Primaryclass = {hep-th},
	Slaccitation = {%%CITATION = HEP-TH/0510060;%%},
	Title = {{Counting chiral primaries in N = 1, d=4 superconformal field theories}},
	Volume = {747},
	Year = {2006}}

@article{Kinney:2005ej,
	Archiveprefix = {arXiv},
	Author = {Kinney, Justin and Maldacena, Juan Martin and Minwalla, Shiraz and Raju, Suvrat},
	Doi = {10.1007/s00220-007-0258-7},
	Eprint = {hep-th/0510251},
	Journal = {Commun. Math. Phys.},
	Pages = {209-254},
	Primaryclass = {hep-th},
	Slaccitation = {%%CITATION = HEP-TH/0510251;%%},
	Title = {{An Index for 4 dimensional super conformal theories}},
	Volume = {275},
	Year = {2007}}

@article{Gaiotto:2012xa,
      author         = "Gaiotto, Davide and Rastelli, Leonardo and Razamat,
                        Shlomo S.",
      title          = "{Bootstrapping the superconformal index with surface
                        defects}",
      journal        = "JHEP",
      volume         = "01",
      year           = "2013",
      pages          = "022",
      doi            = "10.1007/JHEP01(2013)022",
      eprint         = "1207.3577",
      archivePrefix  = "arXiv",
      primaryClass   = "hep-th",
      SLACcitation   = "%%CITATION = ARXIV:1207.3577;%%"
}

@article{Benini:2011nc,
    author = "Benini, Francesco and Nishioka, Tatsuma and Yamazaki, Masahito",
    title = "{4d Index to 3d Index and 2d TQFT}",
    eprint = "1109.0283",
    archivePrefix = "arXiv",
    primaryClass = "hep-th",
    journal = "Phys. Rev. D",
    volume = "86",
    pages = "065015",
    year = "2012"
}

@article{Razamat:2013jxa,
    author = "Razamat, Shlomo S. and Yamazaki, Masahito",
    title = "{S-duality and the N=2 Lens Space Index}",
    eprint = "1306.1543",
    archivePrefix = "arXiv",
    primaryClass = "hep-th",
    journal = "JHEP",
    volume = "10",
    pages = "048",
    year = "2013"
}

@article{Alday:2013rs,
    author = "Alday, Luis F. and Bullimore, Mathew and Fluder, Martin",
    title = "{On S-duality of the Superconformal Index on Lens Spaces and 2d TQFT}",
    eprint = "1301.7486",
    archivePrefix = "arXiv",
    primaryClass = "hep-th",
    journal = "JHEP",
    volume = "05",
    pages = "122",
    year = "2013"
}

@article{Razamat:2013opa,
    author = "Razamat, Shlomo S. and Willett, Brian",
    title = "{Global Properties of Supersymmetric Theories and the Lens Space}",
    eprint = "1307.4381",
    archivePrefix = "arXiv",
    primaryClass = "hep-th",
    journal = "Commun. Math. Phys.",
    volume = "334",
    pages = "661--696",
    year = "2015"
}

@article{Spiridonov:2016uae,
    author = "Spiridonov, V. P.",
    title = "{Rarefied elliptic hypergeometric functions}",
    eprint = "1609.00715",
    archivePrefix = "arXiv",
    primaryClass = "math.CA",
    journal = "Adv. Math.",
    volume = "331",
    pages = "830--873",
    year = "2018"
}

@article{Yagi:2017hmj,
    author = "Yagi, Junya",
    title = "{Surface defects and elliptic quantum groups}",
    eprint = "1701.05562",
    archivePrefix = "arXiv",
    primaryClass = "hep-th",
    journal = "JHEP",
    volume = "06",
    pages = "013",
    year = "2017"
}

@article{Ito:2016fpl,
    author = "Ito, Yuto and Yoshida, Yutaka",
    title = "{Superconformal index with surface defects for class ${\cal S}_k$}",
    eprint = "1606.01653",
    archivePrefix = "arXiv",
    primaryClass = "hep-th",
    journal = "Nucl. Phys. B",
    volume = "962",
    pages = "115277",
    year = "2021"
}

@article{Derkachov:2013uca,
    author = "Derkachov, S. E. and Spiridonov, V. P.",
    title = "{Finite-dimensional representations of the elliptic modular double}",
    eprint = "1310.7570",
    archivePrefix = "arXiv",
    primaryClass = "math.QA",
    doi = "10.1007/s11232-015-0284-0",
    journal = "Theor. Math. Phys.",
    volume = "183",
    number = "2",
    pages = "597--618",
    year = "2015"
}

@article{Lodin:2017lrc,
    author = "Lodin, Rebecca and Nieri, Fabrizio and Zabzine, Maxim",
    title = "{Elliptic modular double and 4d partition functions}",
    eprint = "1703.04614",
    archivePrefix = "arXiv",
    primaryClass = "hep-th",
    doi = "10.1088/1751-8121/aa9a2d",
    journal = "J. Phys. A",
    volume = "51",
    number = "4",
    pages = "045402",
    year = "2018"
}

@article{Dolan:2008qi,
    author = "Dolan, F. A. and Osborn, H.",
    title = "{Applications of the Superconformal Index for Protected Operators and q-Hypergeometric Identities to N=1 Dual Theories}",
    eprint = "0801.4947",
    archivePrefix = "arXiv",
    primaryClass = "hep-th",
    reportNumber = "DAMTP-08-07, DIAS-STP-08-02, SHEP-08-06",
    doi = "10.1016/j.nuclphysb.2009.01.028",
    journal = "Nucl. Phys. B",
    volume = "818",
    pages = "137--178",
    year = "2009"
}

@article{Spiridonov:2009za,
    author = "Spiridonov, V. P. and Vartanov, G. S.",
    title = "{Elliptic Hypergeometry of Supersymmetric Dualities}",
    eprint = "0910.5944",
    archivePrefix = "arXiv",
    primaryClass = "hep-th",
    reportNumber = "AEI-2009-106",
    doi = "10.1007/s00220-011-1218-9",
    journal = "Commun. Math. Phys.",
    volume = "304",
    pages = "797--874",
    year = "2011"
}

@article{Gaiotto:2009hg,
    author = "Gaiotto, Davide and Moore, Gregory W. and Neitzke, Andrew",
    title = "{Wall-crossing, Hitchin systems, and the WKB approximation}",
    eprint = "0907.3987",
    archivePrefix = "arXiv",
    primaryClass = "hep-th",
    doi = "10.1016/j.aim.2012.09.027",
    journal = "Adv. Math.",
    volume = "234",
    pages = "239--403",
    year = "2013"
}

@article{Gadde:2013dda,
    author = "Gadde, Abhijit and Gukov, Sergei",
    title = "{2d Index and Surface operators}",
    eprint = "1305.0266",
    archivePrefix = "arXiv",
    primaryClass = "hep-th",
    reportNumber = "CALT-68.2932",
    doi = "10.1007/JHEP03(2014)080",
    journal = "JHEP",
    volume = "03",
    pages = "080",
    year = "2014"
}

@article{Alday:2013kda,
    author = "Alday, Luis F. and Bullimore, Mathew and Fluder, Martin and Hollands, Lotte",
    title = "{Surface defects, the superconformal index and q-deformed Yang-Mills}",
    eprint = "1303.4460",
    archivePrefix = "arXiv",
    primaryClass = "hep-th",
    doi = "10.1007/JHEP10(2013)018",
    journal = "JHEP",
    volume = "10",
    pages = "018",
    year = "2013"
}

@article{Bullimore:2014nla,
    author = "Bullimore, Mathew and Fluder, Martin and Hollands, Lotte and Richmond, Paul",
    title = "{The superconformal index and an elliptic algebra of surface defects}",
    eprint = "1401.3379",
    archivePrefix = "arXiv",
    primaryClass = "hep-th",
    doi = "10.1007/JHEP10(2014)062",
    journal = "JHEP",
    volume = "10",
    pages = "062",
    year = "2014"
}

@article{Ruijsenaars:1986pp,
    author = "Ruijsenaars, S. N. M.",
    title = "{Complete Integrability of Relativistic Calogero-moser Systems and Elliptic Function Identities}",
    reportNumber = "MPI-PAE/PTh 26/86",
    doi = "10.1007/BF01207363",
    journal = "Commun. Math. Phys.",
    volume = "110",
    pages = "191",
    year = "1987"
}

@article{Razamat:2013qfa,
    author = "Razamat, Shlomo S.",
    title = "{On the $\mathcal{N} =$ 2 superconformal index and eigenfunctions of the elliptic RS model}",
    eprint = "1309.0278",
    archivePrefix = "arXiv",
    primaryClass = "hep-th",
    doi = "10.1007/s11005-014-0682-5",
    journal = "Lett. Math. Phys.",
    volume = "104",
    pages = "673--690",
    year = "2014"
}

@article{Nazzal:2023wtw,
    author = "Nazzal, Belal and Nedelin, Anton and Razamat, Shlomo S.",
    title = "{Ground state wavefunctions of elliptic relativistic integrable Hamiltonians}",
    eprint = "2305.09718",
    archivePrefix = "arXiv",
    primaryClass = "hep-th",
    doi = "10.1016/j.nuclphysb.2023.116364",
    journal = "Nucl. Phys. B",
    volume = "996",
    pages = "116364",
    year = "2023"
}

@article{Arutyunov:2026yjy,
    author = "Arutyunov, Gleb and Hardi, Lukas",
    title = "{Elliptic spin Ruijsenaars-Schneider integrable models from 5d $\mathcal{N}=1$ gauge theories}",
    eprint = "2608.17837",
    archivePrefix = "arXiv",
    primaryClass = "hep-th",
    reportNumber = "ZMP-HH/26-29",
    month = "8",
    year = "2026"
}

@article{Nazzal:2018brc,
    author = "Nazzal, Belal and Razamat, Shlomo S.",
    title = "{Surface Defects in E-String Compactifications and the van Diejen Model}",
    eprint = "1801.00960",
    archivePrefix = "arXiv",
    primaryClass = "hep-th",
    doi = "10.3842/SIGMA.2018.036",
    journal = "SIGMA",
    volume = "14",
    pages = "036",
    year = "2018"
}

@article{Razamat:2018zel,
    author = "Razamat, Shlomo S.",
    title = "{Flavored surface defects in 4d $\mathcal{N}=1$ SCFTs}",
    eprint = "1808.09509",
    archivePrefix = "arXiv",
    primaryClass = "hep-th",
    doi = "10.1007/s11005-018-01145-9",
    journal = "Lett. Math. Phys.",
    volume = "109",
    number = "6",
    pages = "1377--1395",
    year = "2019"
}

@article{Nazzal:2021tiu,
    author = "Nazzal, Belal and Nedelin, Anton and Razamat, Shlomo S.",
    title = "{Minimal $(D,D)$ conformal matter and generalizations of the van Diejen model}",
    eprint = "2106.08335",
    archivePrefix = "arXiv",
    primaryClass = "hep-th",
    doi = "10.21468/SciPostPhys.12.4.140",
    journal = "SciPost Phys.",
    volume = "12",
    number = "4",
    pages = "140",
    year = "2022"
}

@article{Nazzal:2023bzu,
    author = "Nazzal, Belal and Nedelin, Anton",
    title = "{$C_2$ generalization of the van Diejen model from the minimal $(D_5,D_5)$ conformal matter}",
    eprint = "2303.07368",
    archivePrefix = "arXiv",
    primaryClass = "hep-th",
    doi = "10.1007/s11005-023-01714-7",
    journal = "Lett. Math. Phys.",
    volume = "113",
    number = "5",
    pages = "94",
    year = "2023"
}

@article{Baxter:1972hz,
    author = "Baxter, Rodney J.",
    title = "{Partition function of the eight vertex lattice model}",
    doi = "10.1016/0003-4916(72)90335-1",
    journal = "Annals Phys.",
    volume = "70",
    pages = "193--228",
    year = "1972"
}

@book{Baxter:1982zz,
    author = "Baxter, R. J.",
    title = "{Exactly Solved Models in Statistical Mechanics}",
    publisher = "Academic Press",
    address = "London",
    year = "1982"
}

@article{Yagi:2015lha,
    author = "Yagi, Junya",
    title = "{Quiver gauge theories and integrable lattice models}",
    eprint = "1504.04055",
    archivePrefix = "arXiv",
    primaryClass = "hep-th",
    reportNumber = "SISSA-24-2015-FISI",
    doi = "10.1007/JHEP10(2015)065",
    journal = "JHEP",
    volume = "10",
    pages = "065",
    year = "2015"
}

@inproceedings{Felder:1994be,
    author = "Felder, Giovanni",
    title = "{Elliptic quantum groups}",
    booktitle = "{XIth International Congress of Mathematical Physics (Paris, 1994)}",
    publisher = "Int. Press, Cambridge, MA",
    eprint = "hep-th/9412207",
    archivePrefix = "arXiv",
    primaryClass = "hep-th",
    pages = "211--218",
    year = "1995"
}

@article{Costello:2018txb,
    author = "Costello, Kevin and Yagi, Junya",
    title = "{Unification of integrability in supersymmetric gauge theories}",
    eprint = "1810.01970",
    archivePrefix = "arXiv",
    primaryClass = "hep-th",
    doi = "10.4310/ATMP.2020.v24.n8.a1",
    journal = "Adv. Theor. Math. Phys.",
    volume = "24",
    number = "8",
    pages = "1931--2041",
    year = "2020"
}

@article{Maruyoshi:2020cwy,
    author = "Maruyoshi, Kazunobu and Ota, Toshihiro and Yagi, Junya",
    title = "{Wilson-'t Hooft lines as transfer matrices}",
    eprint = "2009.12391",
    archivePrefix = "arXiv",
    primaryClass = "hep-th",
    reportNumber = "OU-HET-1067, RIKEN-iTHEMS-Report-20",
    doi = "10.1007/JHEP01(2021)072",
    journal = "JHEP",
    volume = "01",
    pages = "072",
    year = "2021"
}

@article{Nishioka:2014zpa,
    author = "Nishioka, Tatsuma and Yaakov, Itamar",
    title = "{Generalized indices for $ \mathcal{N} $ = 1 theories in four-dimensions}",
    eprint = "1407.8520",
    archivePrefix = "arXiv",
    primaryClass = "hep-th",
    reportNumber = "PUPT-2468",
    doi = "10.1007/JHEP12(2014)150",
    journal = "JHEP",
    volume = "12",
    pages = "150",
    year = "2014"
}

@article{Jejjala:2022lrm,
    author = "Jejjala, Vishnu and Lei, Yang and van Leuven, Sam and Li, Wei",
    title = "{Modular factorization of superconformal indices}",
    eprint = "2210.17551",
    archivePrefix = "arXiv",
    primaryClass = "hep-th",
    doi = "10.1007/JHEP10(2023)105",
    journal = "JHEP",
    volume = "10",
    pages = "105",
    year = "2023"
}

@article{Pittelli:2024ugf,
    author = "Pittelli, Antonio",
    title = "{Orbifold Indices in Four Dimensions}",
    eprint = "2403.12318",
    archivePrefix = "arXiv",
    primaryClass = "hep-th",
    month = "3",
    year = "2024"
}

@article{Aharony:2013hda,
    author = "Aharony, Ofer and Seiberg, Nathan and Tachikawa, Yuji",
    title = "{Reading between the lines of four-dimensional gauge theories}",
    eprint = "1305.0318",
    archivePrefix = "arXiv",
    primaryClass = "hep-th",
    reportNumber = "WIS-03-13-APR-DPPA, WIS/03/13-APR-DPPA, UT-13-15, IPMU13-0081",
    doi = "10.1007/JHEP08(2013)115",
    journal = "JHEP",
    volume = "08",
    pages = "115",
    year = "2013"
}

@article{Amariti:2019but,
    author = "Amariti, Antonio and Marcassoli, Andrea",
    title = "{Lens space index and global properties for 4d $ \mathcal{N} $ = 2 models}",
    eprint = "1911.13264",
    archivePrefix = "arXiv",
    primaryClass = "hep-th",
    doi = "10.1007/JHEP02(2020)143",
    journal = "JHEP",
    volume = "02",
    pages = "143",
    year = "2020"
}

@article{Amariti:2022cxh,
    author = "Amariti, Antonio and Rota, Simone",
    title = "{Symplectic gauge group on the Lens space}",
    eprint = "2210.12240",
    archivePrefix = "arXiv",
    primaryClass = "hep-th",
    doi = "10.1007/JHEP08(2023)137",
    journal = "JHEP",
    volume = "08",
    pages = "137",
    year = "2023"
}

@article{Spiridonov:2019uuw,
    author = "Spiridonov, V. P.",
    title = "{The rarefied elliptic Bailey lemma and the Yang{\textendash}Baxter equation}",
    eprint = "1904.12046",
    archivePrefix = "arXiv",
    primaryClass = "math-ph",
    doi = "10.1088/1751-8121/ab3358",
    journal = "J. Phys. A",
    volume = "52",
    number = "35",
    pages = "355201",
    year = "2019"
}

@article{Kels:2015bda,
    author = "Kels, Andrew P.",
    title = "{New solutions of the star{\textendash}triangle relation with discrete and continuous spin variables}",
    eprint = "1504.07074",
    archivePrefix = "arXiv",
    primaryClass = "math-ph",
    doi = "10.1088/1751-8113/48/43/435201",
    journal = "J. Phys. A",
    volume = "48",
    number = "43",
    pages = "435201",
    year = "2015"
}

@article{Bazhanov:2010kz,
    author = "Bazhanov, Vladimir V. and Sergeev, Sergey M.",
    title = "{A Master solution of the quantum Yang-Baxter equation and classical discrete integrable equations}",
    eprint = "1006.0651",
    archivePrefix = "arXiv",
    primaryClass = "math-ph",
    doi = "10.4310/ATMP.2012.v16.n1.a3",
    journal = "Adv. Theor. Math. Phys.",
    volume = "16",
    number = "1",
    pages = "65--95",
    year = "2012"
}

@article{Spiridonov:2010em,
    author = "Spiridonov, V. P.",
    title = "{Elliptic beta integrals and solvable models of statistical mechanics}",
    eprint = "1011.3798",
    archivePrefix = "arXiv",
    primaryClass = "hep-th",
    journal = "Contemp. Math.",
    volume = "563",
    pages = "181--211",
    year = "2012"
}

@article{Bazhanov:2011mz,
    author = "Bazhanov, Vladimir V. and Sergeev, Sergey M.",
    title = "{Elliptic gamma-function and multi-spin solutions of the Yang-Baxter equation}",
    eprint = "1106.5874",
    archivePrefix = "arXiv",
    primaryClass = "math-ph",
    doi = "10.1016/j.nuclphysb.2011.10.032",
    journal = "Nucl. Phys. B",
    volume = "856",
    pages = "475--496",
    year = "2012"
}

@article{Franco:2015jna,
    author = "Franco, Sebastian and Hayashi, Hirotaka and Uranga, Angel",
    title = "{Charting Class $\mathcal S_k$ Territory}",
    eprint = "1504.05988",
    archivePrefix = "arXiv",
    primaryClass = "hep-th",
    reportNumber = "IFT-UAM-CSIC-15-036",
    doi = "10.1103/PhysRevD.92.045004",
    journal = "Phys. Rev. D",
    volume = "92",
    number = "4",
    pages = "045004",
    year = "2015"
}

@article{Razamat:2016dpl,
    author = "Razamat, Shlomo S. and Vafa, Cumrun and Zafrir, Gabi",
    title = "{4d $ \mathcal{N}=1 $ from 6d (1, 0)}",
    eprint = "1610.09178",
    archivePrefix = "arXiv",
    primaryClass = "hep-th",
    doi = "10.1007/JHEP04(2017)064",
    journal = "JHEP",
    volume = "04",
    pages = "064",
    year = "2017"
}

@article{Bah:2017gph,
    author = "Bah, Ibrahima and Hanany, Amihay and Maruyoshi, Kazunobu and Razamat, Shlomo S. and Tachikawa, Yuji and Zafrir, Gabi",
    title = "{4d $ \mathcal{N}=1 $ from 6d $ \mathcal{N}=\left(1,0\right) $ on a torus with fluxes}",
    eprint = "1702.04740",
    archivePrefix = "arXiv",
    primaryClass = "hep-th",
    reportNumber = "IPMU17-0013, IMPERIAL-TP-17-AH-02",
    doi = "10.1007/JHEP06(2017)022",
    journal = "JHEP",
    volume = "06",
    pages = "022",
    year = "2017"
}

@article{Spiridonov:2008Essays,
    author = "Spiridonov, V. P.",
    title = "{Essays on the theory of elliptic hypergeometric functions}",
    eprint = "0805.3135",
    archivePrefix = "arXiv",
    primaryClass = "math.CA",
    doi = "10.1070/RM2008v063n03ABEH004533",
    journal = "Russ. Math. Surveys",
    volume = "63",
    number = "3",
    pages = "405--472",
    year = "2008"
}

@article{Bulbul:2025ryz,
    author = {B{\"u}lb{\"u}l, Ya{\u{g}}mur and Gahramanov, Ilmar and Yetkin, Ali Mert and Yumu{\c{s}}ak, Reyhan},
    title = "{Lens Hyperbolic Modular Double}",
    eprint = "2511.06400",
    archivePrefix = "arXiv",
    primaryClass = "hep-th",
    month = "11",
    year = "2025"
}

@article{Kulish:1981gi,
    author = "Kulish, P. P. and Reshetikhin, N. Yu. and Sklyanin, E. K.",
    title = "{Yang-Baxter Equation and Representation Theory. 1.}",
    doi = "10.1007/BF02285311",
    journal = "Lett. Math. Phys.",
    volume = "5",
    pages = "393--403",
    year = "1981"
}

@article{Kim:2024mnp,
    author = "Kim, Hee-Cheol and Nedelin, Anton and Razamat, Shlomo S.",
    title = "{On Ruijsenaars-Schneider spectrum from superconformal indices and ramified instantons}",
    eprint = "2407.08776",
    archivePrefix = "arXiv",
    primaryClass = "hep-th",
    doi = "10.1007/JHEP02(2025)010",
    journal = "JHEP",
    volume = "02",
    pages = "010",
    year = "2025"
}

@article{Belavin:1981ix,
    author = "Belavin, A. A.",
    title = "{Dynamical Symmetry of Integrable Quantum Systems}",
    doi = "10.1016/0550-3213(81)90414-4",
    journal = "Nucl. Phys. B",
    volume = "180",
    pages = "189--200",
    year = "1981"
}

@article{Hasegawa:1997RS,
    author = "Hasegawa, Koji",
    title = "{Ruijsenaars' commuting difference operators as commuting transfer matrices}",
    eprint = "q-alg/9512029",
    archivePrefix = "arXiv",
    primaryClass = "q-alg",
    doi = "10.1007/s002200050137",
    journal = "Commun. Math. Phys.",
    volume = "187",
    pages = "289--325",
    year = "1997"
}

@article{Fluder:2017oxm,
    author = "Fluder, Martin and Song, Jaewon",
    title = "{Four-dimensional Lens Space Index from Two-dimensional Chiral Algebra}",
    eprint = "1710.06029",
    archivePrefix = "arXiv",
    primaryClass = "hep-th",
    doi = "10.1007/JHEP07(2018)073",
    journal = "JHEP",
    volume = "07",
    pages = "073",
    year = "2018"
}

@article{Kels:2018xnf,
    author = "Kels, Andrew P. and Yamazaki, Masahito",
    title = "{Lens Generalisation of $\tau$-Functions for the Elliptic Discrete Painlev{\'e} Equation}",
    eprint = "1810.12103",
    archivePrefix = "arXiv",
    primaryClass = "nlin.SI",
    reportNumber = "IPMU-18-0170",
    doi = "10.1093/imrn/rnz063",
    journal = "Int. Math. Res. Not.",
    volume = "2021",
    number = "1",
    pages = "754--765",
    year = "2021"
}

@article{Sarkissian:2018ppc,
    author = "Sarkissian, Gor and Spiridonov, Vyacheslav P.",
    title = "{From rarefied elliptic beta integral to parafermionic star-triangle relation}",
    eprint = "1809.00493",
    archivePrefix = "arXiv",
    primaryClass = "hep-th",
    doi = "10.1007/JHEP10(2018)097",
    journal = "JHEP",
    volume = "10",
    pages = "097",
    year = "2018"
}

@article{Tachikawa:2014mna,
    author = "Tachikawa, Yuji",
    title = "{Magnetic discrete gauge field in the confining vacua and the supersymmetric index}",
    eprint = "1412.2830",
    archivePrefix = "arXiv",
    primaryClass = "hep-th",
    reportNumber = "IPMU-14-0350, UT-14-47",
    doi = "10.1007/JHEP03(2015)035",
    journal = "JHEP",
    volume = "03",
    pages = "035",
    year = "2015"
}

@article{Imamura:2017wdh,
    author = "Imamura, Yosuke",
    title = "{Orbifold Schur Index and IR formula}",
    eprint = "1710.08853",
    archivePrefix = "arXiv",
    primaryClass = "hep-th",
    reportNumber = "TIT-HEP-661, TIT/HEP-661",
    doi = "10.1093/ptep/pty025",
    journal = "PTEP",
    volume = "2018",
    number = "4",
    pages = "043B01",
    year = "2018"
}

@article{Chen:2020jla,
    author = "Chen, Jin and Haghighat, Babak and Kim, Hee-Cheol and Sperling, Marcus",
    title = "{Elliptic quantum curves of class $ {\mathcal{S}}_k $}",
    eprint = "2008.05155",
    archivePrefix = "arXiv",
    primaryClass = "hep-th",
    doi = "10.1007/JHEP03(2021)028",
    journal = "JHEP",
    volume = "03",
    pages = "028",
    year = "2021"
}

@article{Cherednik:1995DAHA,
    author = "Cherednik, Ivan",
    title = "{Double affine Hecke algebras and Macdonald's conjectures}",
    doi = "10.2307/2118632",
    journal = "Ann. Math.",
    volume = "141",
    number = "1",
    pages = "191--216",
    year = "1995"
}

@article{Faddeev:1999fe,
    author = "Faddeev, L. D.",
    title = "{Modular double of quantum group}",
    eprint = "math/9912078",
    archivePrefix = "arXiv",
    primaryClass = "math.QA",
    journal = "Math. Phys. Stud.",
    volume = "21",
    pages = "149--156",
    year = "2000"
}

@article{Takebe:1992pv,
    author = "Takebe, T.",
    title = "{Generalized Bethe Ansatz with the general spin representations of the Sklyanin algebra}",
    doi = "10.1088/0305-4470/25/5/015",
    journal = "J. Phys. A",
    volume = "25",
    pages = "1071--1083",
    year = "1992"
}

@article{Takebe:1995BA,
    author = "Takebe, Takashi",
    title = "{Bethe ansatz for higher spin eight-vertex models}",
    eprint = "q-alg/9504027",
    archivePrefix = "arXiv",
    primaryClass = "q-alg",
    doi = "10.1088/0305-4470/28/23/019",
    journal = "J. Phys. A",
    volume = "28",
    pages = "6675--6706",
    year = "1995"
}

\end{document}